\documentclass[aps,prx,twocolumn,superscriptaddress,longbibliography]{revtex4-2}

\usepackage{graphicx} 
\usepackage{physics} 
\usepackage{listings} 
\usepackage[english]{babel}
\usepackage{amsthm}
\usepackage{amsmath}
\usepackage[colorlinks=true,breaklinks=true,linkcolor=magenta,citecolor=magenta,urlcolor=magenta]{hyperref} 
\usepackage[colorlinks=true]{hyperref} 
\usepackage{placeins} 
\usepackage{enumitem} 

\usepackage{subcaption} 
\usepackage{amsfonts} 
\usepackage{todo}
\usepackage{titlesec}
\usepackage{bbm}

\usepackage{array}
\usepackage{multirow}
\usepackage{makecell}
\usepackage{tabularx}
\usepackage{tikz}
\usetikzlibrary{quantikz}
\usepackage{tikz-cd}
\tikzcdset{scale cd/.style={every label/.append style={scale=#1},
    cells={nodes={scale=#1}}}}
\usepackage{adjustbox}
\usetikzlibrary{arrows}
\tikzset{
operator/.append style={fill=purple!20},
my label/.append style={above right,xshift=0.3cm},
phase label/.append style={label position=above}
}

\newcommand\tikznode[3][]%
   {\tikz[remember picture,baseline=(#2.base)]
      \node[minimum size=0pt,inner sep=0pt,#1](#2){#3};%
   }
\tikzset{>=stealth}

\newcommand*{\gateStyle}[1]{{\textsf{\itshape #1}}}
\newcommand*{\hGate}{\gateStyle{H}}
\newcommand*{\xGate}{\gateStyle{X}}

\newrobustcmd{\B}{\bfseries}

\newcommand*{\circuitH}{\gate[style={fill=teal!20},label style=black]{\textnormal{\hGate{}}}}
\newcommand*{\circuitX}{\gate[style={fill=blue!20},label style=black]{\textnormal{\xGate}}}

\usepackage[title]{appendix}

\usepackage[capitalise,noabbrev]{cleveref}

\makeatletter
\def\l@subsubsection#1#2{}
\makeatother

\begin{document}

\title{Do Quantum Neural Networks have Simplicity Bias?}
\author{Jessica Pointing} \thanks{contact@jessicapointing.com}
\affiliation{Department of Physics, University of Oxford, Oxford, OX1 3PU, United Kingdom}

\begin{abstract}
One hypothesis for the success of deep neural networks (DNNs) is that they are highly expressive, which enables them to be applied to many problems, and they have a strong inductive bias towards solutions that are simple, known as simplicity bias, which allows them to generalise well on unseen data because most real-world data is structured (i.e. simple). In this work, we explore the inductive bias and expressivity of quantum neural networks (QNNs), which gives us a way to compare their performance to those of DNNs. Our results show that it is possible to have simplicity bias with certain QNNs, but we prove that this type of QNN limits the expressivity of the QNN. We also show that it is possible to have QNNs with high expressivity, but they either have no inductive bias or a poor inductive bias and result in a worse generalisation performance compared to DNNs. We demonstrate that an artificial (restricted) inductive bias can be produced by intentionally restricting the expressivity of a QNN. Our results suggest a bias-expressivity tradeoff. Our conclusion is that the QNNs we studied can not generally offer an advantage over DNNs, because these QNNs either have a poor inductive bias or poor expressivity compared to DNNs. 
\end{abstract}

\maketitle

\section{Introduction}

Classical deep neural networks (DNNs) are successful in a wide range of tasks because they are firstly highly expressive, leading to their ability to express the functions they are trying to learn, and secondly they have a good inductive bias, meaning that their inherent bias towards certain types of functions before seeing any data aligns well with the types of functions they are trying to learn. These two properties are key for any learning agent.

Quantum neural networks (QNNs) aim to combine the power of neural network models and quantum mechanics. The first idea integrating neural networks and quantum mechanics was in 1995 published by Kak \cite{kak_quantum_1995}, just one year after Shor published the famous Shor's algorithm for quantum prime factorisation, proving the potential of quantum computing \cite{shor_algorithms_1994}. Since then, there have been numerous proposals of QNNs \cite{schuld_quest_2014}.

In order for a QNN to be able to be a good learning agent, it should also have high expressivity and a good inductive bias. There are numerous papers looking at the expressivity of QNNs \cite{abbas_power_2021,du_expressive_2020,sim_expressibility_2019,wu_expressivity_2021}. Kubler et al. studied the inductive bias of quantum kernels \cite{kubler_inductive_nodate}. This research paper aims to provide further understanding on the inductive bias of QNNs and how it relates to expressivity. Through doing this, we provide insight on how the inductive bias of QNNs compares to the inherent inductive bias of DNNs, which has been shown to be a simplicity bias. This simplicity bias is beneficial for DNNs because real-world data tends to have descriptions that are simple rather than complex. Therefore, it has been suggested DNNs can generalise well on real-world data because the DNN's inductive bias towards simple functions matches the simple descriptions of real-world data \cite{valle-perez_deep_2019}. Understanding the inductive bias of QNNs is crucial as it tells us whether QNNs are a good learning agent and it opens the pathway to seeing where QNNs have a disadvantage or advantage over classical neural networks. In particular, understanding whether QNNs have a simplicity bias can shed light on whether they will perform well on real-world data.

Several works have explored the inductive bias of DNNs by studying their output when given Boolean data \cite{perez_deep_2019}. In this work, we explore the inductive bias of QNNs by studying their output when given Boolean data. We study how the inductive bias changes when using different encoding methods, which describes the way the input data should be encoded into the QNN. QNNs are closely related to kernel methods, which map the input data into a higher dimensional space \cite{schuld_supervised_2021}. Quantum kernels have been used in analysing quantum advantage in quantum machine learning \cite{huang_power_2021} and the inductive bias of quantum kernels has been explored in \cite{kubler_inductive_nodate}. We also use our methods to investigate the inductive bias of quantum kernels and offer comparisons to the corresponding QNNs. 


\subsection{Inductive bias and expressivity}
The inductive bias is the set of assumptions that a learning algorithm makes to predict outputs for unseen inputs. Having some form of an inductive bias is essential to a learning algorithm, as without it, a network wouldn't be able to prefer one solution to choose over another. Learning algorithms that are highly expressive (i.e. can express many functions) need an inductive bias to generalise well. 

In order to quantify the inductive bias in neural networks, one can look at the prior over functions $P(f)$, which is the probability that a neural network, with parameters randomly sampled from a distribution, will produce a particular function before it has been trained on any data.

The function $f$ refers to how the neural network maps the inputs to the outputs. A neural network aims to find the function $f$ that can map the inputs $x$ to outputs $y$, i.e. $y = f(x)$. The expressivity of a neural network refers to how many functions the neural network can express. If the neural network can express more functions, it has a higher chance of finding the true function.

The samples for calculating the probability $P(f)$ are the functions produced after randomly sampling parameters chosen from some parameter initialisation distribution. To see which types of functions the neural network has a bias towards, we choose a way to quantify types of functions. A common metric is the complexity of the function. This will show us whether a neural network has any bias towards high complexity or low complexity (i.e. simple) functions. 

\subsection{Simplicity bias}
Several papers in the deep neural network literature have shown that deep neural networks have a bias towards low complexity (i.e. simple) functions \cite{perez_deep_2019, mingard_deep_2023, mingard_neural_2020, shah_pitfalls_2020, mingard_is_2021, de_palma_random_2019}. Vallé-Perez et al. showed that a DNN's inductive bias towards simple functions is what enables it to generalise well \cite{perez_deep_2019}. They use Boolean functions and random neural networks (untrained neural networks with randomly sampled parameters) to demonstrate practically that DNNs have a bias towards simple Boolean functions. They argue that as real-world datasets are expected to be highly structured (therefore having a lower-complexity), DNNs generalise well on these problems, because of its simplicity bias. 

Real-world data tends to have descriptions that are simple rather than complex \cite{schmidhuber_discovering_1994}, known by the principle of Occam's razor. As DNNs have an inductive bias towards simple functions, they can generalise well on real-world data as the simple functions a DNN finds with higher probabilities are closer to the simple descriptions underlying real-world data compared to complex functions. 

\subsection{Quantum neural networks}
\label{sec:qnn_overview}
\begin{figure}[h]
\centering
\includegraphics[width=0.6\columnwidth]{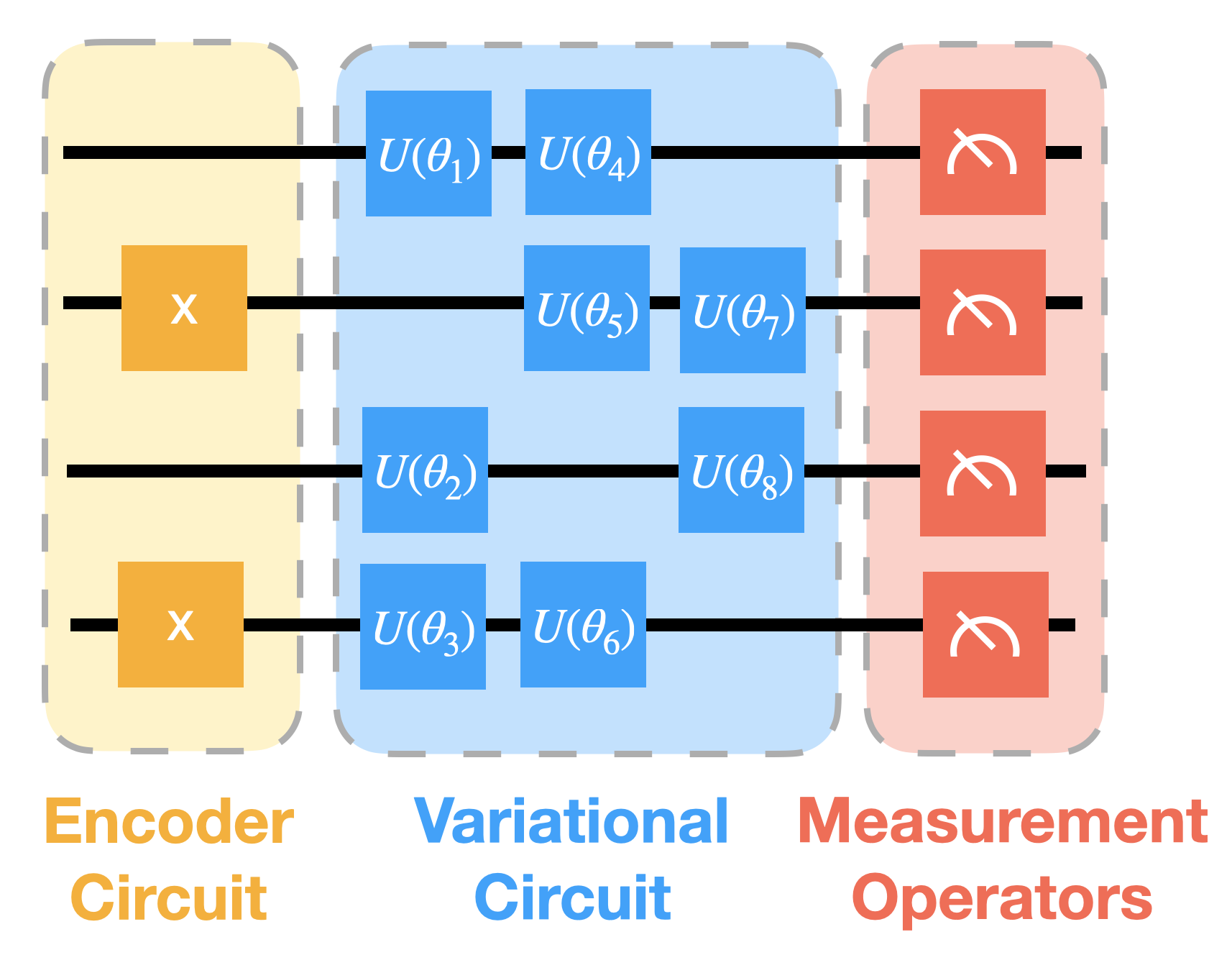}
\caption{\textbf{Quantum neural network with three distinct parts:} the quantum neural network consists of the (1) Encoder circuit, which encodes the data into the QNN (2) Variational circuit, which is parametrised and has its parameters optimised (3) Measurement operators, which retrieve classical information from the QNN.}
\label{fig:qnn_parts}
\end{figure}

We introduce QNNs and their different components, which we use in this work. In a QNN, there are three distinct components \cite{schuld_supervised_2021} as shown in Figure \ref{fig:qnn_parts}:
\begin{enumerate}[noitemsep]
    \item \textbf{Encoder circuit}: this circuit encodes the data into the QNN.
    \item \textbf{Variational circuit}: this circuit is parametrised; the parameters are updated during the learning process of the QNN. The variational circuit starts as an ansatz, which is defined as an educated guess of the solution to a problem and is used as a starting point. In the context of QNNs, the ansatz is the parametrised circuit that is used as a starting point or trial state which is iteratively updated during the optimisation process. 
    \item \textbf{Measurement operators}: information from the QNN is extracted via measurements. From these measurements, gradients and a loss function can be calculated on a classical computer and the parameters are updated via some classical optimiser, such as Adam \cite{kingma_adam_2017}. 
\end{enumerate}

\begin{figure}[h]
\centering
\includegraphics[width=0.8\columnwidth]{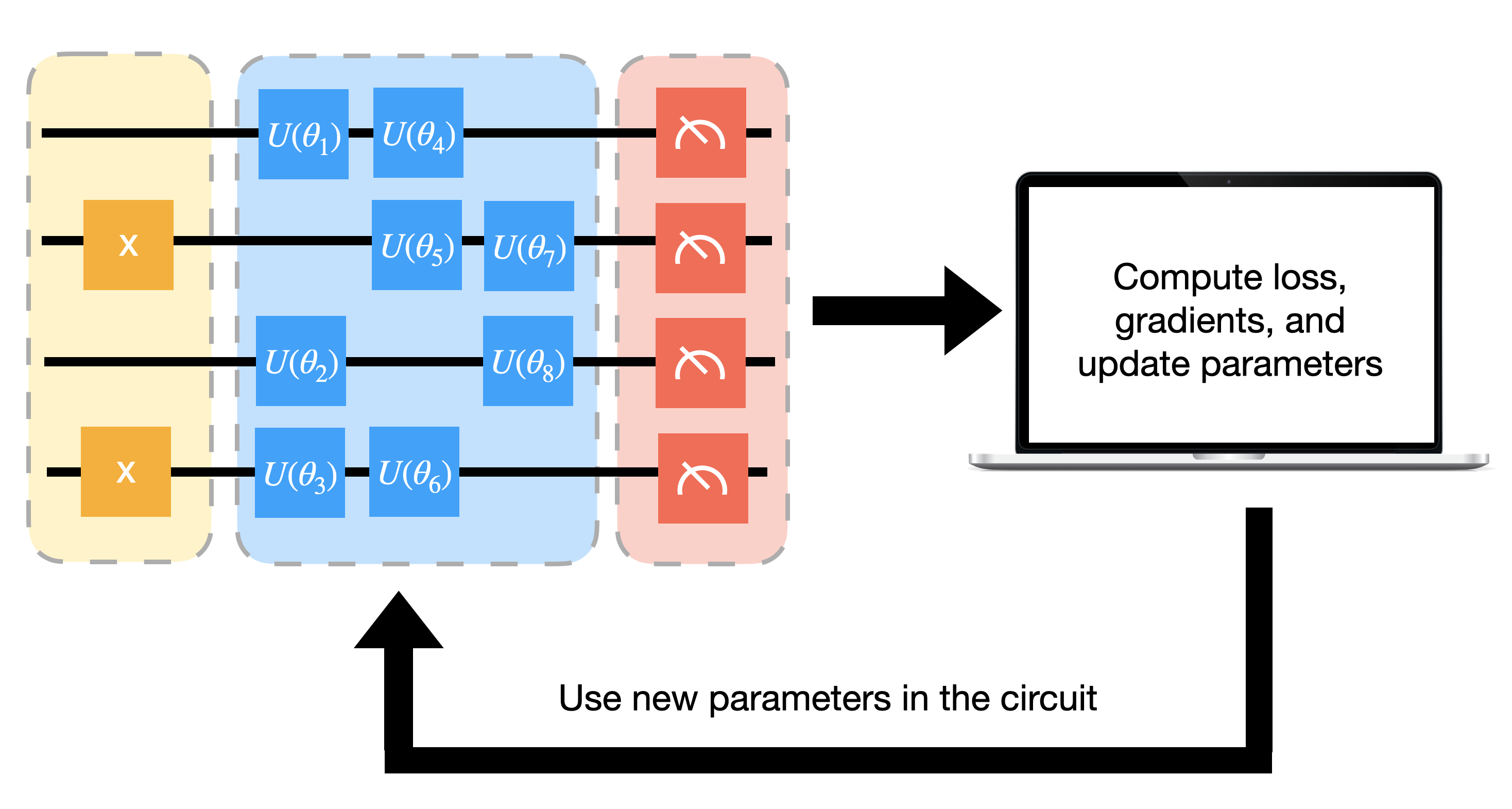}
\caption{\textbf{Learning process for a quantum neural network:} the classical information retrieved from the measurement operators are fed into a classical optimiser to compute the loss, gradients, and update the parameters. These new parameters are updated in the parametrised quantum gates in the variational circuit and the process is repeated until the output of the QNN converges or the process satisfies a stopping condition.}
\label{fig:learning_process_intro}
\end{figure}

During the learning process, the measurement outcomes of the QNN are fed into a classical computer, which makes calculations, including the loss function and gradients. Then the parameters are updated via some classical optimiser. These parameters are updated in the variational circuit in the QNN and the process is repeated until the desired loss is reached. Figure \ref{fig:learning_process_intro} shows this process.

\subsection{Encoding methods}
\label{sec:encoding_methods}
There are multiple ways to encode data into a quantum circuit. We introduce common encoding methods that we analyse in this work:
\paragraph{Basis encoding: }
This maps the input data (in the form of a binary string) into the computational basis of a quantum state. If our vector $\vec{x}$ is in the binary format, then basis encoding makes the following transformation: 
$\vec{x} = b_{n-1} b_{n-2} \cdots b_0 \rightarrow \ket{x} = \ket{b_{n-1} b_{n-2} \cdots b_0}$. For example, basis encoding transforms the input data $\vec{x} = (1,0)$ into $\ket{10}$. This requires $O(n)$ qubits. Further details are in Appendix \ref{sec:basis}.
\paragraph{Amplitude encoding: }
This encodes the data into the amplitudes of the quantum state, so that our quantum state becomes $\ket{\psi} = \frac{1}{\norm{\vec{x}}} \sum_{i=0}^{n-1} x_i \ket{i}$ where our input data is $\vec{x} = x_{n-1} x_{n-2} \cdots x_0$ and $x_i$ is the $i^{th}$ element of $\vec{x}$ and $\ket{i}$ is the $i^{th}$ computational basis state. For example, if our vector is $\vec{x} = (1,0,1)$ then the norm is $\sqrt{1^2 + 0^2 + (1)^2} = \sqrt{2}$, so our encoded quantum state becomes $\frac{1}{\sqrt{2}} \ket{00} + \frac{1}{\sqrt{2}}\ket{10}$. Extra computational basis states can be encoded as zero. This requires $O(logn)$ qubits. Further details are in Appendix \ref{sec:amplitude_encoding}.
\paragraph{ZZ Feature Map: } 
The ZZ Feature Map is based on the paper titled \textit{Supervised learning with quantum-enhanced feature spaces} \cite{havlicek_supervised_2019} and encodes the input data into the angles of the unitary matrices in a particular way. This encoding method is designed to be biased towards the parity function. They define the feature map as $\mathcal{U}_{\Phi}(\vec{x})=U_{\Phi(\vec{x})} H^{\otimes n} U_{\Phi(\vec{x})} H^{\otimes n}$ where
$U_{\phi(x)}=\exp \left(i \sum_{S \subseteq[n]} \phi_S(x) \prod_{i \in S} Z_i\right)$, where $H$ is the Hadamard gate and is equivalent to the matrix $\frac{1}{\sqrt{2}} \begin{psmallmatrix}
    1 & 1 \\
    1 & - 1
\end{psmallmatrix}$, $Z$ is the Pauli-Z gate and is equivalent to the matrix $\frac{1}{\sqrt{2}} \begin{psmallmatrix}
    1 & 0 \\
    0 & - 1
\end{psmallmatrix}$, $S$ is the set of data, $n$ is the number of qubits, $\phi_i(x)=x_i, \phi_{\{i, j\}}(x)=\left(\pi-x_0\right)\left(\pi-x_1\right)$. This feature map can be seen more clearly in its circuit diagram for two qubits in Figure \ref{fig:zz_feature_map}. Further details are in Appendix \ref{sec:zz_feature_map}.

\begin{figure}[h]
    \centering
    \adjustbox{scale=0.7}{%
     \begin{tikzcd}
        \lstick{$q_0$} & \circuitH & \gate{U_1(2x_0)} & \ctrl{1} & \qw & \ctrl{1} & \qw \\
        \lstick{$q_1$} & \circuitH & \gate{U_1(2x_0)} & \targ{} & \gate{U_1(2(\pi-x_0)(\pi-x_1))} & \targ{} & \qw \\
    \end{tikzcd}
    }
\caption{\textbf{ZZ feature map for two qubits:} this feature map is used for encoding data into a QNN. The feature map for two qubits consists of Hadamard gates, followed by $U_1$ gates where the first datapoint $x_0$ is encoded in the parameter for the $U_1$ gate. A CNOT gate follows, then another $U_1$ gate on the second qubit, which encodes both datapoints $x_0$ and $x_1$ in the parameter for the $U_1$ gate. Another CNOT gate follows.}
\label{fig:zz_feature_map}
\end{figure}
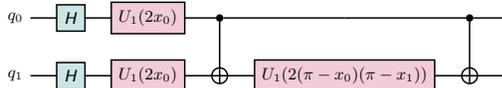
\paragraph{Random Relu Transform:} although the other encoding methods are common, we designed this encoding method to act as a random non-unitary transformation to the data. This is equivalent to the data being transformed through a single hidden layer of a classical neural network. For this encoding method, we encode the data using basis encoding, apply a random unitary matrix, and then apply the relu function, which introduces nonlinearity, to the resulting statevector and normalise it. The transformation is as follows: $\psi=\frac{1}{\|\sigma(Ux)\|} \sum_{i=1}^N \sigma(Ux)_i|i\rangle$ where our input data is $\vec{x} = x_{n-1} x_{n-2} \cdots x_0$ and $x_i$ is the $i^{th}$ element of $\vec{x}$ and $\ket{i}$ is the $i^{th}$ computational basis state, $\sigma(Ux)_i$ is the $i^{th}$ output of a single hidden layer neural network, $\sigma$ is the ReLU function, and $U$ is the random unitary matrix. Further details are in Appendix \ref{sec:random_relu_transform}.
\section{Results}
There are different methods for encoding data into a QNN and our work shows that each method changes the inductive bias of the QNN. Our work sheds light on why some encoding methods perform better than others as we show the effect of the encoding method on the inductive bias. In particular, we look at four major types of encoding methods: Basis encoding, ZZ Feature Map, Amplitude Encoding, and Random Relu encoding. 

To categorise the inductive bias produced by the encoding methods, we choose two ways to quantify the types of functions the QNN produces: entropy and Lempel-Ziv (LZ) complexity. The functions we generate from the QNNs are binary strings. The LZ complexity counts the number of distinct substrings that appear in the string. In effect, this measures how well the string can be compressed by identifying and encoding repeated patterns or substrings. The entropy is min$\{p, n-p\}$ where $p$ is the number of $0$s in the string and $n$ is the length of the string. Measuring the LZ complexity tells us whether a neural network has a bias towards low-complexity functions, known as simplicity bias. Measuring the entropy tells us whether a neural network has a bias towards low-entropy functions, which is a trivial bias because such a neural network is just predicting the output of the neural network (which is either 0 or 1) with the highest frequency. 

Many low-complexity functions are also low-entropy, so if we were just to measure the LZ complexity of the functions, it would seem that many of the encoding methods have a bias towards low-complexity functions, but by measuring entropy as well, we can distinguish the encoding methods that have a bias towards low-complexity and those that have a bias towards low-entropy. 

\subsection{Summary of the bias and expressivity of different encoding methods}

In Table \ref{tab:bias_summary}, we summarise the findings of our results for the inductive bias and expressivity of QNNs with different encoding methods. 

\begingroup \renewcommand{\arraystretch}{3}
\begin{table*}
\begin{tabularx}{\textwidth}{|p{5.8cm} | p{5.8cm} | p{5.85cm}|}
 \hline
    \centering \textbf{Encoding method} & \centering \textbf{Inductive Bias} & \makecell{\textbf{Expressivity (for $n=3$ system)}}\\
 \hline
  \makecell{Basis encoding} &  \makecell{\textbf{Very Poor} \\ No Inductive Bias} & \makecell{\textbf{Good} \\ Full expressivity} \\
  \hline
  \makecell{ZZ feature map} & \makecell{\textbf{Poor} \\ Low-entropy bias \\ (and bias towards parity function)} & \makecell{\textbf{Good} \\ Full expressivity} \\
  \hline
  \makecell{Random relu transform} & \makecell{\textbf{Poor} \\ Low-entropy bias} & \makecell{\textbf{Good} \\ Full expressivity} \\
  \hline
  \makecell{Amplitude encoding} & \makecell{\textbf{Good} \\ Low-complexity bias \\ (i.e. simplicity bias)} & \makecell{\textbf{Poor} \\ Can not express parity function} \\
    \hline
\end{tabularx}
 \caption{Summary of inductive bias and expressivity produced by different encoding methods. A low-entropy bias means that the neural network produces functions with a low entropy with a greater probability than functions with a high entropy. Similarly, a low-complexity bias means that the neural network produces functions with a low Lempel-Ziv (LZ) complexity with a greater probability than functions with a high LZ complexity. \textit{Full expressivity} means that the QNN with that encoding method can express all of the possible Boolean functions. See Appendix \ref{sec:expressivity} for more information about the expressivity of QNNs. For the $n=3$ system, there are $2^{2^3} = 256$ possible Boolean functions so \textit{Full expressivity} means that all 256 of these functions can be expressed. The expressivity of basis and amplitude encoding have been proven (see Appendix \ref{sec:fully_expressive_QNNs} and \ref{sec:proof_expressive} respectively). The expressivity of the ZZ feature map and Random relu transform were verified with $10^7$ and $10^4$ samples respectively from the QNN.}\label{tab:bias_summary}
\end{table*}
\endgroup

Our results show that basis encoding has no inductive bias. This result is also shown by various papers in the literature \cite{kubler_inductive_2021, schuld_supervised_2021}. We also have come up with a proof-by-induction that arrives at the same conclusion. We prove that any arbitrary QNN with basis encoding has no inductive bias. Our proof holds for any QNN and therefore does not depend on the ansatz one is using for the variational circuit. We prove this by induction by decomposing a QNN into layers that depend on previous layers. Details of this proof are in Appendix \ref{sec:proof_qnns_basis_encoding_no_bias}. 

Our experimental results also reveal that other encoding methods can induce various types of inductive bias into the QNN. The ZZ feature map and Random relu transform have a trivial bias, that is a bias towards low-entropy functions. These encoding methods perform poorly (i.e. have high generalisation error) on functions with low complexity but high entropy, showing that the QNN with these encoding methods have a low-entropy bias but not a low-complexity bias. Low complexity, high entropy functions, such as `01' repeated (i.e. $0101 \dots 1010$) are representative of patterns in classical data and therefore this shows a disadvantage on these types of classical boolean data compared to the DNN, which has a bias towards low-complexity functions and can therefore generalise well on simple functions. We show, however, that the QNN with the ZZ feature map has a strong inductive bias towards the parity function, even though the parity function has high complexity and high entropy.

Amplitude encoding has the closest bias to that of a DNN, a simplicity bias. Using this encoding method, however, limits the expressivity of the QNN -- in particular, the parity function can not be expressed for the $n=3$ system and more functions can not be expressed for larger systems. Therefore, there seems to be a bias-expressivity tradeoff. An encoding method with a good inductive bias has lower expressivity, and encoding methods with poor inductive bias have higher expressivity. 

\subsection{Summary of results}
\begin{enumerate}[noitemsep]
    \item Amplitude encoding has a simplicity bias but limits the expressivity of the QNN, which we prove can not express the parity function (see Section \ref{sec:results_inductive_bias_in_qnns} for the inductive bias results and Appendix \ref{sec:proof_expressive} for the proof).
    \item ZZ feature map has a low-entropy bias, which does not match the inherent inductive bias of a DNN (see Section \ref{sec:results_inductive_bias_in_qnns}).
    \item Random relu transform also has a low-entropy bias (see Section \ref{sec:results_inductive_bias_in_qnns}).
    \item QNNs with basis encoding produce no inductive bias on the Boolean system when fully expressive, which we prove by induction (see Appendix \ref{sec:proof_qnns_basis_encoding_no_bias}) and show numerically (see Section \ref{sec:results_inductive_bias_in_qnns}).
    \item QNNs can exhibit an `artificial' inductive bias on the Boolean system when not fully expressive and using basis encoding (see Section \ref{sec:results_inductive_bias_in_qnns}).
    \item Inducing a certain inductive bias into a QNN can negatively affect its ability to train and generalise (see Appendix \ref{sec:training_qnns}). 
    \item A good generalisation performance could be obtained for a QNN by constructing an encoding method that induces an inductive bias towards the function one is learning -- this can be seen for the ZZ feature map which has an inductive bias towards the parity function and low generalisation error (see Section \ref{sec:generalisation_error_qnns_simple} and Appendix \ref{sec:generalisation_error_qnns})
    \item A QNN will have a poor generalisation performance on functions it has a weak bias towards and these functions can be fairly simple functions that a DNN has a strong bias towards (see Section \ref{sec:generalisation_error_qnns_simple} and Appendix \ref{sec:generalisation_error_qnns}). 
    \item The inductive bias of quantum kernels aligns with the inductive bias of QNNs (see Appendix \ref{sec:results_inductive_bias_in_quantum_kernels}).
    \item There is a correlation between the probability of sampling a function from a QNN and its generalisation error. A function that appears more likely in sampling has a lower generalisation error (see Appendix \ref{sec:generalisation_error_qnns}).  
\end{enumerate}

Appendix \ref{sec:appendix_additional_results} provides details on additional experiments and results.

\subsection{Inductive bias of QNNs}
\label{sec:results_inductive_bias_in_qnns}

\label{sec:pf_vs_k_qnn}
We study the inductive bias of QNNs by studying how the QNN interacts with Boolean data. Background information on the Boolean system can be read in Appendix \ref{sec:boolean_system} and on the QNN can be read in Appendix \ref{sec:qnns}. 

\begin{figure*}
     \centering
     \begin{subfigure}[t]{0.3\textwidth}
         \centering
         \includegraphics[width=\textwidth]{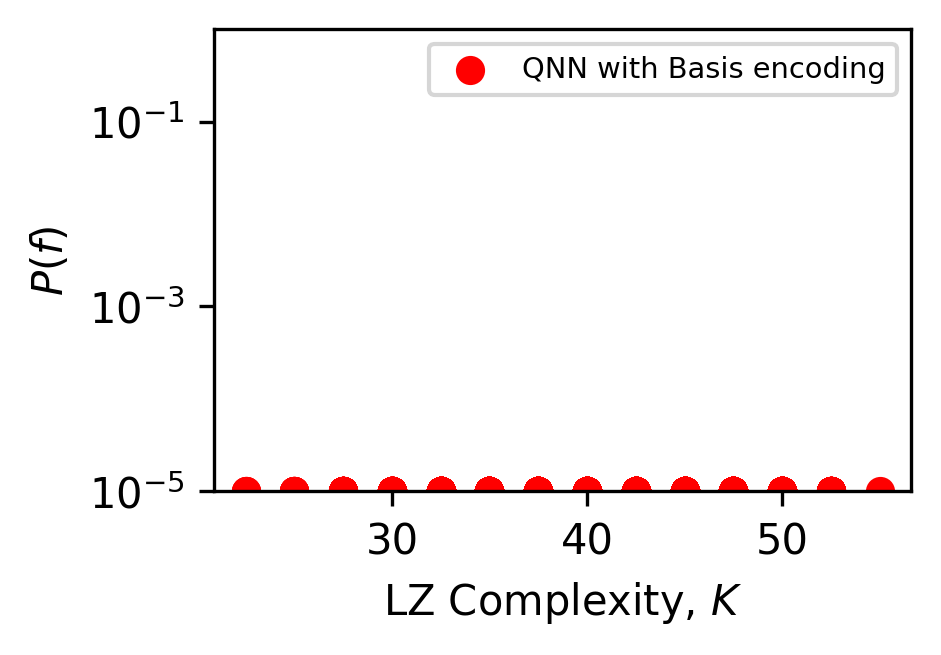}
         \caption{Prior $P(f)$ vs complexity for fully expressive QNN with basis encoding}
         \label{fig:pf_vs_k_e0}
     \end{subfigure}
     \hfill
     \begin{subfigure}[t]{0.3\textwidth}
         \centering
         \includegraphics[width=\textwidth]{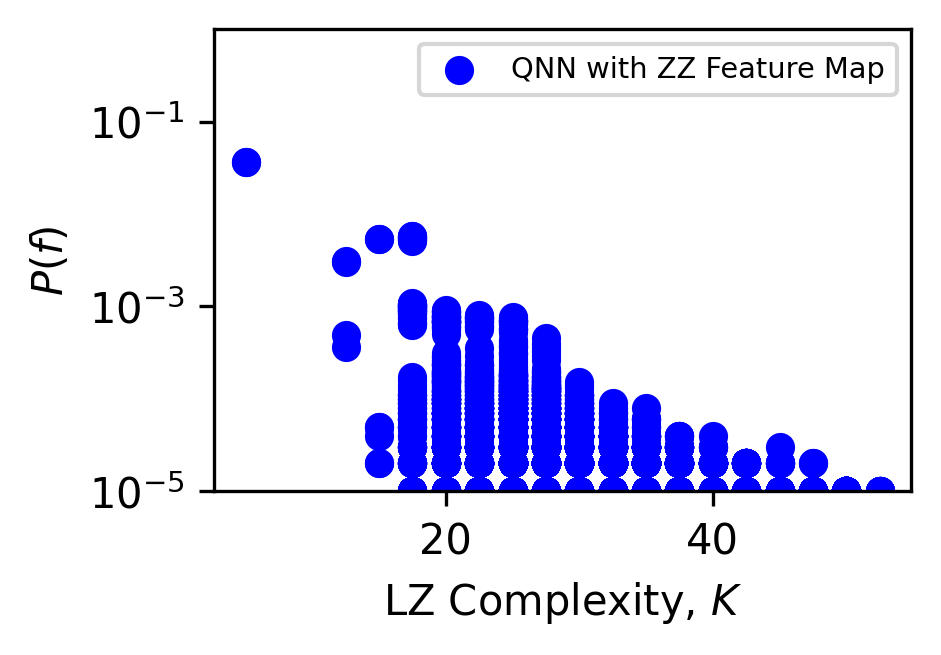}
         \caption{Prior $P(f)$ vs complexity for fully expressive QNN with ZZ Feature Map}
         \label{fig:pf_vs_k_e1}
     \end{subfigure}
     \hfill
     \begin{subfigure}[t]{0.3\textwidth}
         \centering
         \includegraphics[width=\textwidth]{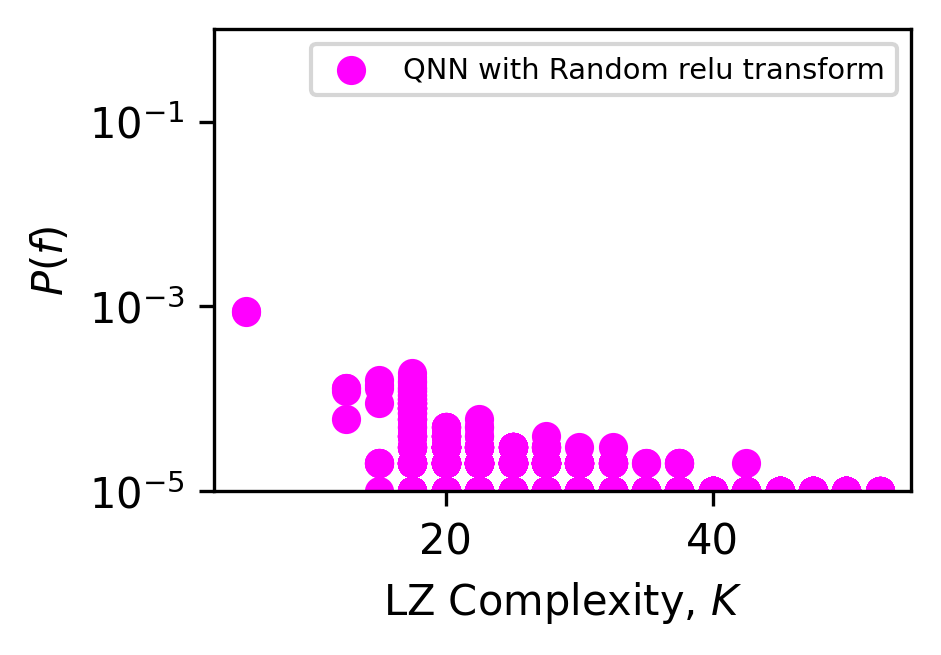}
         \caption{Prior $P(f)$ vs complexity for fully expressive QNN with random relu transform}
         \label{fig:pf_vs_k_e3}
     \end{subfigure}
     \begin{subfigure}[t]{0.3\textwidth}
         \centering
         \includegraphics[width=\textwidth]{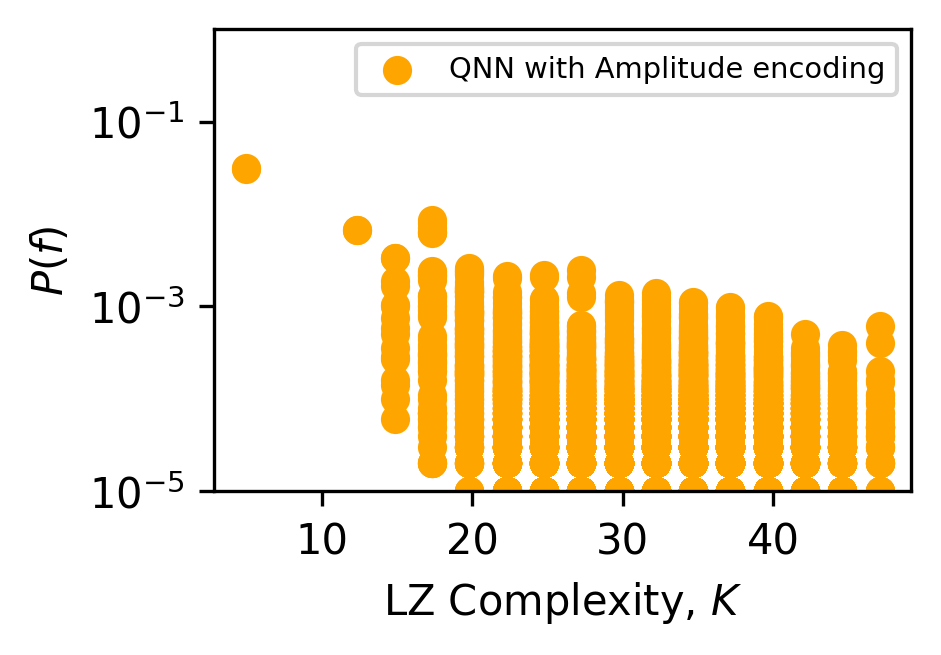}
         \caption{Prior $P(f)$ vs complexity for fully expressive QNN with amplitude encoding}
         \label{fig:pf_vs_k_e4}
     \end{subfigure}
     \hfill
     \begin{subfigure}[t]{0.3\textwidth}
         \centering
         \includegraphics[width=\textwidth]{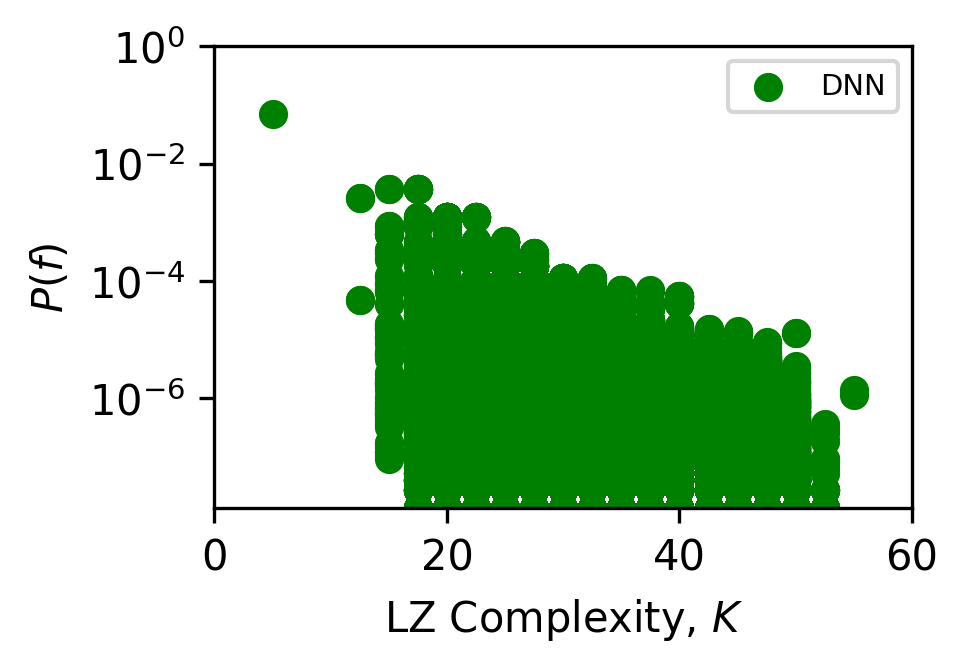}
         \caption{Prior $P(f)$ vs complexity for classical DNN}
         \label{fig:pf_vs_k_dnn}
     \end{subfigure}
     \hfill
     \begin{subfigure}[t]{0.3\textwidth}
         \centering
         \includegraphics[width=\textwidth]{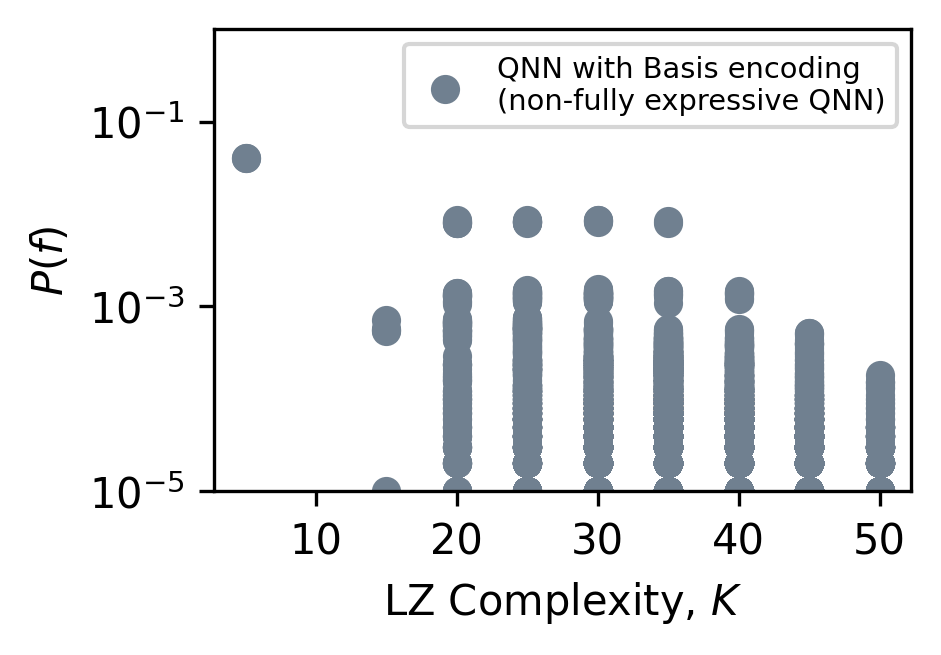}
         \caption{Prior $P(f)$ vs complexity for non-fully expressive QNN with basis encoding}
         \label{fig:pf_vs_k_e0_c21}
     \end{subfigure}
        \caption{\textbf{Probability of a boolean function $\mathbf{(f)}$ versus its complexity for five data qubits and $\mathbf{10^5}$ samples.} $P(f)$ versus Lempel-Ziv complexity, $K$ for a fully expressive QNN with basis encoding (red), ZZ feature map (blue), random relu transform (pink), amplitude encoding (orange), a DNN (green), and a non-fully expressive QNN with basis encoding (grey). $P(f)$ is calculated by generating $10^5$ samples of functions from the QNN by using random samples of parameters $\Theta$ over a uniform distribution.}
        \label{fig:pf_vs_k}
\end{figure*}

We are interested in how the probability of generating a function from an untrained fully-expressive QNN changes as the complexity of that function changes. Additional details about the fully expressive QNN are in Appendix \ref{sec:qnn_details}. In order to generate such data, we encode the Boolean dataset (i.e. the input), via an encoding method, into a fully expressive QNN with randomly sampled parameters and measure the QNN's output. The mapping from the input to output is the Boolean function $f$. We generate many Boolean functions (i.e. the samples) from the QNN by changing the randomly sampled parameters. Then, we can calculate the probability of a particular Boolean function occurring in the samples, which we denote as $P(f)$ in Figure \ref{fig:pf_vs_k}. We measure the complexity of the Boolean function, using the Lempel-Ziv (LZ) complexity measure and plot the probability of a boolean function versus its LZ complexity. Further details about this method can be read in the Methods Section \ref{sec:inductive_bias_in_qnns_method}.

Figure \ref{fig:pf_vs_k} shows $P(f)$ vs $K$ where $K$ is the Lempel-Ziv complexity of the boolean function, for different encoding methods. Figure \ref{fig:pf_vs_k_dnn} is the plot for a classical DNN for comparison. The DNN has a bias towards low-complexity Boolean functions, which is the `simplicity bias' as proposed in \cite{perez_deep_2019}.

We can see in Figure \ref{fig:pf_vs_k_e0} that the probability of generating a particular function does not change with the complexity function, implying that such a QNN is generating boolean functions randomly. This means that encoding the data using basis encoding produces no bias in the system. As a result, using basis encoding results in a poor learning agent, which corresponds with our proof for these numerical results in Appendix \ref{sec:proof_qnns_basis_encoding_no_bias}. 

For the other encoding methods, functions with a lower LZ complexity have a higher probability of occurring, implying a bias towards low-complexity functions. As mentioned earlier in this text, however, many low-complexity functions are also low-entropy functions. Therefore, additional experiments are needed, as shown in Section \ref{sec:generalisation_error_qnns_simple} and Appendix \ref{sec:entropy_vs_complexity_qnns}, to reveal that some of these encoding methods actually have a bias towards low-entropy functions. 

Figure \ref{fig:pf_vs_k_e0_c21} shows the inductive bias in a non-fully expressive QNN with basis encoding. This QNN was proposed in \cite{farhi_classification_2018}, which is described in further detail in Appendix \ref{sec:circuit_for_classification}. We can see that in contrast to the fully expressive quantum circuit with basis encoding, we seem to see a bias towards simple functions. We prove that fully expressive QNNs with basis encoding have no inductive bias in Appendix \ref{sec:proof_qnns_basis_encoding_no_bias}. Therefore, the reason we see a bias in Figure \ref{fig:pf_vs_k_e0_c21} is because the QNN can not express all functions. As the QNN is restricted, it can only produce certain functions which leads to this `artifical' inductive bias. The problem, however, with inducing a bias this way is that the QNN will not be able to learn effectively on functions it can not express, which reduces the generalisation performance of the QNN.

\subsection{Generalisation error of QNNs}
\label{sec:generalisation_error_qnns_simple}

The generalisation error informs us of how good a learning agent the neural network is, as it is a measure of the trained neural network's error on unseen data. A low generalisation error means that the neural network can predict unseen data well. Therefore, looking at the QNN's generalisation error on unseen random boolean functions informs us of how good a learning agent the QNN is and how it compares to a classical DNN. Looking at the generalisation error can also illuminate the inductive bias of the QNNs with different encoding methods.

To study this, we train the QNN with an encoding method to learn a particular boolean function, which is the target function. To bypass barren plateau issues \cite{mcclean_barren_2018} with traditional training methods, which make it difficult to obtain zero error on the training set, we instead train the QNNs by finding a set of parameters that obtain zero error on the training set. To do this, we randomly sample a set of parameters and use them in the QNN and then run the QNN with the training inputs and obtain the predicted $y_{\text{predict}}$ label. We calculate the training error and repeat this process until we obtain a set of parameters where the training error is zero. Once we have obtained that set of parameters (which ensures that that the QNN is trained to zero training error) we use these parameters (i.e. the trained QNN) to calculate the error (i.e. the generalisation error) on the test set. More details about the training method can be read in Section \ref{sec:training_qnns}. 

We find the generalisation error for target functions with different complexities. We have constructed certain target functions so that they have the properties we want to study, in particular functions with low-entropy and low-complexity, low-entropy and high-complexity, high-entropy and low-complexity, and high-entropy and high-complexity. We also generate random target functions. More details about the target functions and further results on the generalisation error can be read in Appendix \ref{sec:generalisation_error_qnns}.

Figure \ref{fig:qnn:posterior:n=5:actual} shows how the generalisation error (i.e. test error) varies by the Lempel-Ziv (LZ) complexity and entropy of the function across different encoding methods. 

For basis encoding, the test error is around 0.5 for all values of complexity and entropy, which is to be expected, as it has no inductive bias and acts as a random learner. A random learner randomly predicts outputs on the test set and has a 50\% chance of obtaining the correct output. Looking at the generalisation error further confirms this inductive bias. 

For amplitude encoding, the test error increases as the LZ complexity of the function increases, which implies an inductive bias towards low-complexity functions. Low-complexity functions have a low test error, as such a QNN is most likely to predict simple functions when given the test set given its bias towards low-complexity functions. There is no straightforward pattern of the test error as the entropy increases, however, implying there is no bias towards low-entropy functions. We see that high-entropy functions can have a low test error. 

For the ZZ feature map and random relu transform, the test error generally increases as the complexity increases, but this pattern is stronger as the entropy increases, implying a bias towards low-entropy functions. For the ZZ feature map, we can see that there are a couple of target functions that have a low-moderate complexity (below 25) but generalise poorly, with a test error around 0.5. Further inspection into these target functions reveal that they have a high entropy, showing that the ZZ feature map does not have a bias towards low-complexity functions when the entropy is high. 

\begin{figure*}
    \centering
\includegraphics[width=\textwidth]{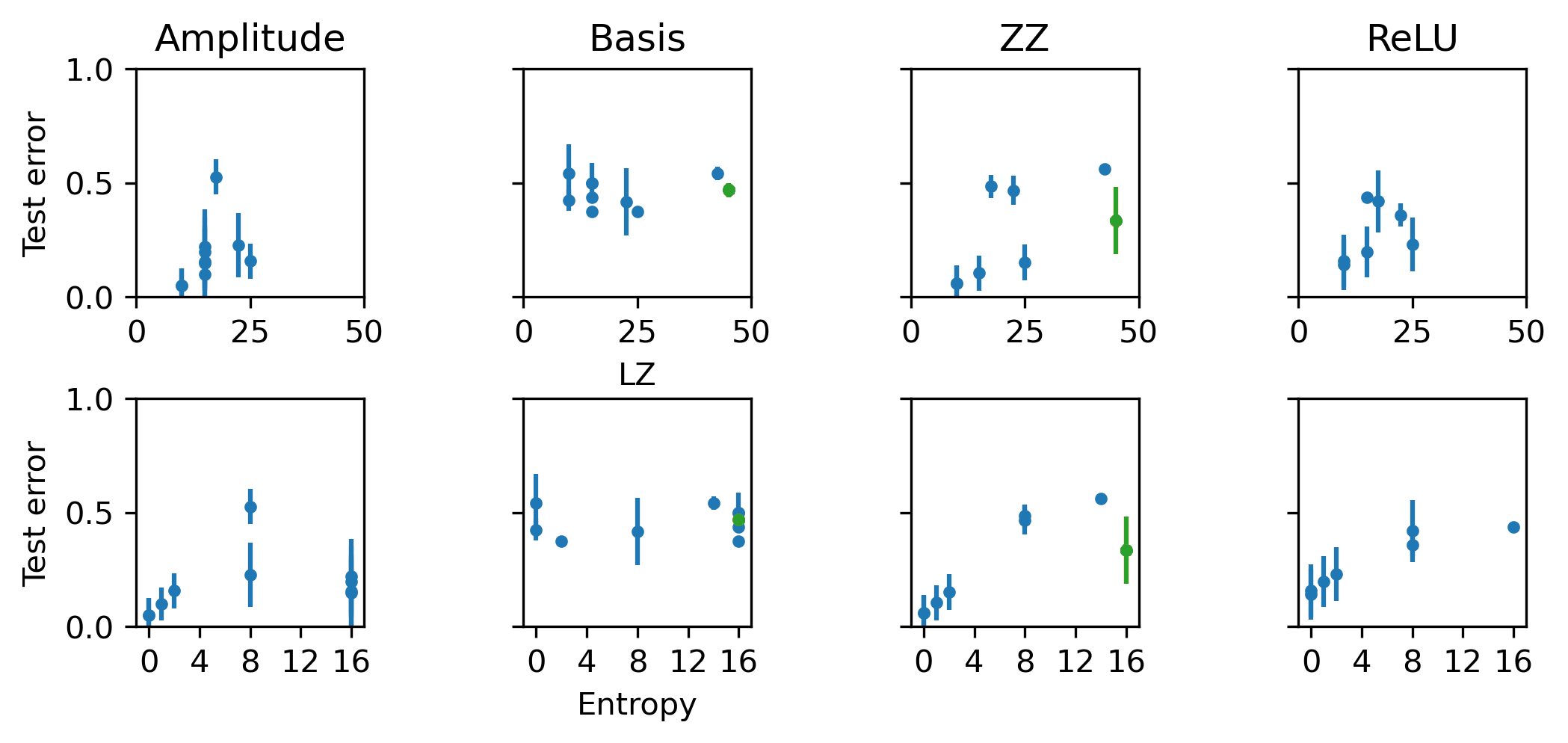}
    \caption{\textbf{Generalisation error versus Lempel-Ziv (LZ) complexity (top figures) and entropy (bottom figures) for five data qubits for a fully expressive QNN with different encoding methods.} The QNN is trained to zero training error on a training set $S$ of size $m = 16$ and the generalisation error is calculated on the remaining $16$ functions. The QNNs are trained by randomly sampling parameters $10^5$ times and finding the generalisation error on the set of parameters that obtain zero training error. Error bars are one standard deviation. The datapoints show the error on the test dataset for different target functions with different LZ complexties and entropies. The maximum entropy is 16, as the boolean functions have length $2^5 = 32$ and entropy is the minimum number of 0s or 1s in the function. The green datpoint is the parity function.}
    \label{fig:qnn:posterior:n=5:actual}

\end{figure*}
\subsection{Expressivity of QNNs}
A stronger inductive bias seems to imply a weaker expressivity and a weaker inductive bias seems to imply a stronger expressivity for the encoding methods we have studied. We have shown that amplitude encoding has a simplicity bias, similar to the bias of the DNN. Amplitude encoding, however, limits the expressivity of the QNN. This can be a problem because we would need to know something about our function to know whether it could be expressed by a QNN with amplitude encoding. 

In 1969, Minksy published a book \textit{Perceptrons} \cite{marvin_perceptrons_1969} that highlighted one of the limitations of the perceptron, which was that it was not fully expressive; it could not express the XOR function, which is the parity function for two inputs. The parity function's value is one if and only if the input vector has an odd number of ones. Similarly, we highlight that the QNN with amplitude encoding has a limitation, which is that it is not fully expressive and can not express the parity function. The details of the proof are in Appendix \ref{sec:proof_expressive}. As the system gets larger, the QNN is unable to express more functions. For the $n=3$ system, the QNN can not express a minimum of two Boolean functions; they are the parity function ($01101001$) and its reverse ($10010110$). For the $n=4$ system, the QNN with amplitude encoding can not express a minimum of 5612 Boolean functions, including the parity function. 

On the other hand, a QNN with basis encoding is fully expressive as is shown in Appendix \ref{sec:qnn_details}. A QNN with the ZZ feature map and random relu transform are also highly expressive, as we verify for the $n=3$ system that they are able to express all of the possible Boolean functions by generating $10^7$ and $10^4$ samples from the QNN respectively. These encoding methods, however, suffer from a weaker inductive bias, which is a low-entropy bias. 

\section{Methods}
\subsection{Finding inductive bias in QNNs}
\label{sec:inductive_bias_in_qnns_method}
In order to find the inductive bias in a QNN, we implement these steps:

\textbf{(a) Generate Boolean data}: We generate all possible Boolean data given $n$ data qubits. There are $2^n$ datapoints. For example, for $n=2$, our input data $\vec{x}$ is $x_0 = (0,0)$, $x_1 = (0,1)$, $x_2 = (1,0)$, $x_3 = (1,1)$. 

\textbf{(b) Create a random set of parameters}: We create a random set of parameters by sampling from the uniform distribution. These set of parameters are used in the variational circuit of the QNN. 

\textbf{(c) Encode input data into the QNN}: We implement an encoder circuit using one of the encoding methods in Section \ref{sec:encoding_methods} to encode the data into the QNN.

\textbf{(d) Construct QNN}: We construct a QNN that consists of the encoder circuit in Step \textbf{(c)} and the variational circuit mentioned in Appendix \ref{sec:qnn_details}. The QNN has $n$ data qubits and 1 readout qubit $q_r$ which is the output of the QNN.

\textbf{(e) Calculate the expectation value of the measurement operator}: We calculate the expectation value $\bra{\psi}\sigma_z\ket{\psi}$ where $\sigma_z = I \otimes \cdots Z_{q_r} \cdots \otimes I$ where $Z$ is applied only to the readout qubit $q_r$ and $I$ is applied to all other qubits. $\ket{\psi} = U(\vec{\theta}_j)\ket{x_i}$ where $U(\vec{\theta_j})$ is the variational circuit with the parameters $\vec{\theta}$ drawn from the uniform distribution for parameter iteration $j$ and $\ket{x_i}$ is the output state of the encoder circuit for the $i^{th}$ data. 

\textbf{(f) Threshold the expectation value}: Given that we are working with the Boolean system, we threshold the expectation value. We will obtain an expectation value $\expval{\sigma_z}$ in the range $[-1,1]$. If $\expval{\sigma_z} < 0$, then the classification of the QNN is $y = 1$. If $\expval{\sigma_z} \geq 0$, $y = 0$.

\textbf{(g) Construct the Boolean function}: We can now construct the Boolean function by repeating Steps \textbf{(c) - (f)} with all of the input data. For each input data, we map it to its output of the QNN to obtain the Boolean function. For example, for $n=2$, if $x_0 = (0,0)$ produces $y_0 = 0$, $x_1 = (0,1)$ produces $y_1 = 1$, $x_2 = (1,0)$ produces $y_2 = 1$, and $x_3 = (1,1)$ produces $y_3 = 0$, then our boolean function = $0110$.

\textbf{(h) Calculate the probabilities of obtaining certain Boolean functions}: We now repeat steps \textbf{(b) - (h)} for different random sets of parameters. If we use $m$ sets of parameters, we produce $m$ datapoints for the boolean functions and we can calculate the probability of a particular boolean function occurring in the $m$ samples.

\subsection{Finding inductive bias in Quantum Kernels}
\label{sec:inductive_bias_in_quantum_kernels_method}
In order to find the inductive bias in a Quantum Kernel, we implement these steps:

\textbf{(a) Construct the Kernel Matrix}: The kernel matrix is $K_{ij} = \left| \langle \phi^\dagger(\vec{x}_j)| \phi(\vec{x}_i) \rangle \right|^{2}$. $\phi(\vec{i})$ is the statevector from the encoder circuit with input data $x_i$. Therefore, we can construct the Kernel matrix by calculating each entry for all pairs of input data.

\textbf{(b) Sample from the Kernel Matrix}: The multivariate normal distribution formula is shown in Equation \ref{eq:multivariate_distribution}
\begin{equation}
f_{\mathbf{X}}\left(x_1, \ldots, x_k\right)=\frac{\exp \left(-\frac{1}{2}(\mathbf{x}-\boldsymbol{\mu})^{\mathrm{T}} \mathbf{K}^{-1}(\mathbf{x}-\boldsymbol{\mu})\right)}{\sqrt{(2 \pi)^k|\mathbf{K}|}} 
\label{eq:multivariate_distribution}
\end{equation}
where $\mathbf{x}$ is the input data, $\mathbf{K}$ is the kernel matrix, and the mean $\boldsymbol{\mu} = 0$. 

\textbf{(c) Threshold the sample} We threshold $f_{\mathbf{X}}$ so that if $f_{\mathbf{X}_i} < 0$, where $i$ is the $i^{th}$ bit in the sample, then the $i^{th}$ bit of the thresholded sample $y$ is $y_i = 1$. If  $f_{\mathbf{X}_i} \geq 0$, then $y_i = 0$.

\textbf{(d) Calculate the probabilities of obtaining certain Boolean functions} We sample $m$ times from the kernel, which allows us to calculate the probability of a boolean function from the $m$ samples, similarly to Step \textbf{(h)} in Section \ref{sec:inductive_bias_in_qnns_method}.

\subsection{Training QNNs}
\label{sec:training_qnns_simple}
We train the QNN to learn a particular boolean function, which is the target function. In order to construct the training and test set, we randomly select half of the indices of the bits in the target function for the training set and the other half for the test set. The training inputs then consist of the Boolean input data at the indices of the bits of the training set. The test inputs consist of the Boolean input data at the indices of the bits of the test set. The training labels are the values of the bits at the indices of the training set in the target function and the test labels are the values of the bits at the indices of the test set in the target function. Further details and an example are in Appendix \ref{sec:training_qnns}. 
\section{Discussion}
The success of deep neural networks (DNNs) has prompted research into what makes DNNs good learning agents. Recently, it has been argued that their natural bias towards simple functions enables them to generalise well on real-world data, which also has simple descriptions \cite{perez_deep_2019}. The aim of this paper was to investigate the inductive bias of QNNs and answer whether they also have simplicity bias, to give us a clue into their performance on real-world data and how they compare to DNNs. 

We have demonstrated that by randomly sampling parameters and seeing which boolean functions are generated by a QNN, we can see which functions are more likely to be produced and therefore which functions the QNN has a biased towards. Unlike DNNs, which naturally have simplicity bias, the inductive bias in QNNs is produced via the encoding method of the QNN. 

Our experimental results, along with a proof-by-induction, show that a fully-expressive QNN with basis encoding produces no inductive bias, which makes it essentially a random learner. An inductive bias can be induced into the circuit by using a different encoding method. Our results have shown that a simplicity bias is possible by using amplitude encoding, but such an encoding method limits the expressivity of the QNN, which is a disadvantage compared to DNNs which are still highly expressive with their natural simplicity bias. If one is using amplitude encoding, then one would need to understand the dataset to understand whether it could be expressed. This requires some manual intervention and is a limitation compared to DNNs that can assumed to express the functions in one's dataset.

Our results also show that it is possible to have encoding methods that do not limit the expressivity of the QNN, such as the ZZ feature map and Random relu transform, but then the inductive bias is poor, as it has an inductive bias towards low-entropy functions. This inductive bias is poor because low complexity, high entropy functions are representative of patterns in classical data and therefore this shows a disadvantage on these types of classical boolean data compared to the DNN. These results imply a bias-expressivity tradeoff. 

Furthermore, an inductive bias can also be induced into the circuit by restricting the expressivity of the circuit -- however, this is an `artificial' inductive bias and can negatively impact the performance of the QNN if a solution function can not be expressed by the QNN. In addition, we show that if an encoding method has a strong bias towards a particular function (as is the case with the ZZ feature map on the parity function), then it can have strong generalisation performance on that function. This, however, requires manual intervention and is somewhat equivalent to `hard-coding' the bias into the QNN. DNNs, however, generally don't need to be `hard-coded' and can be used as they are. Moreover, constructing such encoding methods with a bias towards a particular function leads to a weak bias towards other functions which the DNN has a strong bias towards, even simple functions such as the `$01$' repeated boolean function. 

Overall, we show that there is a fundamental difference between QNNs and DNNs: that DNNs have an inherent inductive bias towards simple functions, whereas the inductive bias of QNNs depends on the encoding method and whilst a simplicity bias can be obtained with amplitude encoding, the expressivity of the QNN becomes limited, implying a bias-expressivity tradeoff. Further work could explore this bias-expressivity tradeoff in the context of QNNs. With the limitations of inductive bias and expressivity in the current framework of QNNs, further work could explore alternative frameworks of QNNs.

\section{Acknowledgements}
I am grateful to Lewis Anderson for suggesting the density matrix transformation integral, which helped me come up with the basis encoding proof in Appendix \ref{sec:proof_qnns_basis_encoding_no_bias}; to Mustafa Bakr for suggesting an improvement to the structure of the paper and for his support throughout the process of publishing the paper; to Nikita Gourianov for suggesting an improvement to the abstract; to Ard Louis for responding with interest after I proposed the idea to him after I read his work on simplicity bias in DNNs; to Christopher Mingard and Ard Louis for various discussions on the work, including, but not limited to, helping me understand more deeply their previous work on simplicity bias in DNNs, discussions on generating data and plots, and helping make the narrative of the bias-expressivity tradeoff clearer.

I also thank Sonia Contera for her unwavering support, the Department of Physics for their support and funding and, finally, the EPSRC Doctoral Training Partnerships for funding. I am grateful for my family and friends for their love and support throughout this process. Most of all, I give thanks to God the Father, the Son Jesus Christ, and the Holy Spirit for all of the gifts He richly blesses us with, including the gift to do this research, write, and publish this work.

\FloatBarrier
\bibliographystyle{IEEEtran}
\bibliography{references} 

\onecolumngrid

\tableofcontents

\appendix

\begin{appendices}
\label{sec:detailed_results}

\section{Proofs}
\subsection{Proof of expressivity of amplitude encoding}
\label{sec:proof_expressive}
In this section, we prove that an arbitrary QNN with amplitude encoding can not express the parity function for $n=3$. 

\textbf{Step 1: Calculate the initial statevector.} The first step is to obtain the initial encoded statevector, which is the statevector of the QNN after the Boolean data has been encoded into the QNN, but before the variational quantum circuit has been applied; this statevector is $\ket{\psi_i}$. 

\textbf{Step 2: Calculate the final statevector.} We wish to obtain $\ket{\psi_f} = U \ket{\psi_i}$ where $\ket{\psi_f}$ is the final statevector after applying the variational quantum circuit $U$ to the initial encoded statevector. For the proof, we choose an arbitrary variational quantum circuit $U$.

\textbf{Step 3: Calculate the expectation value.} We want to find the expectation value, which is $\expval{\sigma_z} = \bra{\psi_f} \sigma_z \ket{\psi_f}$ where $\sigma_z = I \otimes I \otimes Z$ for three qubits, as $Z$ is applied to the readout qubit $q_r$ and the identity matrix $I$ is applied to the data qubits. 

\textbf{Step 4: Constrain the expectation value to match the function. } We now put a constraint on the expectation value, so that it matches the function we are checking that the QNN can express. Our constraint is $\expval{\sigma_z}_i = f_i$ where $f$ is the function and $i$ is the index of the function. If $\expval{\sigma_z}_i \geq 0 $, then $f_i = 0$; if $\expval{\sigma_z}_i < 0 $, then $f_i = 1$. We use the Z3 solver \cite{de_moura_z3_2008} to check the satisfiability of these formulas. 

\subsubsection{Example}
This example shows the calculation of the above steps to prove that an arbitrary QNN with amplitude encoding can not express the parity function for $n=3$. 

\textbf{Step 1: Calculate the initial statevector. } For the $n=3$ boolean dataset, we need two qubits to encode the Boolean data using amplitude encoding. Table \ref{tab:proof_amplitude_encode} shows how we encode the $n=3$ dataset into two qubits and the process for calculating $\ket{\psi_i}$.

\begingroup \renewcommand{\arraystretch}{2}
\begin{table}
\begin{tabularx}{\columnwidth}{ 
  | p{2cm} | p{5cm}| p{4cm} |p{3cm} | p{3.15cm}|}
 \hline
    \textbf{Boolean Data} & \textbf{Process of encoding the data} & \textbf{Normalised encoded state} & \textbf{Statevector of data qubits $\ket{\phi_i}$} & \textbf{Statevector with readout qubit $\ket{\psi_i}$}\\
 \hline
  000 &  N/A & N/A & N/A & N/A \\
  \hline
  001 & $0 \cdot \ket{00} + 0 \cdot \ket{01} + 1 \cdot \ket{10} + 0 \cdot \ket{11}$ & $\ket{10}$ & $\begin{psmallmatrix}
                    0 \\
                    0 \\
                    1 \\ 
                    0 
                \end{psmallmatrix}$ & $\begin{psmallmatrix}
                    0 \\
                    0 \\
                    0 \\
                    0 \\
                    1 \\
                    0 \\
                    0 \\
                    0
                \end{psmallmatrix}$\\
  \hline
  010 & $0 \cdot \ket{00} + 1 \cdot \ket{01} + 0 \cdot \ket{10} + 0 \cdot \ket{11}$ & $\ket{01}$ & $\begin{psmallmatrix}
      0 \\
      1 \\
      0 \\
      0
  \end{psmallmatrix}$ & $\begin{psmallmatrix}
      0 \\
      0 \\
      1 \\
      0 \\
      0 \\
      0 \\
      0 \\
      0
  \end{psmallmatrix}$\\
  \hline
  011 & $0 \cdot \ket{00} + 1 \cdot \ket{01} + 1 \cdot \ket{10} + 0 \cdot \ket{11}$ & $\frac{1}{\sqrt{2}} \ket{01} + \frac{1}{\sqrt{2}} \ket{10}$ & $\frac{1}{\sqrt{2}} \begin{psmallmatrix}
      0 \\
      1 \\
      1 \\
      0
  \end{psmallmatrix}$ & $\frac{1}{\sqrt{2}}\begin{psmallmatrix}
      0 \\
      0 \\
      1 \\
      0 \\
      1 \\
      0 \\
      0 \\
      0
  \end{psmallmatrix}$\\
  \hline
  100 & $1 \cdot \ket{00} + 0 \cdot \ket{01} + 0 \cdot \ket{10} + 0 \cdot \ket{11}$ & $\ket{00}$ & $\begin{psmallmatrix}
      1 \\
      0 \\
      0 \\
      0
  \end{psmallmatrix}$ & $\begin{psmallmatrix}
      1 \\
      0 \\
      0 \\
      0 \\
      0 \\
      0 \\
      0 \\
      0
  \end{psmallmatrix}$ \\
  \hline
  101 & $1 \cdot \ket{00} + 0 \cdot \ket{01} + 1 \cdot \ket{10} + 0 \cdot \ket{11}$ & $\frac{1}{\sqrt{2}} \ket{00} + \frac{1}{\sqrt{2}} \ket{10}$ & $ \frac{1}{\sqrt{2}}\begin{psmallmatrix}
      1 \\
      0 \\
      1 \\
      0
  \end{psmallmatrix}$ & $\frac{1}{\sqrt{2}}\begin{psmallmatrix}
      1 \\
      0 \\
      0 \\
      0 \\
      1 \\
      0 \\
      0 \\
      0
  \end{psmallmatrix}$\\
  \hline
  110 & $1 \cdot \ket{00} + 1 \cdot \ket{01} + 0 \cdot \ket{10} + 0 \cdot \ket{11}$ & $\frac{1}{\sqrt{2}} \ket{00} + \frac{1}{\sqrt{2}} \ket{01}$ & $\frac{1}{\sqrt{2}}\begin{psmallmatrix}
      1 \\
      1 \\
      0 \\
      0
  \end{psmallmatrix}$ & $\frac{1}{\sqrt{2}}\begin{psmallmatrix}
      1 \\
      0 \\
      1 \\
      0 \\
      0 \\
      0 \\
      0 \\
      0
  \end{psmallmatrix}$\\
  \hline
  111 & $1 \cdot \ket{00} + 1 \cdot \ket{01} + 1 \cdot \ket{10} + 0 \cdot \ket{11}$ & $\frac{1}{\sqrt{3}} \ket{00} + \frac{1}{\sqrt{3}} \ket{01} + \frac{1}{\sqrt{3}} \ket{10}$ & $\frac{1}{\sqrt{3}}\begin{psmallmatrix}
      1 \\
      1 \\
      1 \\
      0
  \end{psmallmatrix}$ & $\frac{1}{\sqrt{3}}\begin{psmallmatrix}
      1 \\
      0 \\
      1 \\
      0 \\
      1 \\
      0 \\
      0 \\
      0
  \end{psmallmatrix}$\\
  \hline
\end{tabularx} 
 \caption{Encoding the $n=3$ boolean dataset into two qubits using amplitude encoding. If $\vec{x}$ is the dataset, the first column shows the $x_i$ data elements. $x_0 = 000$ can not be encoded using amplitude encoding so its following columns are Not Applicable (N/A). The second column shows the process for encoding the data into the two qubits. The data is encoded into the two-qubit state as follows: $x_{i_0} \ket{00} + x_{i_1} \ket{01} + x_{i_2} \ket{10} + 0 \ket{11}$ where $x_{i_j}$ is the $j^{th}$ index in the $x_{i}$ data element. The $\ket{11}$ state is encoded with $0$ as it is not needed for the data encoding. The third column shows the simplified, normalised, encoded state. The fourth column is showing the statevector representation of the third column. The fifth column adds the readout qubit $q_r$ to the statevector, which is achieved by applying the tensor product of the statevector of the data qubits $\ket{\phi_i}$ with the readout qubit $\ket{q_r} = \ket{0}$ so that the encoded statevector of the QNN with the readout qubit is $\ket{\psi_i} = \ket{\phi_i} \otimes \ket{q_r}$.}
 \label{tab:proof_amplitude_encode}
\end{table} \endgroup

\textbf{Step 2: Calculate the final statevector}
Equation \ref{eq:step_2_example} provides an example of obtaining $\ket{\psi_f}$, as described in Step 2, for the $x_1 = 001$ boolean data, which we indicate as $\ket{\psi_{f_1}}$.  

\begin{equation}
    \ket{\psi_{f_1}} = U \ket{\psi_{i_1}} = \begin{pmatrix}
    a_1 & a_2 & a_3 & a_4 & a_5 & a_6 & a_7 & a_8 \\
    b_1 & b_2 & b_3 & b_4 & b_5 & b_6 & b_7 & b_8 \\
    c_1 & c_2 & c_3 & c_4 & c_5 & c_6 & c_7 & c_8 \\
    d_1 & d_2 & d_3 & d_4 & d_5 & d_6 & d_7 & d_8 \\
    e_1 & e_2 & e_3 & e_4 & e_5 & e_6 & e_7 & e_8 \\
    f_1 & f_2 & f_3 & f_4 & f_5 & f_6 & f_7 & f_8 \\
    g_1 & g_2 & g_3 & g_4 & g_5 & g_6 & g_7 & g_8 \\
    h_1 & h_2 & h_3 & h_4 & h_5 & h_6 & h_7 & h_8 \\
\end{pmatrix}  \begin{pmatrix}
                    0 \\
                    0 \\
                    0 \\
                    0 \\
                    1 \\
                    0 \\
                    0 \\
                    0
                \end{pmatrix} = \begin{pmatrix}
                    a_5 \\
                    b_5 \\
                    c_5 \\
                    d_5 \\
                    e_5 \\
                    f_5 \\
                    g_5 \\
                    h_5
                \end{pmatrix}
    \label{eq:step_2_example}
\end{equation}

\textbf{Step 3: Calculate the expectation value}. Equation \ref{eq:expectation_value} provides an example for calculating $\sigma_z$ for three qubits and Equation \ref{eq:step_3_example} shows how to use $\sigma_z$ to calculate the expectation value for the $x_1  = 001$ boolean data, which we indicate as $\expval{\sigma_{z_1}}$.

\begin{equation}
    \sigma_z = I \otimes I \otimes Z = \begin{pmatrix}
    1 & 0 \\
    0 & 1 
\end{pmatrix} \otimes \begin{pmatrix}
    1 & 0 \\
    0 & 1 
\end{pmatrix} \otimes \begin{pmatrix}
    1 & 0 \\
    0 & -1
\end{pmatrix} = \begin{pmatrix}
    1 & 0 & 0 & 0 & 0 & 0 & 0 & 0 \\
    0 & -1 & 0 & 0 & 0 & 0 & 0 & 0 \\
    0 & 0 & 1 & 0 & 0 & 0 & 0 & 0 \\
    0 & 0 & 0 & -1 & 0 & 0 & 0 & 0 \\
    0 & 0 & 0 & 0 & 1 & 0 & 0 & 0 \\
    0 & 0 & 0 & 0 & 0 & -1 & 0 & 0 \\
    0 & 0 & 0 & 0 & 0 & 0 & 1 & 0 \\
    0 & 0 & 0 & 0 & 0 & 0 & 0 & -1 \\
\end{pmatrix}
\label{eq:expectation_value}
\end{equation}

\begin{equation}
\begin{aligned}
\expval{\sigma_{z_1}} & = \bra{\psi_{f_1}} \sigma_z \ket{\psi_{f_1}} =
    \begin{pmatrix}
    a_5^{\dagger} & b_5^{\dagger} & c_5^{\dagger} & d_5^{\dagger} & e_5^{\dagger} & f_5^{\dagger} & g_5^{\dagger} & h_5^{\dagger}
\end{pmatrix}
\begin{pmatrix}
    1 & 0 & 0 & 0 & 0 & 0 & 0 & 0 \\
    0 & -1 & 0 & 0 & 0 & 0 & 0 & 0 \\
    0 & 0 & 1 & 0 & 0 & 0 & 0 & 0 \\
    0 & 0 & 0 & -1 & 0 & 0 & 0 & 0 \\
    0 & 0 & 0 & 0 & 1 & 0 & 0 & 0 \\
    0 & 0 & 0 & 0 & 0 & -1 & 0 & 0 \\
    0 & 0 & 0 & 0 & 0 & 0 & 1 & 0 \\
    0 & 0 & 0 & 0 & 0 & 0 & 0 & -1 \\
\end{pmatrix}
\begin{pmatrix}
    a_5 \\
    b_5 \\
    c_5 \\
    d_5 \\
    e_5 \\
    f_5 \\
    g_5 \\
    h_5
\end{pmatrix} \\
& = a_5^{\dagger} a_5 - b_5^{\dagger} b_5 + c_5^{\dagger} c_5 - d_5^{\dagger} d_5 + e_5^{\dagger} e_5 - f_5^{\dagger} f_5 + g_5^{\dagger} g_5 - h_5^{\dagger} h_5
\end{aligned}
\label{eq:step_3_example}
\end{equation}

\textbf{Step 4: Constrain the expectation value to match the function. } We repeat the above steps for the whole Boolean dataset $\vec{x}$ and obtain the expectation value formulas $\expval{\sigma_{z_1}}, \expval{\sigma_{z_2}} \cdots \expval{\sigma_{z_{n-1}}}$ for the dataset. In order to simplify these formulas for the Z3 solver, we substitute some of the values in the formula for new variables. In particular, we substitute $\lambda^{\dagger}_{i} \lambda_{j}$ for $\lambda_{ij}$ where $\lambda$ is the variable (in the case of three qubits, $\lambda \in \{a, b, c, d, e, f, g, h\}$) and $i$ is the index of the variable with the dagger, and $j$ is the index of the variable without the dagger (for three qubits, $i,j \in \{1,2,3,4,5,6,7,8\}$). For example, the formula in Equation \ref{eq:step_3_example} becomes $a_55 - b_55 + c_55 - d_55 + e_55 - f_55 + g_55 - h_55$. Table \ref{tab:constraints} shows these simplified formulas and constraints. 

For $n=3$, there are $2^{2^3} = 256$ possible Boolean functions $f$. Given that amplitude encoding can not encode the $x_0 = 000$ boolean function, there are $2^{{2^3}-1} = 128$ Boolean functions that we check that the QNN with amplitude encoding can express. We express the constraints for these 128 Boolean functions using the Z3 solver and find that the QNN with amplitude encoding can express 126 Boolean functions, except two, which are the parity function $1101001$ and its reverse $0010110$.

\begingroup \renewcommand{\arraystretch}{2}
\begin{table}
\begin{tabularx}{\columnwidth}{| p{1.3cm} | p{13cm}| p{1cm} |p{2cm}|}
 \hline
    \textbf{Boolean Data} & \textbf{Expectation value formula $\expval{\sigma_{z_i}}$} & \textbf{$f_i$} & \textbf{Constraint}\\
 \hline
  000 &  N/A & 0 & N/A \\
  \hline
  001 & $a_{55}-b_{55}+c_{55}-d_{55}+e_{55}-f_{55}+g_{55}-h_{55}$ & 1 & $< 0$ \\
  \hline
  010 & $a_{33}-b_{33}+c_{33}-d_{33}+e_{33}-f_{33}+g_{33}- h_{33}$ & 1 & $< 0$ \\
  \hline
  011 & $\frac{1}{2} (a_{33}+ a_{35}+ a_{53}+ a_{55}- b_{33}- b_{35}- b_{53}- b_{55}+ c_{33}+ c_{35}+ c_{53}+ c_{55}- d_{33}- d_{35}- d_{53}- d_{55} + e_{33}+ e_{35}+ e_{53}+ e_{55}- f_{33}- f_{35}- f_{53}- f_{55}+ g_{33}+ g_{35}+ g_{53}+ g_{55}- h_{33}- h_{35}- h_{53}- h_{55})$ & 0 & $\geq 0$ \\
  \hline
  100 & $ a_{11}- b_{11}+ c_{11}- d_{11}+ e_{11}- f_{11}+ g_{11}- h_{11}$ & 1 & $< 0$ \\
  \hline
  101 & $\frac{1}{2}(a_{11}+ a_{15}+ a_{51}+ a_{55}- b_{11}- b_{15}- b_{51}- b_{55}+ c_{11}+ c_{15}+ c_{51}+ c_{55}- d_{11}- d_{15}- d_{51}- d + e_{11}+ e_{15}+ e_{51}+ e_{55}- f_{11}- f_{15}- f_{51}- f_{55}+ g_{11}+ g_{15}+ g_{51}+ g_{55}- h_{11}- h_{15}- h_{51}- h_{55})$ & 0 & $\geq 0$ \\
  \hline
  110 & $\frac{1}{2}(a_{11}+ a_{13}+ a_{31}+ a_{33}- b_{11}- b_{13}- b_{31}- b_{33}+ c_{11}+ c_{13}+ c_{31}+ c_{33}- d_{11}- d_{13}- d_{31}- d + e_{11}+ e_{13}+ e_{31}+ e_{33}- f_{11}- f_{13}- f_{31}- f_{33}+ g_{11}+ g_{13}+ g_{31}+ g_{33}- h_{11}- h_{13}- h_{31} - h_{33})$ & 0 & $\geq 0$ \\
  \hline
  111 & $\frac{1}{3}(a_{11}+ a_{13}+ a_{15}+ a_{31}+ a_{33}+ a_{35}+ a_{51}+ a_{53}+ a_{55}- b_{11}- b_{13}- b_{15}- b_{31}- b_{33}- b_{35} - b_{51}- b_{53}- b_{55}+ c_{11}+ c_{13}+ c_{15}+ c_{31}+ c_{33}+ c_{35}+ c_{51}+ c_{53}+ c_{55}- d_{11}- d_{13}- d_{15} - d_{31}- d_{33}- d_{35}- d_{51}- d_{53}- d_{55}+ e_{11}+ e_{13}+ e_{15}+ e_{31}+ e_{33}+ e_{35}+ e_{51}+ e_{53}+ e_{55} - f_{11}- f_{13}- f_{15}- f_{31}- f_{33}- f_{35}- f_{51}- f_{53}- f_{55}+ g_{11}+ g_{13}+ g_{15}+ g_{31}+ g_{33}+ g_{35} + g_{51}+ g_{53}+ g_{55}- h_{11}- h_{13}- h_{15}- h_{31}- h_{33}- h_{35}- h_{51}- h_{53}- h_{55})$ & 1 & $< 0$ \\
 \hline
\end{tabularx} 
 \caption{The formulas and constraints used for the satisfiability tests. The first column shows the Boolean dataset. The second column shows the simplified formula for the expectation value $\expval{\sigma_{z_i}}$ for the Boolean data $x_i$. The third column shows the parity function $f$ and the fourth column shows the constraint that needs to be made on the formula in the second column in order to meet the value of the function in the third column. The $x_0 = 000$ data can not be encoded using amplitude encoding and therefore its columns are Not Applicable (N/A).}
 \label{tab:constraints}
 \end{table} \endgroup

We repeat these steps for the $n=4$ system and find the QNN with amplitude encoding can not express a minimum of 5612 Boolean functions, including the parity function.
\FloatBarrier
\subsection{Proof that fully-expressive QNNs with basis encoding have no bias}
\label{sec:proof_qnns_basis_encoding_no_bias}

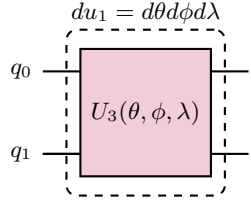
\begin{figure}[h]
    \centering
    \adjustbox{scale=1}{%
     \begin{tikzcd}
        \lstick{$q_0$} & \gate[2]{U_3(\theta, \phi, \lambda)}\gategroup[2,steps=1,style={dashed, rounded corners, inner xsep=2pt}, background]{$du_1 = d\theta d\phi d\lambda$} & \qw \\
        \lstick{$q_1$} & \qw & \qw 
    \end{tikzcd}}
\caption{$u_1$ gate}
\label{fig:u1_gate}
\end{figure}

We prove that any arbitrary fully-expressive QNN with basis encoding has no inductive bias. Our proof holds for any QNN and therefore does not depend on the ansatz one is using for the variational circuit. We prove this by induction by decomposing a QNN into layers that depend on previous layers.
\\
\\
\textbf{Step 1: Base Case for $n = 1$}
\\
$D$ is a channel that transforms a density matrix $\rho \stackrel{D}{\rightarrow} \rho^{\prime}=\sum p_{i} U_{i} \alpha U_{i}^\dagger$ where $p_{i}$ is the probability of picking $U_i$ from some set. We state that $U_i \in U_3$ where 

$$U_3 = \begin{bmatrix} \cos(\theta/2) & -e^{i \lambda} \sin(\theta/2) \\
e^{i\phi} \sin(\theta/2) & e^{i(\phi + \lambda)} \cos(\theta/2) \end{bmatrix}$$ 

which implements the most general single-qubit rotation. We denote the $U_3$ gate as $u_1$ in this proof as the $U_3$ gate acts on one qubit and has three parameters as shown in Figure \ref{fig:u1_gate}. Our set of $U$s are infinite and therefore we integrate over all of the possible angles from $0$ to $2\pi$, so our channel makes the following transformation: $p \rightarrow \rho^{\prime}=\int d\rho U \alpha U^{\dagger}$where $d\rho=d \theta d \varphi d \lambda$ and $\int d \rho=1$.

$\alpha$ is the initial density matrix and its general form can be written as $\alpha = \begin{bmatrix}
    \alpha_1 & \alpha_2 \\
    \alpha_3 & \alpha_4
\end{bmatrix}$

Therefore we calculate:
$$ \begin{aligned}
& \int_{0}^{2 \pi} \int_{0}^{2 \pi} \int_{0}^{2 \pi} u_1 \alpha u_1^{\dagger} d \theta d \varphi d \lambda \\
= & \int_{0}^{2 \pi} \int_{0}^{2 \pi} \int_{0}^{2 \pi}\left[\begin{array}{l}
\cos (\theta / 2),-e^{i \lambda} \sin (\theta / 2) \\
e^{i \phi} \sin (\theta / 2), e^{i(\phi+\lambda)} \cos (\theta / 2)
\end{array}\right] \cdot\left[\begin{array}{ll}
\alpha_{1} & \alpha_{2} \\
\alpha_{3} & \alpha_{4}
\end{array}\right] \cdot\left[\begin{array}{cc}
\cos (\theta / 2), e^{-i \phi}\sin(\theta / 2) \\
-e^{-i \lambda}\sin(\theta / 2), e^{-i(\phi+\lambda)}(\theta / 2)
\end{array}\right] d \theta d \varphi d \lambda \\
= & 4 \pi^{3}\left[\begin{array}{cc}
\alpha_{1}+\alpha_{4} & 0 \\
0 & \alpha_{1}+\alpha_{4}
\end{array}\right]
\end{aligned} $$

Because we are using basis encoding, the initial state can either be $\alpha=|0\rangle\langle 0|=\left(\begin{array}{ll}1 & 0 \\ 0 & 0\end{array}\right)$ or $\left.\alpha=|1\right\rangle\langle 1|=\left(\begin{array}{ll}0 & 0 \\ 0 & 1\end{array}\right)$. Therefore $4 \pi^{3}\left[\begin{array}{cc}\alpha_{1}+\alpha_{4} & 0 \\ 0 & \alpha_{1}+\alpha_{4}\end{array}\right]=4 \pi^{3}\left[\begin{array}{ll}1 & 0 \\ 0 & 1\end{array}\right]$

$\begin{aligned} & \int_{0}^{2 \pi} \int_{0}^{2 \pi} \int_{0}^{2 \pi}  d \theta d \varphi d \lambda = 8\pi^3 \end{aligned}$ so we normalise the matrix to: $\frac{4\pi^3}{8\pi^3} = \begin{bmatrix}
    \frac{1}{2} & 0 \\
    0 & \frac{1}{2}
\end{bmatrix}$

\begin{figure*}[h]
    \centering
    \adjustbox{scale=1}{%
    \adjustbox{scale=1}{%
     \begin{tikzcd}
        \lstick{$q_0$} & \gate[2]{U_2(\theta_1, \cdots \theta_{15} )}\gategroup[2,steps=1,style={dashed, rounded corners, inner xsep=2pt}, background]{$du_2 = d\theta_1 \cdots d\theta_{15}$} & \qw \\
        \lstick{$q_1$} & \qw & \qw 
    \end{tikzcd}}
    $=$ \adjustbox{scale=1}{%
     \begin{tikzcd}
        \lstick{$q_0$} & \gate{U_3(\theta_1,\theta_2,\theta_3)} & \targ{} & \gate{R_z(\theta_7)} & \ctrl{1} & \qw & \targ{} & \gate{U_3(\theta_{10}, \theta_{11}, \theta_{12})} \\
        \lstick{$q_1$} & \gate{U_3(\theta_4,\theta_5,\theta_6)} & \ctrl{-1} & \gate{R_y(\theta_8)} & \targ{} & \gate{R_y(\theta_9)} & \ctrl{-1} & \gate{U_3(\theta_{13}, \theta_{14}, \theta_{15})}
    \end{tikzcd}}
    }
\caption{$u_2$ gate: the circuit for a general two-qubit rotation}
\label{fig:u2_gate}
\end{figure*}
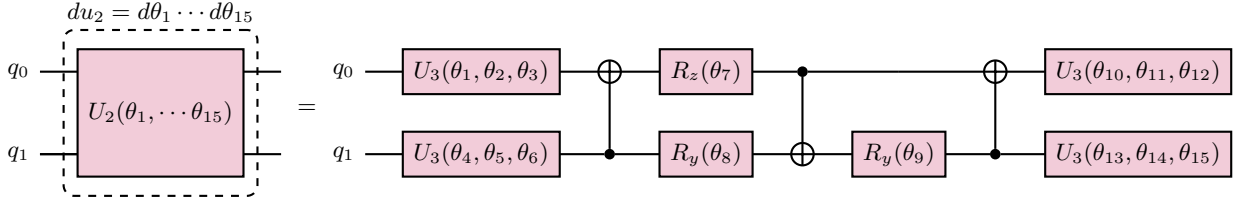

The final density matrix is $\left[\begin{array}{cc}1 / 2 & 0 \\ 0 & 1 / 2\end{array}\right]$ which shows us that the probability of measuring 0 and 1 are the same and therefore there is no bias.

$$
\begin{aligned}
& \text { The base case for $n = 1$ is the following: } \\
& \int_{0}^{2\pi} \int_{0}^{2\pi} \int_{0}^{2\pi} u_1 \alpha u_1^\dagger d \theta d \varphi d \lambda=\left[\begin{array}{cc}
1 / 2 & 0 \\
0 & 1 / 2
\end{array}\right]= I \times \frac{2^{\sigma_{1}} \pi^{\sigma_{1}}}{2^{1}(2 \pi)^{\sigma_{1}}}=I \times \frac{4 \pi^{3}}{8 \pi^{3}}=I \times \frac{1}{2}
\end{aligned}
$$

where $\sigma_1$ is the number of parameters for the circuit on one qubit.
\\
\\
\textbf{Step 2: Base case for $n = 2$}
\\
$D$ is a channel that transforms a density matrix $\rho \stackrel{D}{\rightarrow} \rho^{\prime}=\sum p_{i} U_{i} \alpha U_{i}^\dagger$ where $p_{i}$ is the probability of picking $U_i$ from some set. We state that $U_i \in U_2$ where $U_2$ is a general two-qubit gate. We use the general two-qubit gate proposed in \cite{vatan_optimal_2004} that has 15 parameters. This circuit is shown in Figure \ref{fig:u2_gate}.

We denote the $U_2$ gate as $u_2$ in this proof as the $U_2$ gate acts on two qubits as shown in Figure \ref{fig:u2_gate}. Our set of $U$s are infinite and therefore we integrate over all of the possible angles from $0$ to $2\pi$, so our channel makes the following transformation: $p \rightarrow \rho^{\prime}=\int d\rho U \alpha U^{\dagger}$where $d\rho=d \theta_1 \cdots d \theta_{15}$ (as there are 15 parameters in the $u_2$ gate) and $\int d \rho=1$.

$\alpha$ is the initial density matrix and its general form can be written as $\alpha = \begin{bmatrix}
    \alpha_1 & \alpha_2 & \alpha_3 & \alpha_4 \\
    \alpha_5 & \alpha_6 & \alpha_7 & \alpha_8 \\
    \alpha_9 & \alpha_{10} & \alpha_{11} & \alpha_{12} \\
    \alpha_{13} & \alpha_{14} & \alpha_{15} & \alpha_{16} \\
\end{bmatrix}$.

Therefore we calculate:
$$ \begin{aligned}
& \underbrace{\int_{0}^{2 \pi} \cdots \int_{0}^{2 \pi}}_{15} u_2 \alpha u_2^{\dagger} d \theta_1 \cdots d \theta_{15} \\
= & \underbrace{\int_{0}^{2 \pi} \cdots \int_{0}^{2 \pi}}_{15}
\underbrace{(U_{3} \otimes U_{3}) \cdot (CX_{1,0}) \cdot (I \otimes R_{y}) \cdot (CX_{0,1}) \cdot (R_{z} \otimes R_{y}) \cdot (CX_{1,0}) \cdot (U_{3} \otimes U_{3})}_{u_2} \cdot \alpha \cdot u_2^{\dagger} d\theta_1 \cdots d\theta_{15} \\
= & 8192 \cdot \pi^{15} \begin{bmatrix}
    1 & 0 & 0 & 0 \\
    0 & 1 & 0 & 0 \\
    0 & 0 & 1 & 0 \\
    0 & 0 & 0 & 1    
\end{bmatrix}
\end{aligned} $$

where $CX_{x,y}$ is the CNOT gate where the control qubit is $x$ and the target qubit is $y$.

Because we are using basis encoding, the initial state can either be:

$$\ket{00}\bra{00} = \begin{pmatrix}
    1 & 0 & 0 & 0 \\
    0 & 0 & 0 & 0 \\
    0 & 0 & 0 & 0 \\
    0 & 0 & 0 & 0
\end{pmatrix}, \ket{01}\bra{01} = \begin{pmatrix}
    0 & 0 & 0 & 0 \\
    0 & 1 & 0 & 0 \\
    0 & 0 & 0 & 0 \\
    0 & 0 & 0 & 0
\end{pmatrix}, \ket{10}\bra{10} = \begin{pmatrix}
    0 & 0 & 0 & 0 \\
    0 & 0 & 0 & 0 \\
    0 & 0 & 1 & 0 \\
    0 & 0 & 0 & 0
\end{pmatrix}, \ket{11}\bra{11} = \begin{pmatrix}
    0 & 0 & 0 & 0 \\
    0 & 0 & 0 & 0 \\
    0 & 0 & 0 & 0 \\
    0 & 0 & 0 & 1
\end{pmatrix}$$

As $\begin{aligned} & \underbrace{\int_{0}^{2 \pi} \cdots \int_{0}^{2 \pi}}_{15}  d \theta_1 \cdots d \theta_{15} = (2^{15} \cdot \pi^{15}) \end{aligned}$, we normalise the matrix to: $\frac{8192\pi^{15}}{2^{15}\cdot\pi^{15}} \begin{bmatrix}
    1 & 0 & 0 & 0 \\
    0 & 1 & 0 & 0 \\
    0 & 0 & 1 & 0 \\
    0 & 0 & 0 & 1
\end{bmatrix}= \begin{bmatrix}
    \frac{1}{4} & 0 & 0 & 0\\
    0 & \frac{1}{4} & 0 & 0 \\
    0 & 0 & \frac{1}{4} & 0 \\
    0 & 0 & 0 & \frac{1}{4}
\end{bmatrix}$

The final density matrix is $\begin{bsmallmatrix}
    \frac{1}{4} & 0 & 0 & 0\\
    0 & \frac{1}{4} & 0 & 0 \\
    0 & 0 & \frac{1}{4} & 0 \\
    0 & 0 & 0 & \frac{1}{4}
\end{bsmallmatrix}$ which shows us that the probability of measuring the states $\ket{00}, \ket{01}, \ket{10}, \ket{11}$ are equal and therefore there is no bias.

$$
\begin{aligned}
& \text { The base case for $n=2$ is} \\
& \underbrace{\int_{0}^{2\pi} \cdots \int_{0}^{2\pi}}_{15} u_2 \alpha u_2^\dagger d \theta_1 \cdots d \theta_{15} =\begin{bmatrix}
    \frac{1}{4} & 0 & 0 & 0\\
    0 & \frac{1}{4} & 0 & 0 \\
    0 & 0 & \frac{1}{4} & 0 \\
    0 & 0 & 0 & \frac{1}{4}
\end{bmatrix} = I \times \frac{2^{\sigma_{2}} \pi^{\sigma_{2}}}{2^{2}(2 \pi)^{\sigma_{2}}}=I \times \frac{1}{4}
\end{aligned}
$$

where $\sigma_2$ is the number of parameters for the circuit on two qubits.
\\
\\
\textbf{Step 3: Assume $n=k$ holds}

Assuming $n=k$ holds, then:

$$
\begin{aligned}
& \int \underbrace{u_{k-1}^{\prime} u_{2} u_{k-1}}_{u_{k-1}} \alpha \underbrace{u_{k-1}^{\dagger} u_{2}^{\dagger} u_{k-1}^{\dagger}}_{u_{k}} \underbrace{d u_{k-1}^{\prime} d u_{2} d u_{k-1}}_{d u_{k}} \\
= & \int u_{k} \alpha u_{k}^{\dagger} d u_{k}=I^{\otimes k} \cdot \frac{(2 \pi)^{\sigma_k}}{2^k}
\end{aligned}
$$

and the result is normalised by $(2 \pi)^{\sigma_{k}}$ where $\sigma_{k}=2 \sigma_{k-1}+p$ is the number of parameters for a circuit on $n=k$ qubits where $p = 15$.

\textbf{Step 4: Show $n=k+1$ holds}

\begin{figure*}[h]
$$\adjustbox{scale=1}{%
     \begin{tikzcd}
        \lstick{$q_0$} & \gate[4]{SU(N)}\gategroup[4,steps=1,style={dashed, rounded corners, inner xsep=2pt}, background,label style={label position=below,anchor=
    north,yshift=-0.2cm}]{$dU_N$} & \qw \\
        \lstick{$q_1$} & \qw & \qw \\[0.2cm]
        \lstick{\vdots} & \qw & \qw \\
        \lstick{$q_N$} & \qw & \qw 
    \end{tikzcd}
}  \longleftrightarrow
\adjustbox{scale=1}{%
        \begin{tikzcd}
        \lstick{$q_0$} & \qw & \gate[2]{U_2(\theta, \phi, \lambda)}\gategroup[2,steps=1,style={dashed, rounded corners, inner xsep=2pt}, background]{$dU_2 = d\theta d\phi d\lambda$} & \qw & \qw \\
        \lstick{$q_1$} & \gate[3]{SU(N-1)}\gategroup[3,steps=1,style={dashed, rounded corners, inner xsep=2pt}, background,label style={label position=below,anchor=
    north,yshift=-0.2cm}]{$dU_{N-1}$} & \qw & \gate[3]{SU^{\prime}(N-1)}\gategroup[3,steps=1,style={dashed, rounded corners, inner xsep=2pt}, background,label style={label position=below,anchor=
    north,yshift=-0.2cm}]{$dU^{\prime}_{N-1}$}\\[0.2cm]
        \lstick{\vdots} & \qw & \qw & \qw & \qw \\
        \lstick{$q_N$} & \qw & \qw & \qw & \qw
    \end{tikzcd}
}$$
\caption{$U_N$ gate}
\label{fig:un_gate}
\end{figure*}
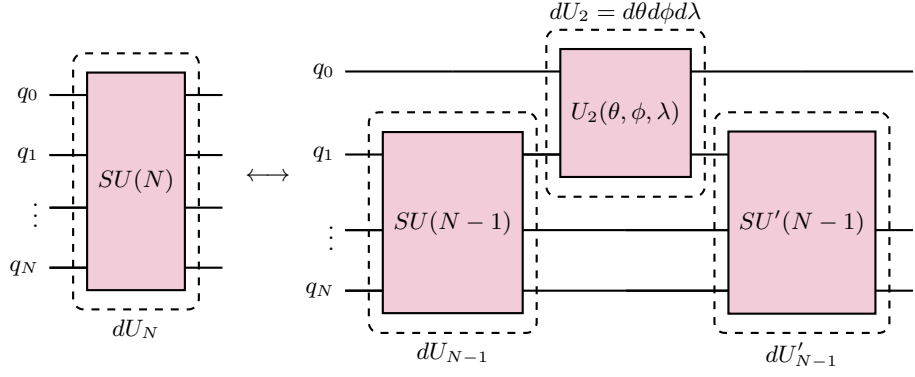

We decompose a QNN acting on $N$ qubits into layers that act on $N-1$ qubits and $2$ qubits \cite{matteo_understanding_2021, de_guise_simple_2018} using the $u_2$ gate as shown in Figure \ref{fig:un_gate}. We use this structure for the inductive step. 

$$
\begin{aligned}
& \text { we want to show: } \int u_{k+1} \alpha u_{k+1}^{\dagger}d u_{k+2} = I^{\otimes k+1} \cdot 2^{\sigma_{k+1}} \frac{\pi^{\sigma_{k+1}}}{2^{k+1}} = 2\sigma_k + p \\
\end{aligned}
$$

$$
\begin{aligned}
    &= \int u_{k}^{\prime} u_{2} u_{k} \alpha u_{k}^{\dagger} u_{2}^{\dagger} u_{k}^{\prime \dagger}d u_{k+1} \\
    &=\int u_{k}^{\prime} u_{2} u_{k} \alpha u_{k}^{\dagger} u_{2}^{\dagger} u_{k}^{\prime \dagger}d u_{k}^{\prime} d\theta d\phi d\lambda du_{k} \\
    &= \frac{(2\pi)^{\sigma_{k}}}{2^{k}} \cdot \frac{(2\pi)^{p}}{2^{2}} \int\left(I \otimes u_{k}^{\prime}\right)\left(I \otimes I^{k}\right)\left(I \otimes u_{k}^{\prime}\right) d u_{k} \\
    & = \frac{(2\pi)^{\sigma_{k}}}{2^{k}} \cdot \frac{(2\pi)^{p}}{2^{2}} \int I \otimes u_{k}^{\prime} I^{k} u_{k}{ }^{\prime} d u_{k} \\
    & = \frac{(2\pi)^{\sigma_{k}}}{2^{k}} \cdot \frac{(2\pi)^{p}}{2^{2}} \int I \otimes I^{k} u_{k}^{\prime} u_{k}^{\dagger \prime} d u_{k}^{\prime}=\int I \otimes I^{k} \cdot I d u_{k}{ }^{\prime}=\int I^{\otimes k+1} d u_{k}{ }^{\prime} \\
    & =\frac{(2 \pi)^{\sigma+k}}{2^{k}} \frac{(2 \pi)^{p}}{2^{2}} \frac{(2 \pi)^{\sigma_{k}}}{2^{k}} I^{\otimes k+1}=\frac{(2 \pi)^{2_{\sigma_k}+p}}{2^{k+1}} I^{\otimes k+1}
\end{aligned}
$$

This concludes the proof that the density matrix of an arbitrary circuit with parameters sampled uniformly is equivalent to $$\int u_{n} \alpha u_{n}^{\dagger} d u_{n}=I^{\otimes n} \cdot 2^{(\sigma_{n}-n)} \pi^{\sigma_{n}}=D$$

$D$ is then normalised by $(2 \pi)^{\sigma_{n}}$ so that $$w=I^{\otimes n} \frac{(2 \pi)^{\sigma_{n}}}{2^{n}(2 \pi)^{\sigma_{n}}}=I^{\otimes n} \cdot 2^{-n}$$ which shows that the outputs have equal probability and therefore no bias.

\newpage
\section{Additional Details}
\subsection{Quantum Neural Networks}
\label{sec:qnn_details}
An overview of Quantum Neural Networks (QNNs) is given in Section \ref{sec:qnn_overview}. As described in that Section, a QNN consists of three parts: the encoding circuit, the variational circuit, and the measurement operators. The encoding methods for the encoding circuit are described in Section \ref{sec:encoding_methods}. In this section, we review the variational circuit and provide details on the fully-expressive and non-fully expressive QNNs used to analyse the inductive bias. An overview of the measurement operators and learning process is given in Appendix \ref{sec:qnns}.

\subsubsection{Expressivity of variational circuits}
\label{sec:expressivity}

The expressivity of a quantum variational circuit is typically defined as what states in the Hilbert space the variational quantum circuit can express. 

In this work, we study the Boolean system (further details are provided in Appendix \ref{sec:boolean_system}) and therefore we define a different type of expressivity, which we call the \textit{Boolean expressivity} -- which is defined as the Boolean functions the quantum variational circuit can express. The Boolean expressivity is less strict than the Hilbert space expressivity, meaning that a variational quantum circuit can have maximum Boolean expressivity (can represent all Boolean functions) but not maximum Hilbert space expressivity (can represent all possible quantum states). If a quantum circuit has maximum Hilbert space expressivity, it must also have maximum Boolean expressivity. If a quantum circuit has maximum Hilbert space expressivity, meaning that it can express all possible quantum states, then it can also express all possible Boolean functions as the Boolean functions are derived by thresholding the outcomes of quantum states.

The quantum circuit for classification in Figure \ref{fig:variational_quantum_circuit} has neither maximum Hilbert space expressivity nor maximum Boolean expressivity. This can pose a problem when studying the Boolean system as in order to understand the QNN's true inherent bias, it should be able to express all possible Boolean functions, otherwise any bias would be a byproduct of its lack of expressivity and not from the system itself. Therefore, we introduce in the next section a quantum variational circuit that is fully expressive in Boolean functions.

\subsubsection{Fully Boolean expressive QNNs}
\label{sec:fully_expressive_QNNs}
In order to construct a QNN that can learn all Boolean functions, which we call a fully expressive QNN, we use the QNN presented in the paper titled \textit{Tunable Quantum Neural Networks for Boolean Functions} \cite{ngoc_tunable_2020}, where they show that a quantum circuit can be designed to express a particular Boolean function. As a result, a general quantum circuit can have parameters that can be modified to learn any Boolean function. 

\textbf{(a) Quantum Boolean Circuit}

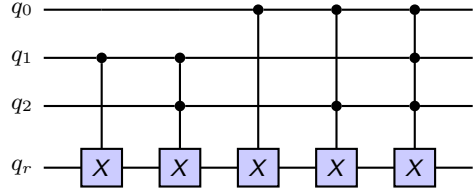
\begin{figure}[h]
    \centering
    \begin{tikzcd}
        \lstick{$q_0$} & \qw      & \qw        & \ctrl{3} & \ctrl{3}   & \ctrl{3}   & \qw \\
        \lstick{$q_1$} & \ctrl{2} & \ctrl{2}   & \qw      & \qw        & \control{} & \qw \\
        \lstick{$q_2$} & \qw      & \control{} & \qw      & \control{} & \control{} & \qw \\
        \lstick{$q_r$} & \circuitX & \circuitX   & \circuitX & \circuitX   & \circuitX   & \qw \\
    \end{tikzcd}
    \caption{\textbf{Quantum Circuit for expressing the Boolean function 00101001}: the Boolean function in Table \ref{tab:quantum_boolean_function} can be expressed as this quantum circuit. The circuit uses multi-controlled-X gates to express the behaviour of the Boolean function.}
    \label{fig:quantum_boolean_circuit}
\end{figure}

An example of a Boolean function is shown in Table \ref{tab:quantum_boolean_function} and Figure \ref{fig:quantum_boolean_circuit} shows the corresponding quantum circuit to express this Boolean function. In general, a quantum Boolean circuit has $n$ data qubits, corresponding to the length of the input data $x$ and one readout qubit $q_r$ which corresponds to $f(x)$. The quantum Boolean circuit consists of multi-controlled X gates where the qubits with the symbol \begin{tikzcd} \control{} \end{tikzcd} are the control qubits and the qubits with the symbol \begin{tikzcd} \gate{X} \end{tikzcd} are the target qubits. If a controlled qubit is equivalent to the state $\ket{0}$, then nothing happens to the target qubit, whereas if all controlled qubits are equivalent to the state $\ket{1}$, the $X$ gate is applied to the target qubit, effectively flipping the target qubit. 

\begin{table}
\begin{tabularx}{0.2\textwidth}{ 
  | >{\centering\arraybackslash}X 
  | >{\centering\arraybackslash}X 
  | }
 \hline
    $x$ & $f(x)$ \\
 \hline
  000 & 0 \\
  \hline
  001 & 0 \\
  \hline
  010 & 1 \\
  \hline
  011 & 0 \\
  \hline
  100 & 1 \\
  \hline
  101 & 0 \\
  \hline
  110 & 0 \\
  \hline
  111 & 1 \\
    \hline
\end{tabularx}
 \caption{\textbf{Example of a Boolean function}: $x$ consists of all the possible combinations of a binary string with length three. Each input $x_i$ is mapped to an output $f(x_i)$, which is the Boolean function $00101001$.}\label{tab:quantum_boolean_function}
\end{table}

As examples, one can see that if all the qubits are equivalent to $\ket{0}$ then the readout qubit will also remain in the $\ket{0}$ state -- this corresponds to the first row in Table \ref{tab:quantum_boolean_function} where $x = 000$ and $f(x) = 0$. Then, we can see that if the first qubit is in the state $\ket{0}$, the second qubit in the state $\ket{1}$ and the third qubit in the state $\ket{0}$, then this corresponds to the first gate in the quantum circuit, which has a control on the second qubit and target on the readout qubit. The readout qubit will be flipped to the $\ket{1}$ state and this is equivalent to the third row in the Table \ref{tab:quantum_boolean_function} where $x = 010$ and $f(x) = 1$. Finally, if all the qubits are equivalent to $\ket{1}$, then the readout qubit will be flipped and be in the state $\ket{1}$ corresponding to the last row in Table \ref{tab:quantum_boolean_function} where $x = 111$ and $f(x)  = 1$. 

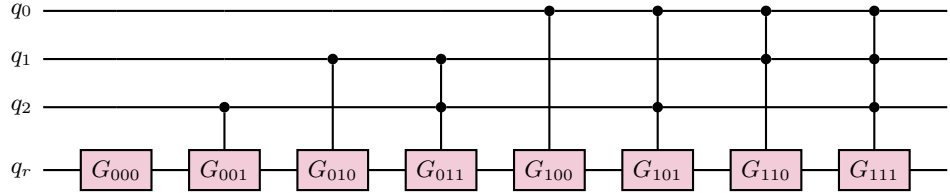
\begin{figure}[h!]
    \centering
    \begin{tikzcd}
        \lstick{$q_0$} & \qw            & \qw            & \qw             & \qw            & \ctrl{3}       & \ctrl{3}       & \ctrl{3}       & \ctrl{3}       & \qw\\
        \lstick{$q_1$} & \qw            & \qw            & \ctrl{2}        & \ctrl{2}       & \qw            & \qw            & \control{}     & \control{}     & \qw \\
        \lstick{$q_2$} & \qw            & \ctrl{1}       & \qw             & \control{}     & \qw            & \control{}     & \qw            & \control{}     & \qw \\
        \lstick{$q_r$} & \gate{G_{000}} & \gate{G_{001}} & \gate{G_{010}}  & \gate{G_{011}} & \gate{G_{100}} & \gate{G_{101}} & \gate{G_{110}} & \gate{G_{111}} & \qw \\
    \end{tikzcd}
    \caption{\textbf{Boolean QNN for three inputs:} this QNN can express any Boolean function with length three.}
    \label{fig:quantum_boolean_circuit_n_3}
\end{figure}

\textbf{(b) Boolean QNN}
In order to extend this Quantum Boolean circuit into a Boolean QNN, the paper \cite{ngoc_tunable_2020} introduces $G_u$, a tunable quantum gate, whose value can either be $I$, the identity gate or $X$, the $X$ gate, which flips a qubit and is equivalent to the matrix $\begin{psmallmatrix}
    0 & 1 \\
    1 & 0
\end{psmallmatrix}$, that is $G_u \in \{I, X\}$ where $u = u_0 \cdots u_{n-1} \in \mathbb{B}^n$ and $\mathbb{B} = \{0, 1\}$ is the set of the Boolean values and $\mathbb{B}^{\mathbb{B}^n}$ the set of functions from $\mathbb{B}^n$ to $\mathbb{B}$. $C_u$ is the multi-controlled X gate. The gates $C_u$ in the quantum Boolean circuit commute together and $I$ commutes with all matrices, so the order of the gates does not change the overall circuit. A QNN with $n$ data qubits contains $2^n$ gates. Each gate has two possible values, so there exists in total $2^{2^n}$ different circuits. The fully expressive QNN for $n = 3$ is shown in Figure \ref{fig:quantum_boolean_circuit_n_3}.

\textbf{(c) Parametrised Boolean QNN}
In this work, we are interested in how the functions change given randomly sampled parameters. Therefore we replace the $G_u$ gates with $U_3$ gates where $U_3$ is a general single qubit rotation gate.

$$U_3 = \begin{pmatrix} \cos(\theta/2) & -e^{i \lambda} \sin(\theta/2) \\
e^{i\phi} \sin(\theta/2) & e^{i(\phi + \lambda)} \cos(\theta/2) \end{pmatrix}$$

$U_3$ can represent any quantum gate, including the Identity Gate and the $X$ gate. Therefore, this parametrised boolean QNN can be equivalent to the boolean QNN when the unitary gates are selected to the be the Identity and $X$ gates. In particular, we can see that $U_3(\theta = 0, \phi = \pi, \lambda = \pi) = I$ where $I = \begin{psmallmatrix}
    1 & 0 \\
    0 & 1
\end{psmallmatrix}$. $U_3(\theta = \pi, \lambda = \pi, \phi = 0) = X$ where $X = \begin{psmallmatrix}
    0 & 1 \\
    1 & 0
\end{psmallmatrix}$.

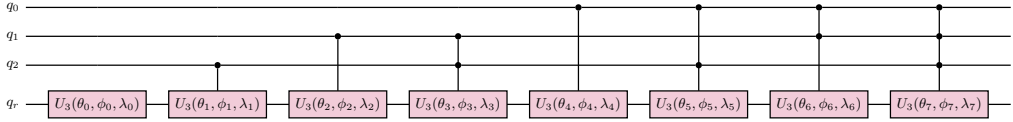
\begin{figure}[h!]
    \centering
    \adjustbox{scale=0.6}{%
    \begin{tikzcd}
        \lstick{$q_0$} & \qw            & \qw            & \qw             & \qw            & \ctrl{3}       & \ctrl{3}       & \ctrl{3}       & \ctrl{3}       & \qw\\
        \lstick{$q_1$} & \qw            & \qw            & \ctrl{2}        & \ctrl{2}       & \qw            & \qw            & \control{}     & \control{}     & \qw \\
        \lstick{$q_2$} & \qw            & \ctrl{1}       & \qw             & \control{}     & \qw            & \control{}     & \qw            & \control{}     & \qw \\
        \lstick{$q_r$} & \gate{U_3(\theta_0,\phi_0,\lambda_0)} & \gate{U_3(\theta_1,\phi_1,\lambda_1)} & \gate{U_3(\theta_2,\phi_2,\lambda_2)}  & \gate{U_3(\theta_3,\phi_3,\lambda_3)} & \gate{U_3(\theta_4,\phi_4,\lambda_4)} & \gate{U_3(\theta_5,\phi_5,\lambda_5)} & \gate{U_3(\theta_6,\phi_6,\lambda_6)} & \gate{U_3(\theta_7,\phi_7,\lambda_7)} & \qw \\
    \end{tikzcd}}
    \caption{\textbf{Boolean QNN for three inputs:} this QNN can express any Boolean function of length three and is overparametrised as it contains more parameters needed to express any Boolean function of length three.}
    \label{fig:parametrised_boolean_qnn}
\end{figure}

Therefore, the Boolean QNN in Figure \ref{fig:quantum_boolean_circuit_n_3} is transformed to the parametrised Boolean QNN in Figure \ref{fig:parametrised_boolean_qnn}.

\subsubsection{Non-fully expressive QNNs}
\label{sec:circuit_for_classification}
We introduce a variational quantum circuit based on the paper titled \textit{Classification with Quantum Neural Networks on Near Term Processors} \cite{farhi_classification_2018}; this variational quantum circuit is used for classification and has been implemented in Tensorflow Quantum \cite{broughton_tensorflow_2021} to classify simplified MNIST data \cite{tensorflow_quantum_mnist_nodate}. For classification, the QNN consists of $n+1$ qubits where $n$ is the the number of data qubits (that encode the classical data) and an extra qubit is added as the readout qubit $q_r$, whose measurement is equivalent to the classification output.

They set the unitaries in the form $e^{i\theta \sum}$ where $\sum$ is a tensor product of operators from the Pauli matrices set: $\{\sigma_x, \sigma_y, \sigma_z\}$ acting on multiple qubits. In the Tensorflow Quantum example, they choose $\sum$ to be $XX$ and $ZZ$ in which the second qubit is the readout qubit and the first qubit is one of the other qubits. $XX$ is the tensor product of two $X$ gates. In particular $$XX = \begin{pmatrix} 0 & 1 \\ 1 & 0 \end{pmatrix} \otimes \begin{pmatrix} 0 & 1 \\ 1 & 0 \end{pmatrix} = \begin{pmatrix} 0 & 0 & 0 & 1 \\ 0 & 0 & 1 & 0 \\ 0 & 1 & 0 & 0 \\ 1 & 0 & 0 & 0 \end{pmatrix}$$

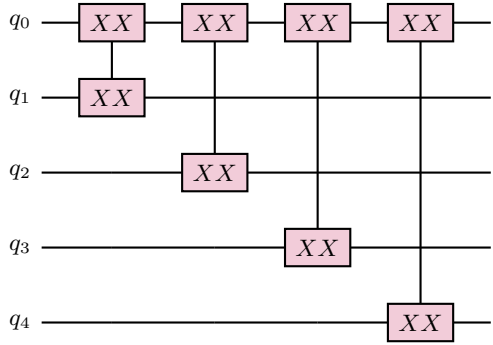
\begin{figure}[h]
    \centering
    \begin{tikzcd}
        \lstick{$q_0$} & \gate{XX} \vqw{1}& \gate{XX} \vqw{2}& \gate{XX} \vqw{3}&\gate{XX} \vqw{4} & \qw\\
        \lstick{$q_1$} & \gate{XX}        & \qw              & \qw              & \qw & \qw \\
        \lstick{$q_2$} & \qw              & \gate{XX}        & \qw              & \qw & \qw\\
        \lstick{$q_3$} & \qw              & \qw              & \gate{XX}        & \qw & \qw\\
        \lstick{$q_4$} & \qw              & \qw              & \qw              & \gate{XX} &\qw \\
    \end{tikzcd}
    \caption{\textbf{Variational Quantum Circuit:} this circuit consists of controlled-XX gates applied to each qubit.}
    \label{fig:variational_quantum_circuit}
\end{figure}

They construct layers of the variational quantum circuit. Each layer consists of $n$ instances of the same gate, where $n$ is the number of data qubits in the circuit. Figure \ref{fig:variational_quantum_circuit} shows an example of this variational quantum circuit for $n=4$. We can see that qubit $q_0$ is the readout qubit, as it is always being acted upon and qubits $q_1, q_2, q_3, q_4$ are the data qubits.

\FloatBarrier
\subsection{Training and test datasets for Boolean data}
\label{sec:training_qnns}
We describe the general process for generating the training and test datasets in Section \ref{sec:training_qnns_simple}. In this section, we provide an example. If the target function $t = 10100011$, then $t_0 = 1, t_1 = 0, t_2 = 1$, etc... We randomly select half of these indices to be the training set, for example $0,2,4,5$ and the other half is the test set, which is $1,3,6,7$.

Table \ref{tab:boolean_data_n3} shows the Boolean data inputs and their indices for $n=3$.

\begin{table}
\begin{tabularx}{0.2\textwidth}{ 
  | >{\centering\arraybackslash}X 
  | >{\centering\arraybackslash}X 
  | }
 \hline
    Input & Index \\
 \hline
  000 & 0 \\
  \hline
  001 & 1 \\
  \hline
  010 & 2 \\
  \hline
  011 & 3 \\
  \hline
  100 & 4 \\
  \hline
  101 & 5 \\
  \hline
  110 & 6 \\
  \hline
  111 & 7 \\
    \hline
\end{tabularx}
 \caption{Boolean input data for $n=3$} 
 \label{tab:boolean_data_n3}
\end{table}

The training inputs then are the Boolean data inputs $X$ at indices $0,2,4,5$. This corresponds to training inputs of $X_0 = (0,0,0), X_2 = (0,1,0), X_4 = (1,0,0), X_5 = (1,0,1)$. The test inputs are at indices $1,3,6,7$ corresponding to test inputs of $X_1 = (0,0,1), X_3 = (0,1,1), X_6 = (1,1,0), X_7 = (1,1,1)$.

The training labels are the values of the bits at the training indices of the target function. For example, for the target function $t = 10100011$, the training labels are $t_0 = 1, t_2 = 1, t_4 = 0, t_5 = 0$. Similarly, the test labels are the values of the bits at the test indices of the target function, so the test labels are $t_1 = 0, t_3 = 0, t_6 = 1, t_7 = 1$.

\newpage
\section{Additional Experiments and Results}
\label{sec:appendix_additional_results}
\FloatBarrier
\subsection{Inductive bias in QNNs}
\subsubsection{Probability of complexity vs complexity}
\label{sec:qnn_pk}
\begin{figure*}
     \centering
     \begin{subfigure}[t]{0.3\textwidth}
         \centering
         \includegraphics[width=\textwidth]{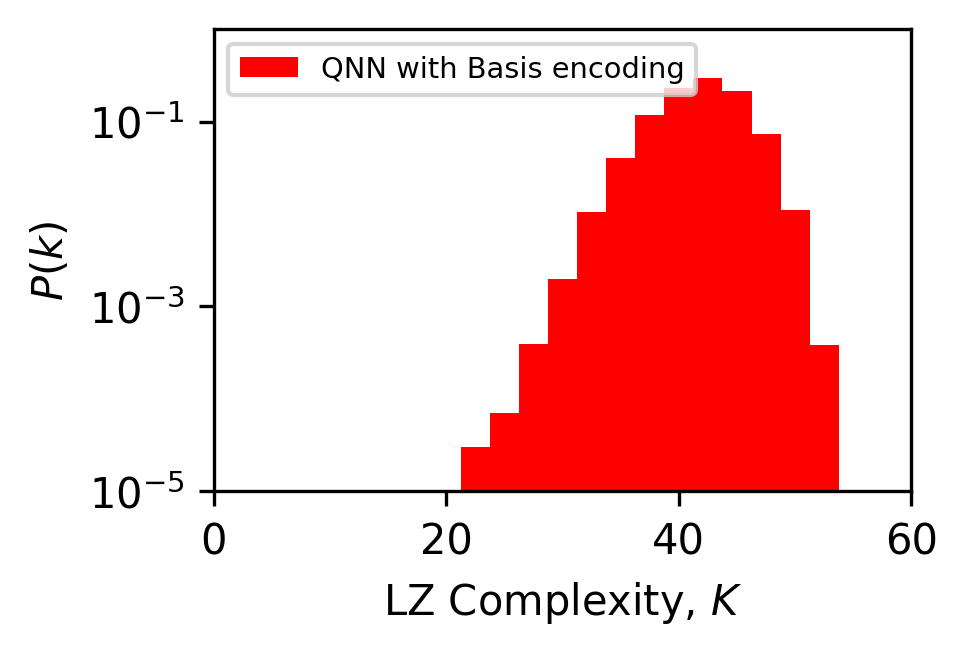}
         \caption{Prior $P(K)$ for QNN with basis encoding}
         \label{fig:qnn_pk_e0}
     \end{subfigure}
     \begin{subfigure}[t]{0.3\textwidth}
         \centering
         \includegraphics[width=\textwidth]{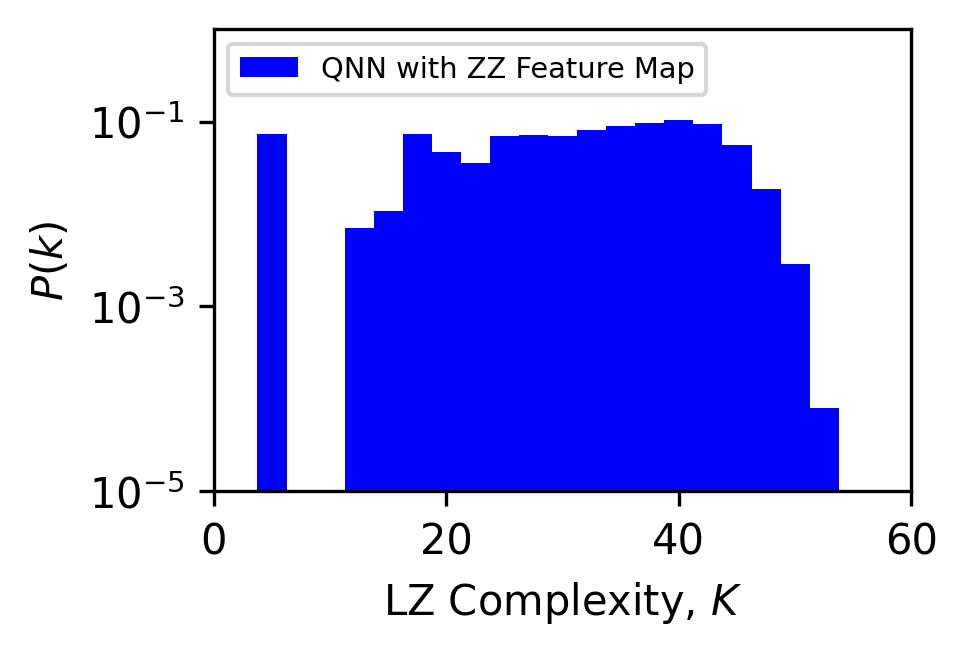}
         \caption{Prior $P(K)$ for QNN with ZZ Feature Map}
         \label{fig:qnn_pk_e1}
     \end{subfigure}
     \begin{subfigure}[t]{0.3\textwidth}
         \centering
         \includegraphics[width=\textwidth]{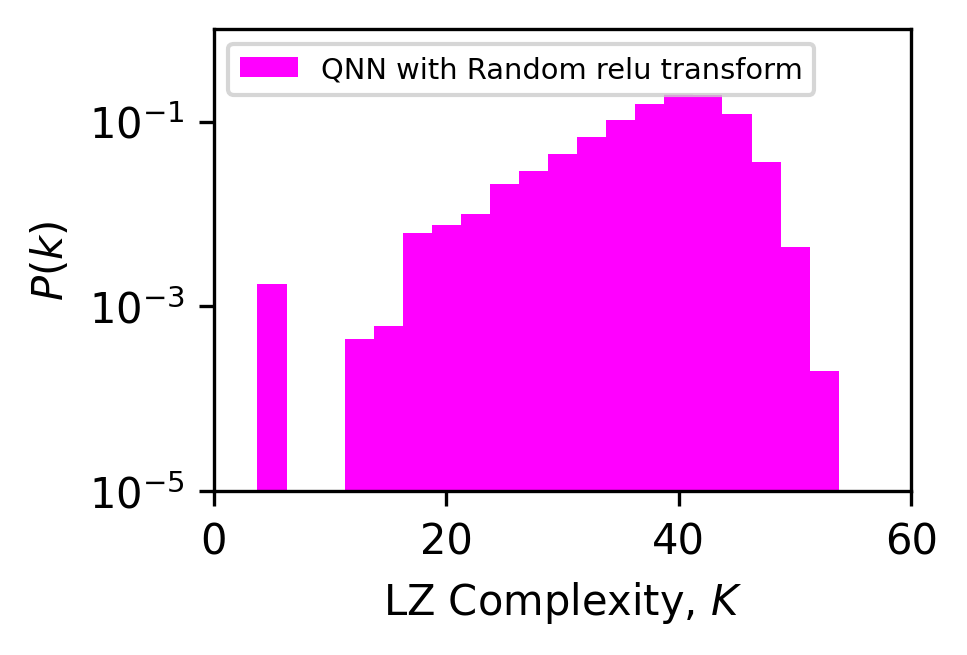}
         \caption{Prior $P(K)$ for QNN with relu transform}
         \label{fig:qnn_pk_e3}
     \end{subfigure}
     \begin{subfigure}[t]{0.3\textwidth}
         \centering
         \includegraphics[width=\textwidth]{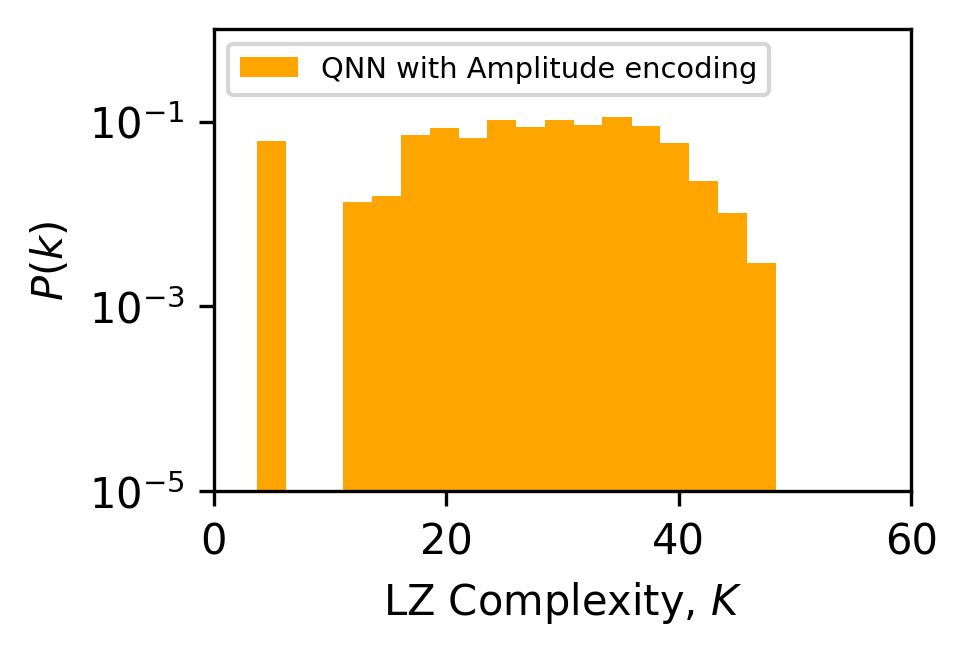}
         \caption{Prior $P(K)$ for QNN with amplitude encoding}
         \label{fig:qnn_pk_e4}
     \end{subfigure}
     \begin{subfigure}[t]{0.3\textwidth}
         \centering
         \includegraphics[width=\textwidth]{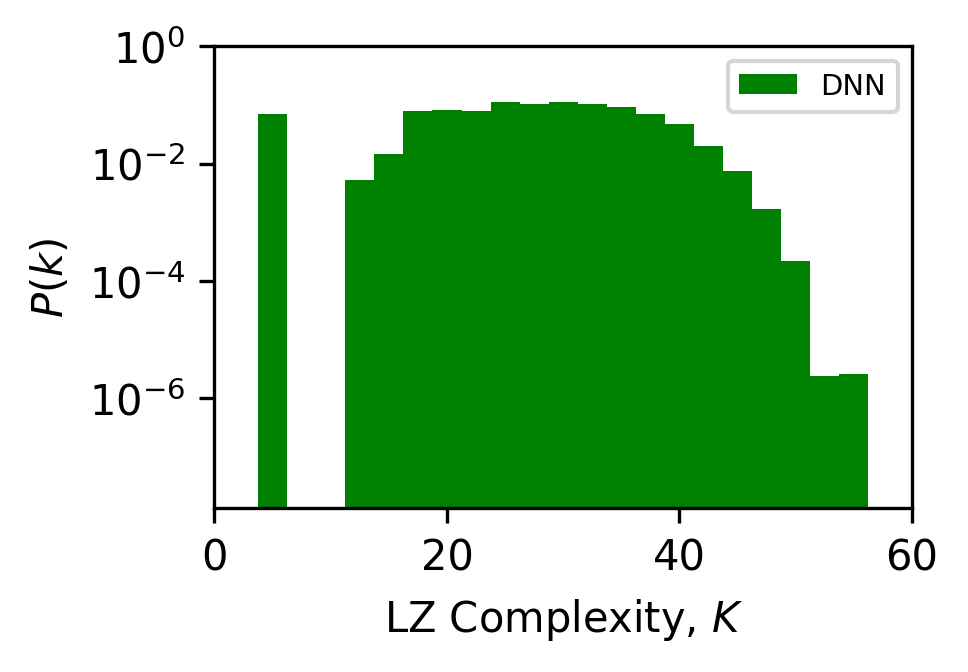}
         \caption{Prior $P(K)$ for classical DNN}
         \label{fig:qnn_pk_dnn}
     \end{subfigure}
     \begin{subfigure}[t]{0.3\textwidth}
         \centering
         \includegraphics[width=\textwidth]{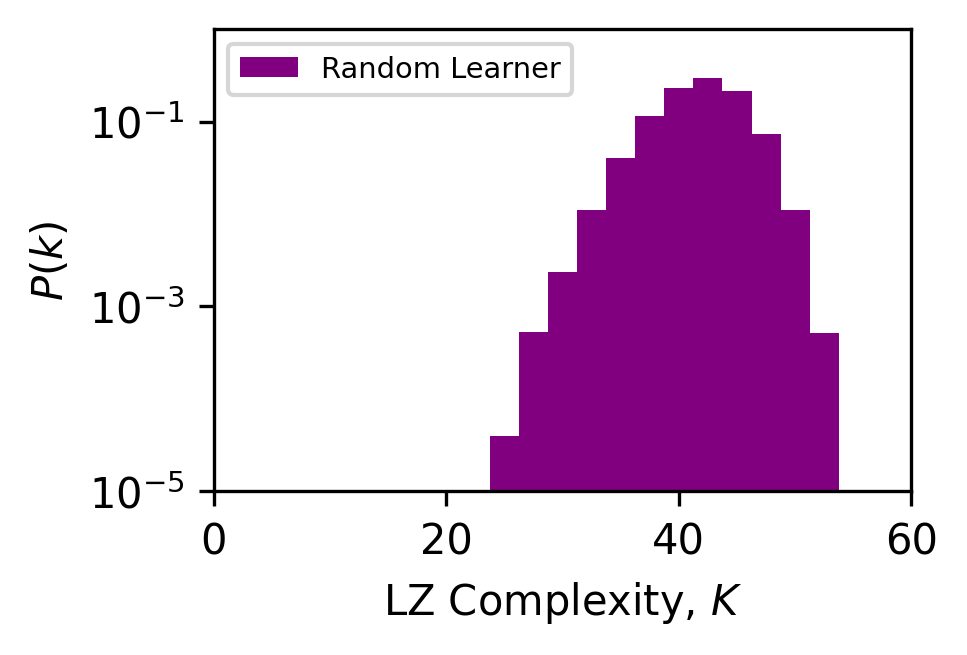}
         \caption{Prior $P(K)$ for random learner}
         \label{fig:qnn_pk_random_learner}
     \end{subfigure}
        \caption{\textbf{Probability to obtain a function of Lempel-Ziv complexity $\mathbf{K}$ versus its complexity for five data qubits and $\mathbf{10^5}$ samples.} $P(f)$ versus Lempel-Ziv complexity, $K$ for a QNN with basis encoding (red), ZZ feature map (blue), random relu transform (pink), amplitude encoding (orange), and a DNN (green). $P(K)$ is calculated by generating $10^5$ samples of functions from the QNN by using random samples of parameters $\Theta$ over a uniform distribution and $10^7$ samples for the DNN.}
        \label{fig:qnn_pk}
\end{figure*}

We look at how the probability of functions with a certain complexity $K$ changes with the complexity $K$ of those functions. There are exponentially more complex functions than simple functions, so for a random learner with no bias, we'd expect that functions with higher complexity have a higher probability, as shown in Figure \ref{fig:qnn_pk_random_learner}. This Figure \ref{fig:qnn_pk_random_learner} is generated by generating random boolean functions and measuring their complexity.  

When there is a bias towards simple functions as the DNN exhibits as shown in Figure \ref{fig:qnn_pk_dnn}, the probability of the complexity flattens across complexities as the DNN's exponential bias towards simple functions counteracts the exponentially more complex functions. 

The QNN with basis encoding has a $P(K)$ vs $K$ plot approximately like the random learner, showing that such QNNs have no bias. This corresponds with the results in Figure \ref{fig:pf_vs_k_e0}. The QNNs with other encoding methods look similar to the DNN and flatten out the $P(K)$ probabilities. 

\subsubsection{Entropy vs complexity}
\label{sec:entropy_vs_complexity_qnns}
\begin{figure*}
     \centering
     \begin{subfigure}[t]{0.49\textwidth}
         \centering
         \includegraphics[width=0.7\textwidth]{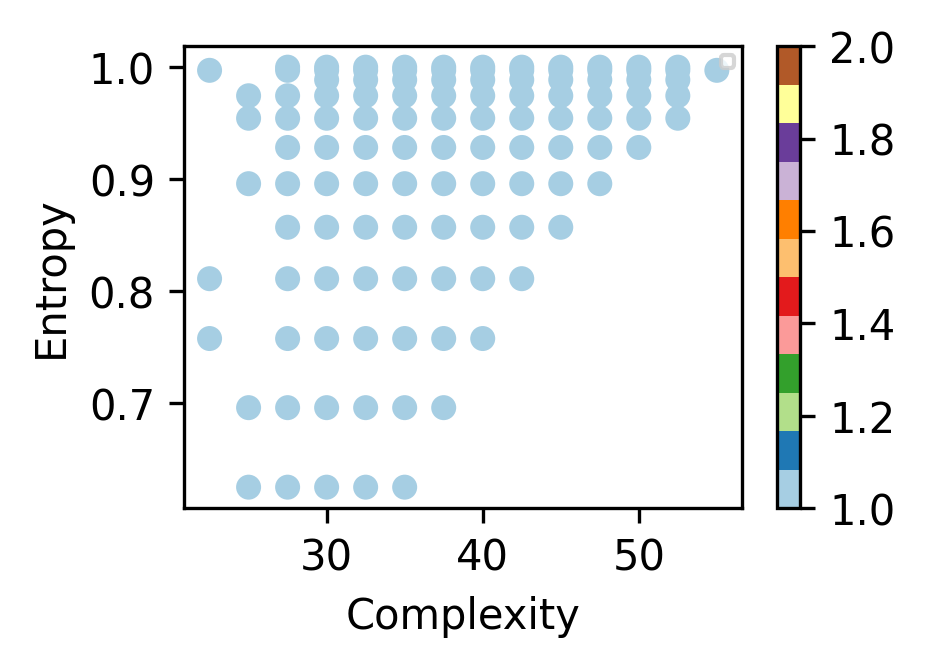}
         \caption{Entropy vs complexity for QNN with basis encoding}
         \label{fig:qnn_entropy_vs_complexity_basis_encoding}
     \end{subfigure}
     \begin{subfigure}[t]{0.49\textwidth}
         \centering
         \includegraphics[width=0.7\textwidth]{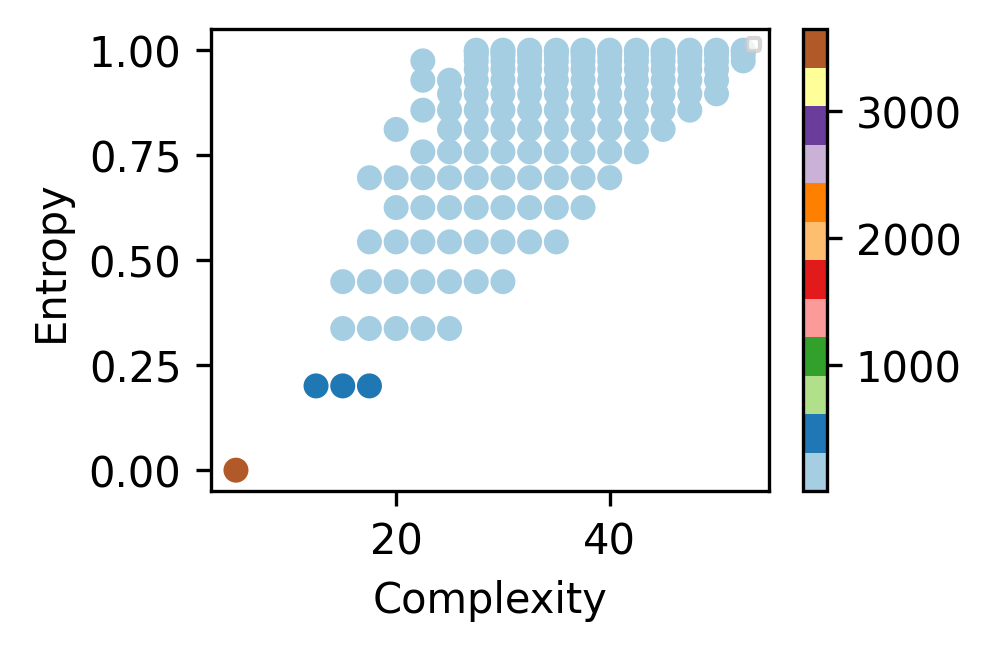}
         \caption{Entropy vs complexity for QNN with ZZ feature map}
         \label{fig:qnn_entropy_vs_complexity_zz_feature_map}
     \end{subfigure}
     \begin{subfigure}[t]{0.49\textwidth}
         \centering
         \includegraphics[width=0.7\textwidth]{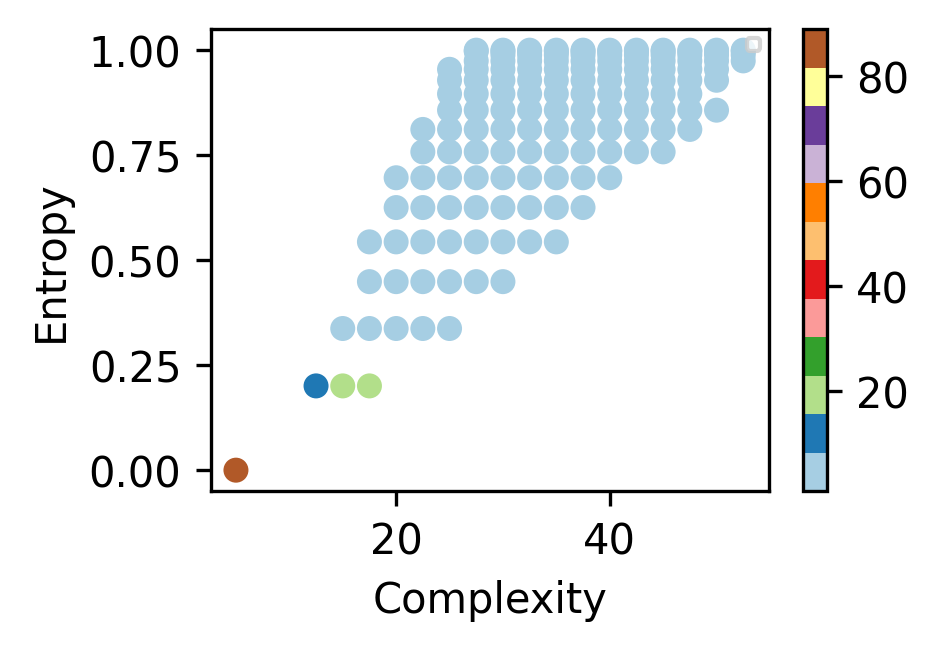}
         \caption{Entropy vs complexity for QNN with classical transform}
         \label{fig:qnn_entropy_vs_complexity_classical_transform}
     \end{subfigure}
     \begin{subfigure}[t]{0.49\textwidth}
         \centering
         \includegraphics[width=0.7\textwidth]{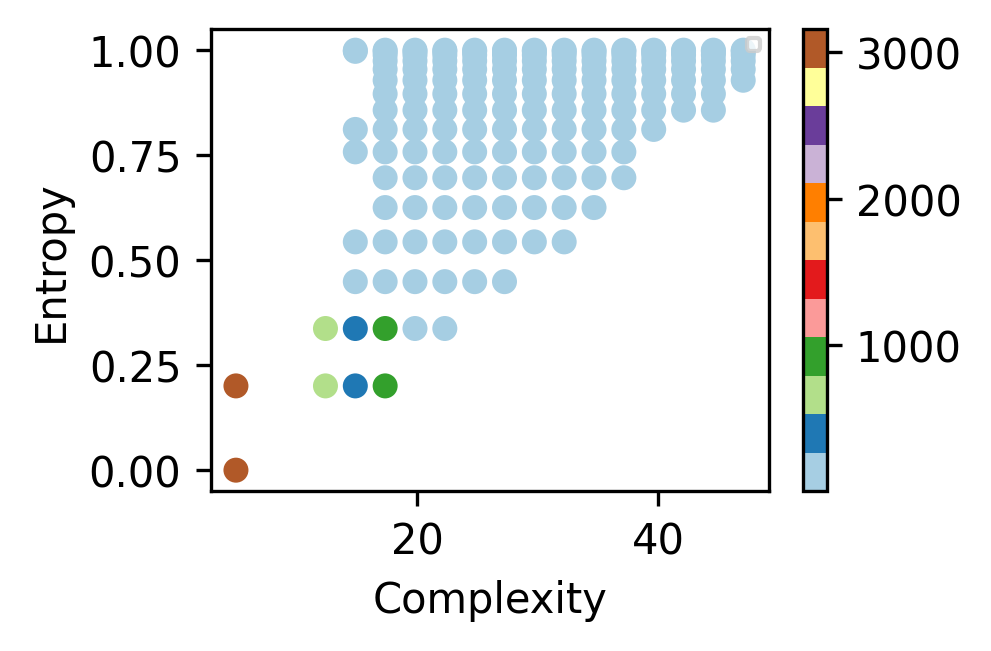}
         \caption{Entropy vs complexity for QNN with amplitude encoding}
         \label{fig:qnn_entropy_vs_complexity_amplitude_encoding}
     \end{subfigure}
        \caption{\textbf{Entropy of a boolean function $\mathbf{(f)}$ versus its complexity for five data qubits and $\mathbf{10^6}$ samples.} The colour of the datapoint corresponds to the maximum count of a function that has that complexity and entropy. Therefore, a datapoint with a colour that has a higher count indicates that the QNN has a stronger bias towards functions with that entropy and complexity. The boolean functions produced by the QNN were grouped and for each entropy-complexity pair, there were multiple functions, so the maximum count of a function is displayed.}
        \label{fig:qnn_entropy_vs_complexity}
\end{figure*}

To see more clearly the simplicity bias and the entropy bias for the different encoding methods, we plot the entropy vs complexity of the functions produced from the QNN in Figure \ref{fig:qnn_entropy_vs_complexity}. 

The plots show an entropy or simplicity bias in several ways: first, we look at low-complexity, high-entropy functions. We would expect an encoding method with a simplicity bias to have a strong bias towards these types of functions and therefore produce many of them. On the other hand, we would expect an encoding method with an entropy bias to not have a bias towards these types of functions and therefore not produce them. As $n = 5$, there are $4.3 \cross 10^9$ possible functions. As we produce $10^5$ samples, we would expect that the functions that are not produced by the QNN (and therefore have no datapoint on the plots) are functions that the QNN does not have a bias towards. 

Figure \ref{fig:qnn_entropy_vs_complexity_basis_encoding} shows that the QNN with amplitude encoding produces functions with low-complexity and high-entropy. For example, at complexity at around 20, the datapoints produce a vertical line from entropy 0.25 up until 1. For the QNN with the ZZ feature map and the Random relu transform, however, we see that such QNNs are only able to produce functions with low-entropy when there is low-complexity. Datapoints for low-complexity, high-entropy functions are missing; this is because these QNNs have a bias towards low-entropy functions. Furthermore, we can look at the colours of the datapoints. The datapoints with higher-count colours for the QNNs with the ZZ feature map and Random relu transform are below entropy of 0.25 and complexity of 20 (the three dark blue datapoints and the brown datapoint). For the amplitude encoding, however, in addition to those datapoints, it also produces datapoints with higher-count colours for functions with entropy above 0.25 and complexity below 20 as well as an additional brown datapoint with entropy higher than the brown datapoint shared with the encoding methods. This further demonstrates the simplicity bias of amplitude encoding and the entropy bias of the other encoding methods. The figure shows that for basis encoding, there is neither entropy nor complexity bias as all of the datapoints remain at a similar count, therefore indicating no bias. 

\subsubsection{Probability of boolean function vs rank}
\label{sec:qnn_rank}
\begin{figure*}
     \centering
     \begin{subfigure}[t]{0.49\textwidth}
         \centering
         \includegraphics[width=0.6\textwidth]{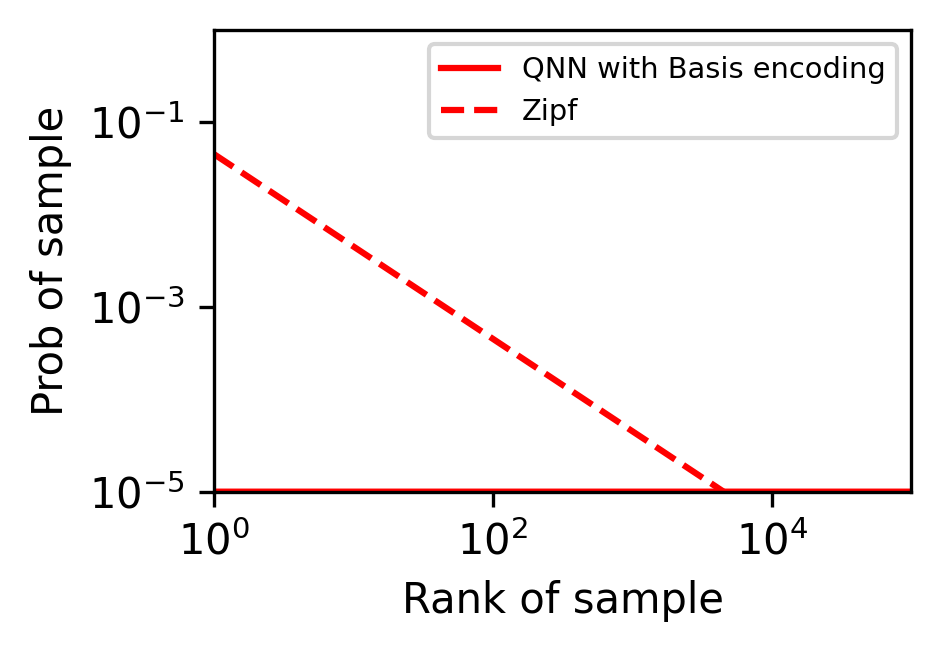}
         \caption{$P(f)$ vs Rank for QNN with basis encoding}
         \label{fig:qnn_rank_e0}
     \end{subfigure}
     \begin{subfigure}[t]{0.49\textwidth}
         \centering
         \includegraphics[width=0.6\textwidth]{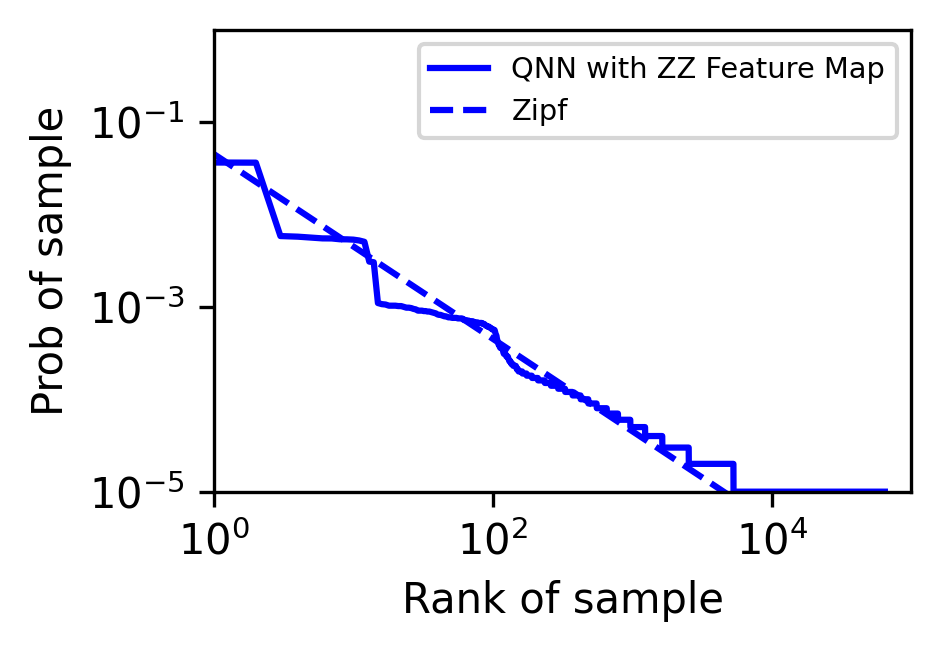}
         \caption{$P(f)$ vs Rank for QNN with ZZ Feature Map}
         \label{fig:qnn_rank_e1}
     \end{subfigure}
     \begin{subfigure}[t]{0.49\textwidth}
         \centering
         \includegraphics[width=0.6\textwidth]{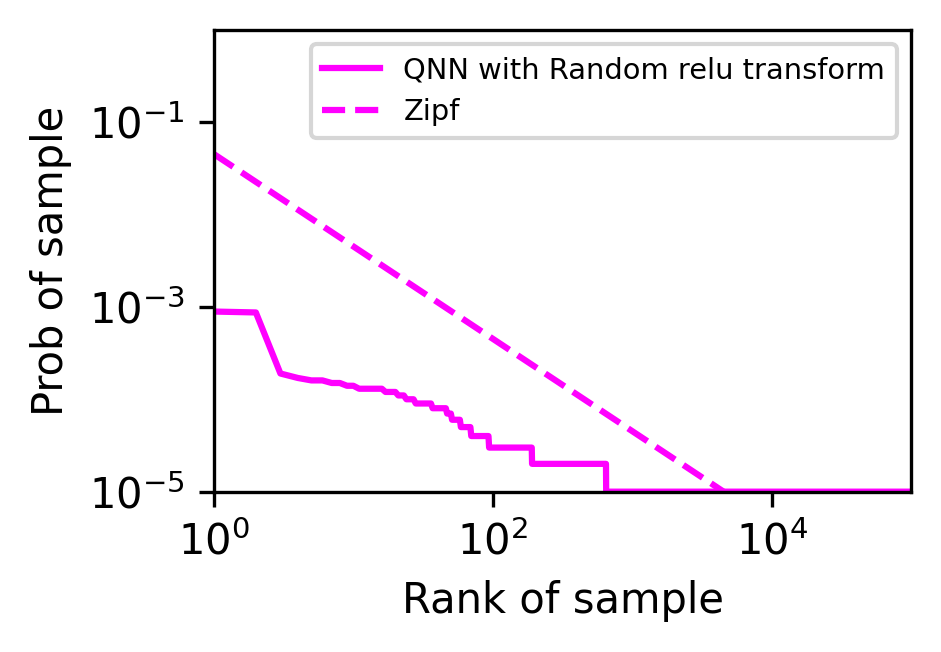}
         \caption{$P(f)$ vs Rank for QNN with relu transform}
         \label{fig:qnn_rank_e3}
     \end{subfigure}
     \begin{subfigure}[t]{0.49\textwidth}
         \centering
         \includegraphics[width=0.6\textwidth]{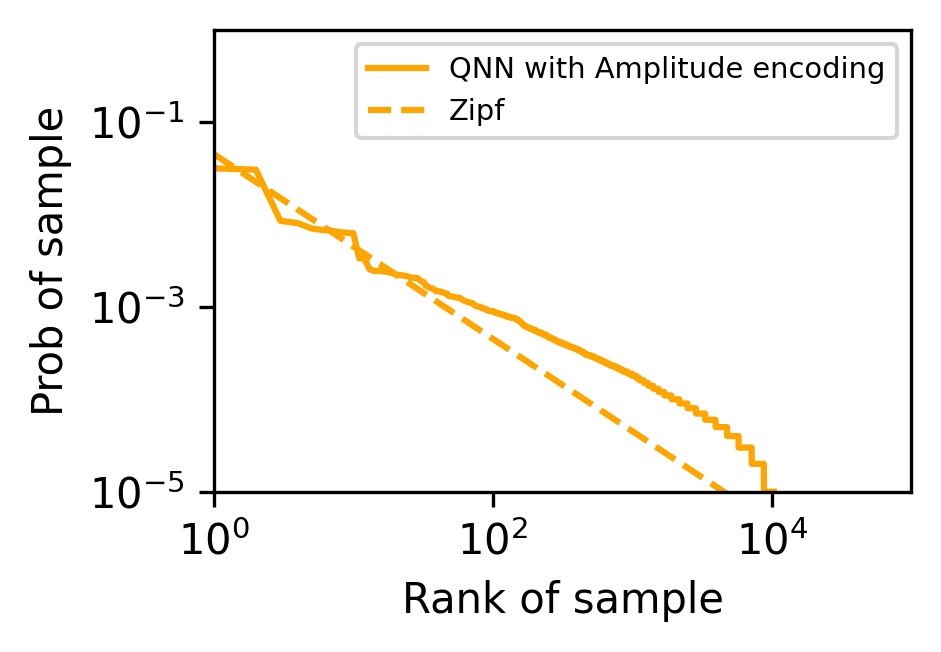}
         \caption{$P(f)$ vs Rank for QNN with amplitude encoding}
         \label{fig:qnn_rank_e4}
     \end{subfigure}
     \caption{\textbf{Probability of a boolean function versus its rank for five data qubits and $\mathbf{10^5}$ samples}. $P(f)$ ranked by probability of individual functions, generated from $10^5$ samples of parameters $\Theta$ over a uniform distribution. The dotted line is Zipf's law $P(f) = \frac{1}{(32 ln 2) \text{Rank}(f)}$}
        \label{fig:qnn_rank}
\end{figure*}

We plot the probability of a boolean function versus the rank of the sample in Figure \ref{fig:qnn_rank}. We compare the QNN data with the normalized Zipf law $P(r) = \frac{1}{ln(N_O)r}$ where $r$ is the rank and $N_O = 2^{2^n}$ where $n$ is the number of data qubits. Zipf's law tells us that the frequency of a sample is inversely proportional to its rank. In Figure \ref{fig:qnn_rank} we plot the data for $n=5$ so $N_O = 2^{32}$.

Figure \ref{fig:qnn_rank_e0} shows the rank plot for basis encoding - its probability versus rank is uniform at $10^{-s}$ where $s$ is the number of samples. Figure \ref{fig:qnn_rank_e1} and \ref{fig:qnn_rank_e4} fit Zipf's law.

\FloatBarrier
\subsection{Inductive bias in Quantum Kernels}
\label{sec:results_inductive_bias_in_quantum_kernels}

In this section, we explain quantum kernels and how they relate to QNNs. We show additional experiments that demonstrate that the inductive bias in QNNs corresponds to the inductive bias in quantum kernels. 

\subsubsection{The kernel matrix}
A kernel function $k$ maps the input data into a higher dimensional space $k(\vec{x}_i, \vec{x}_j) = \langle f(\vec{x}_i), f(\vec{x}_j) \rangle$ where $\vec{x}_i, \vec{x}_j$ are $n$ dimensional inputs, $f$ is a map from $n$-dimension to $m$ dimension space. A kernel matrix $K$ can be constructed from all the input data: $K_{ij} = k(\vec{x}_i,\vec{x}_j)$.

A quantum feature map $\phi(\vec{x})$ maps classical data, represented as a vector $\vec{x}$, to a quantum Hilbert space and the Kernel matrix is $K_{ij} = \left| \langle \phi^\dagger(\vec{x}_j)| \phi(\vec{x}_i) \rangle \right|^{2}$.

To obtain the kernel matrix from a QNN, we encode the $x_i$ data into the QNN and obtain the final statevector $\ket{\phi}$. We can then compute $\left| \langle \phi^\dagger(\vec{x}_j)| \phi(\vec{x}_i) \rangle \right|^{2}$ for all pairs of $i,j$. Further details on how to do this are in Appendix \ref{sec:quantum_kernels}.

\subsubsection{Kernel matrix plots}
\begin{figure*}
     \centering
     \begin{subfigure}[t]{0.3\textwidth}
         \centering
         \includegraphics[width=\textwidth]{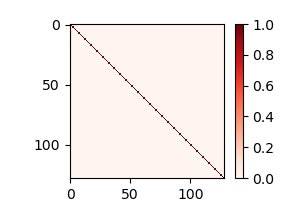}
         \caption{Basis encoding}
         \label{fig:kernel_e0}
     \end{subfigure}
     \begin{subfigure}[t]{0.3\textwidth}
         \centering
         \includegraphics[width=\textwidth]{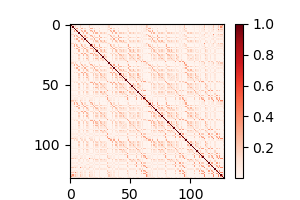}
         \caption{ZZ feature map}
         \label{fig:kernel_e1}
     \end{subfigure}
     \begin{subfigure}[t]{0.3\textwidth}
         \centering
         \includegraphics[width=\textwidth]{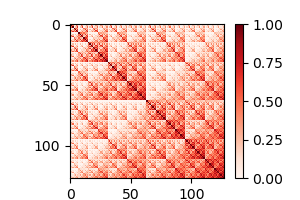}
         \caption{Amplitude encoding}
         \label{fig:kernel_e4}

     \end{subfigure}
     \begin{subfigure}[t]{0.3\textwidth}
         \centering
         \includegraphics[width=\textwidth]{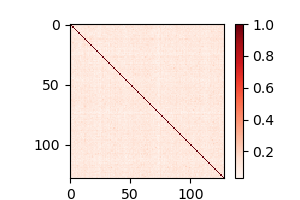}
         \caption{Random relu transform}
         \label{fig:kernel_e3}
     \end{subfigure}
     \begin{subfigure}[t]{0.3\textwidth}
         \centering
         \includegraphics[width=\textwidth]{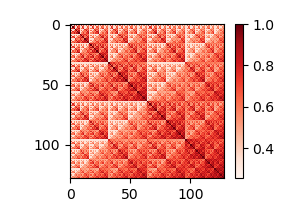}
         \caption{Classical FCN (fully convolutional network)}
         \label{fig:emp:kernel_e4}
     \end{subfigure}
     \caption{Kernels matrices for different encodings for seven qubits ($n=7$) resulting in $128 \cross 128$ dimension matrices.}
     \label{fig:kernel}
\end{figure*}

We plot the kernel matrices for different encoding methods in Figure \ref{fig:kernel}. One can see in Figure \ref{fig:kernel_e0} that for the basis encoding, there is no correlation between datapoints. The ZZ feature map (Figure \ref{fig:kernel_e1}) and Random relu transform (Figure \ref{fig:kernel_e3}) show some correlation, and the amplitude encoding (Figure \ref{fig:kernel_e4}) shows stronger correlation. The DNN kernel (Figure \ref{fig:emp:kernel_e4}) is included for comparison. 

\FloatBarrier

\subsubsection{Kernel eigenvalues}
\begin{figure}[h]
     \centering
         \centering
         \includegraphics[width=0.3\textwidth]{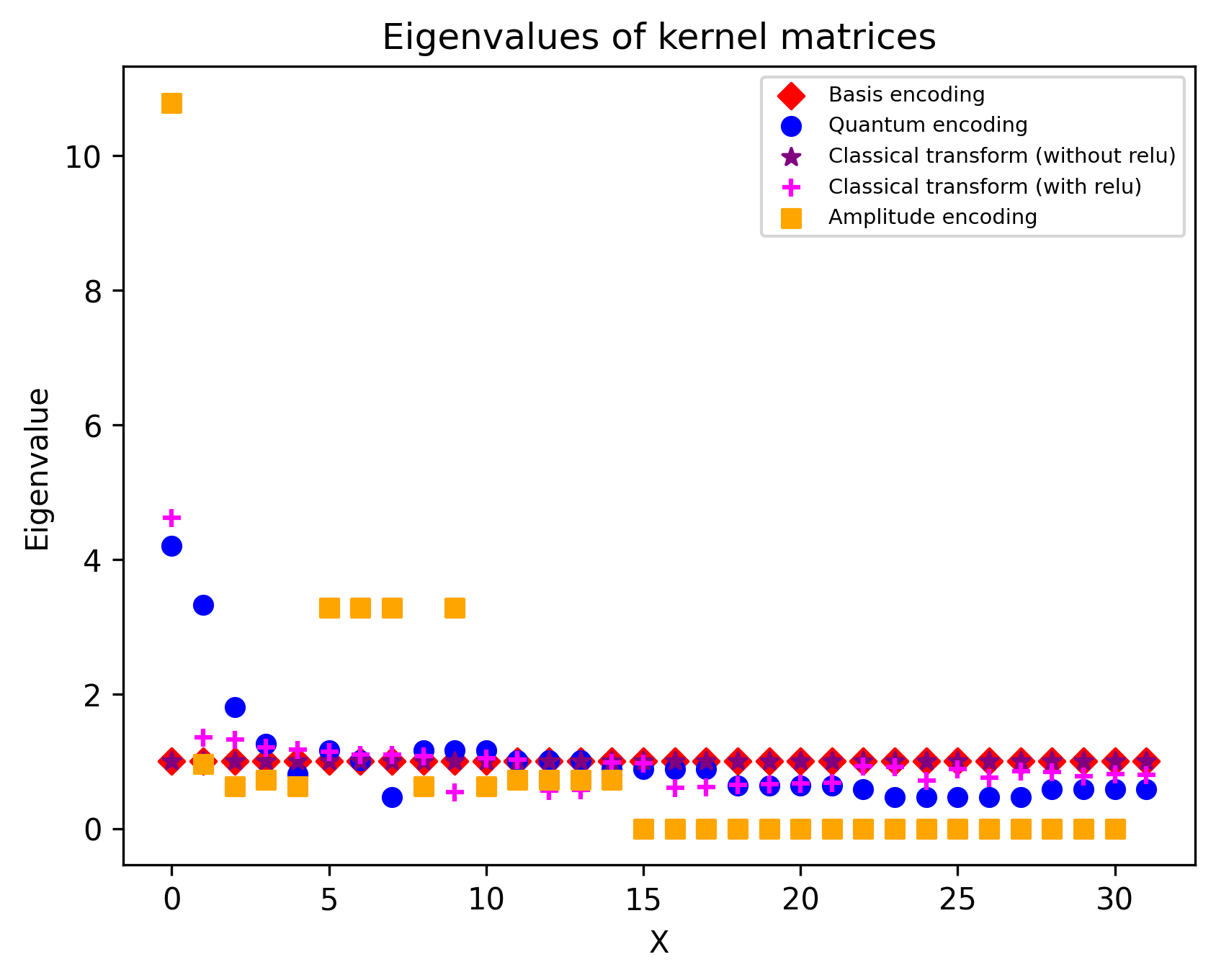}
        \caption{Kernel eigenvalues for different encoding methods for five qubits ($n=5$)}
        \label{fig:kernel_eigenvalues}
\end{figure}

The eigenvalues of the kernel matrices are plotted in Figure \ref{fig:kernel_eigenvalues}. Looking at the eigenvalues also provides insight into the type of bias. 

The plot shows that for the kernel with basis encoding, its eigenvalues are flat, which is not the case for the eigenvalues for the other encoding methods. Kubler et al. provide a detailed analysis of the eigenvalues of kernel matrices and how it relates to inductive bias \cite{kubler_inductive_nodate}.

\FloatBarrier

\subsubsection{Probability of boolean function vs complexity}
\begin{figure*}
     \centering
     \begin{subfigure}[t]{0.49\textwidth}
         \centering
         \includegraphics[width=0.75\textwidth]{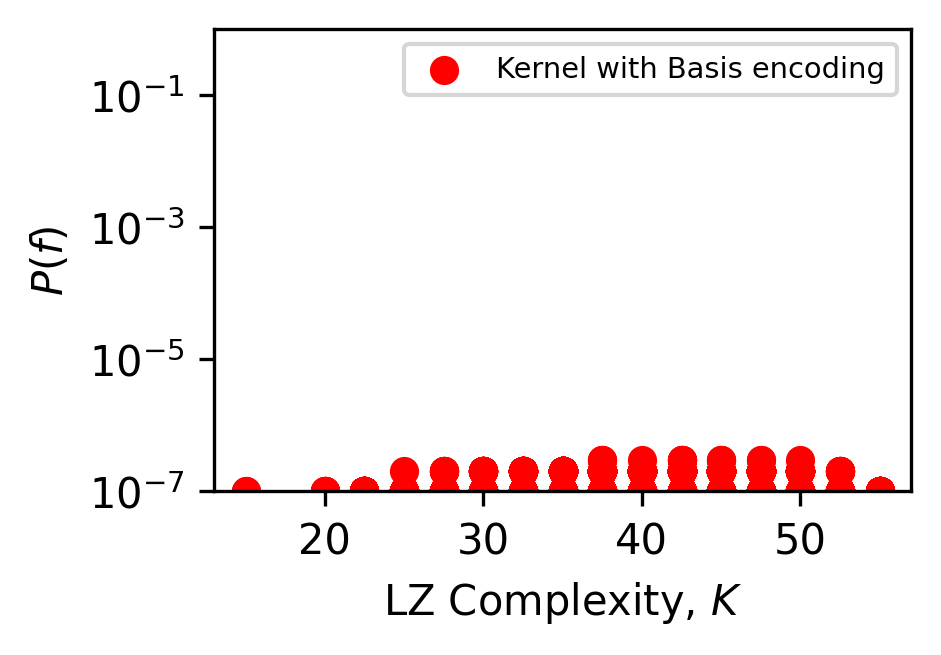}
         \caption{Prior $P(f)$ vs complexity for the quantum kernel with basis encoding}
         \label{fig:kernel_pf_e0}
     \end{subfigure}
     \begin{subfigure}[t]{0.49\textwidth}
         \centering
         \includegraphics[width=0.75\textwidth]{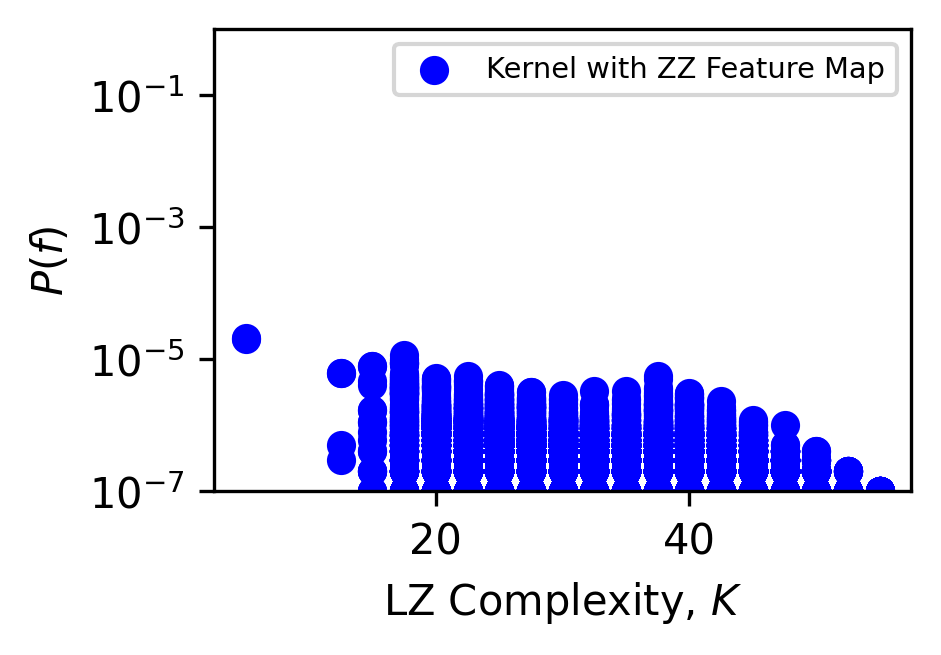}
         \caption{Prior $P(f)$ vs complexity for the quantum kernel with ZZ feature map}
         \label{fig:kernel_pf_e1}
     \end{subfigure}
     \begin{subfigure}[t]{0.49\textwidth}
         \centering
         \includegraphics[width=0.75\textwidth]{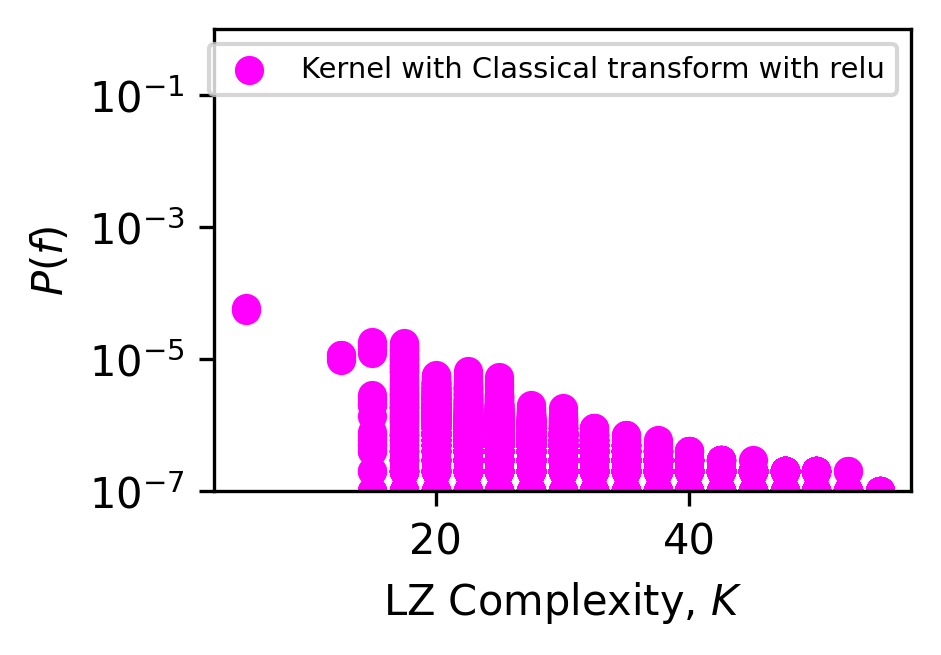}
         \caption{Prior $P(f)$ vs complexity for the quantum kernel with random relu transform}
         \label{fig:kernel_pf_e3}
     \end{subfigure}
     \begin{subfigure}[t]{0.49\textwidth}
         \centering
         \includegraphics[width=0.75\textwidth]{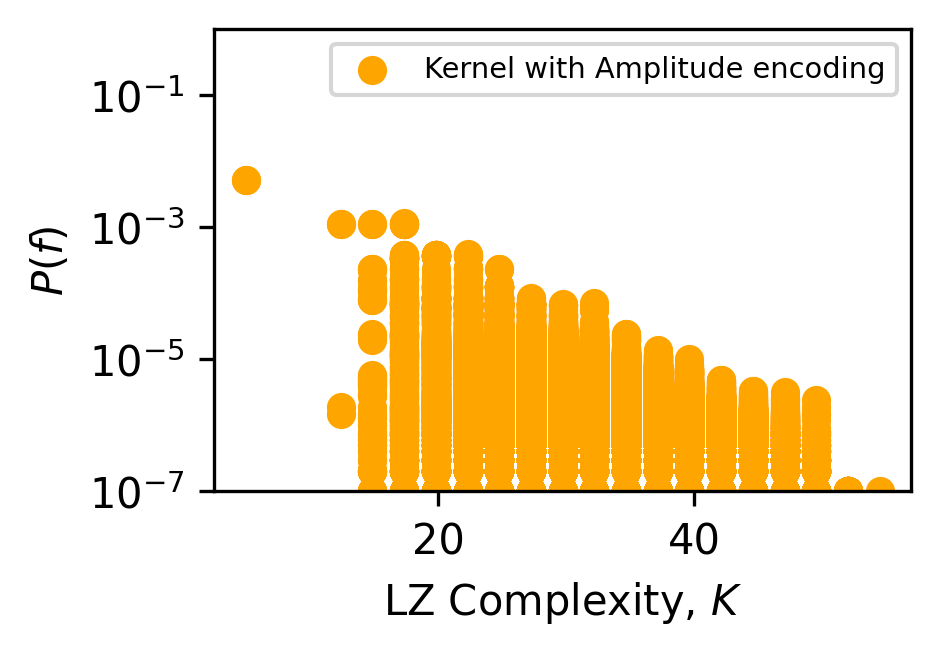}
         \caption{Prior $P(f)$ vs complexity for the quantum kernel with amplitude encoding}
         \label{fig:kernel_pf_e4}
     \end{subfigure}
        \caption{\textbf{Probability of a boolean function $\mathbf{(f)}$ versus its complexity for five data qubits and $\mathbf{10^7}$ samples.} $P(f)$ versus Lempel-Ziv complexity, $K$ for a Quantum Kernel with basis encoding (red), ZZ Feature Map (blue), random relu transform (pink), and amplitude encoding (orange). $P(f)$ is calculated by generating $10^7$ samples of functions from the quantum kernel by using multivariate normal in numpy.}
        \label{fig:kernel_pf_vs_k}
\end{figure*}

We plot the probability of the boolean function vs complexity as described in Section \ref{sec:pf_vs_k_qnn}. The trends in Figure \ref{fig:kernel_pf_vs_k} are similar to those of the QNN shown in Section \ref{sec:pf_vs_k_qnn}, demonstrating the alignment between the QNN and quantum kernel with regards to how the encoding method changes their inductive bias.

\FloatBarrier
\subsubsection{Probability of complexity vs complexity}
\begin{figure*}
     \centering
     \begin{subfigure}[t]{0.3\textwidth}
         \centering
         \includegraphics[width=\textwidth]{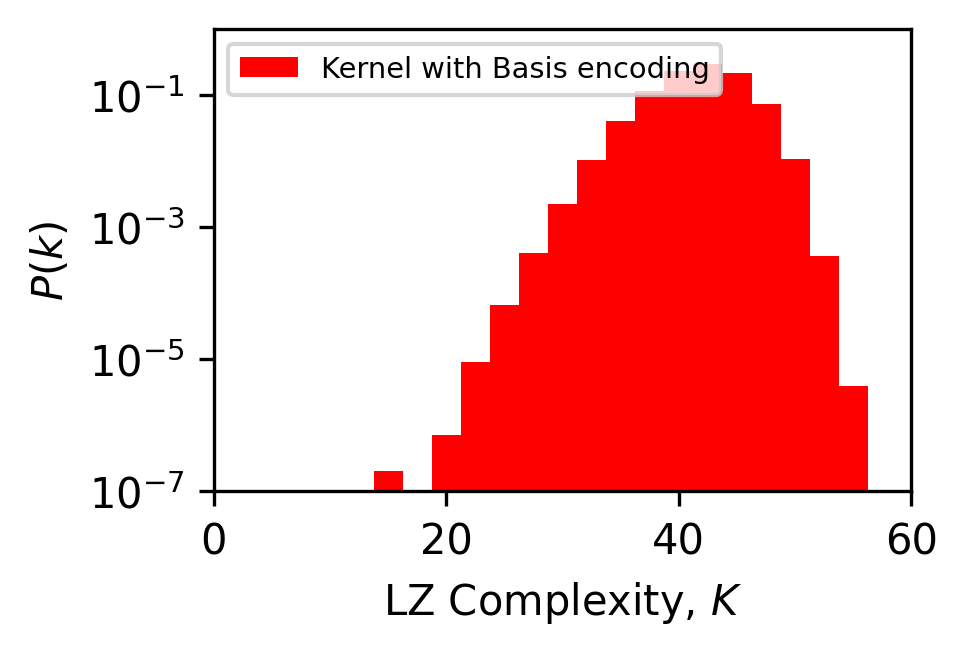}
         \caption{Prior $P(K)$ for the quantum kernel with basis encoding}
         \label{fig:pk_e0}
     \end{subfigure}
     \hfill
     \begin{subfigure}[t]{0.3\textwidth}
         \centering
         \includegraphics[width=\textwidth]{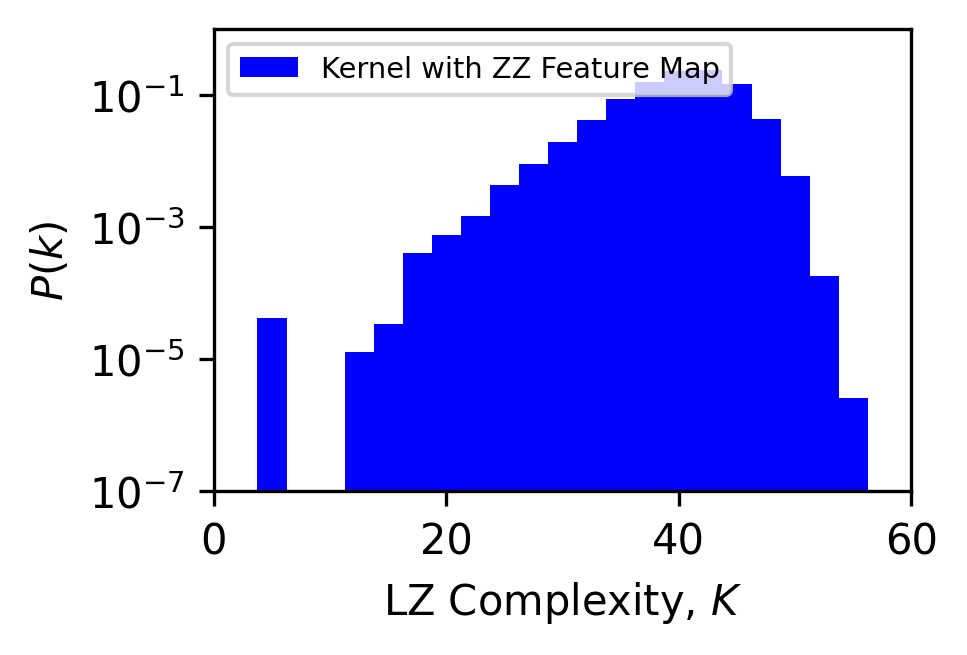}
         \caption{Prior $P(K)$ for the quantum kernel with ZZ feature map}
         \label{fig:kernel_pk_e1}
     \end{subfigure}
     \hfill
     \begin{subfigure}[t]{0.3\textwidth}
         \centering
         \includegraphics[width=\textwidth]{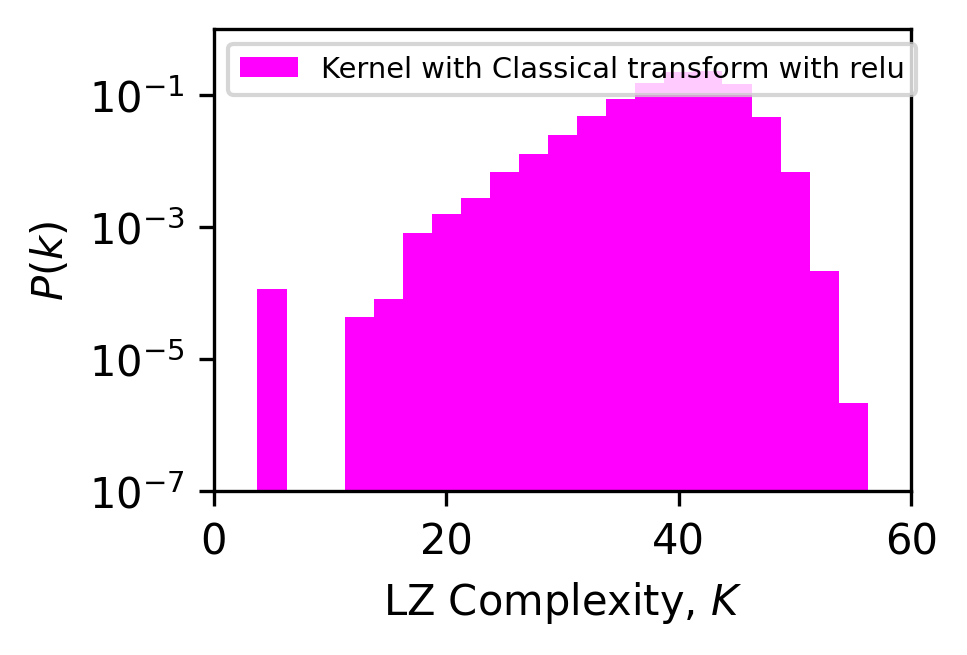}
         \caption{Prior $P(K)$ for the quantum kernel with random relu transform}
         \label{fig:kernel_pk_e3}
     \end{subfigure}

     \begin{subfigure}[t]{0.3\textwidth}
         \centering
         \includegraphics[width=\textwidth]{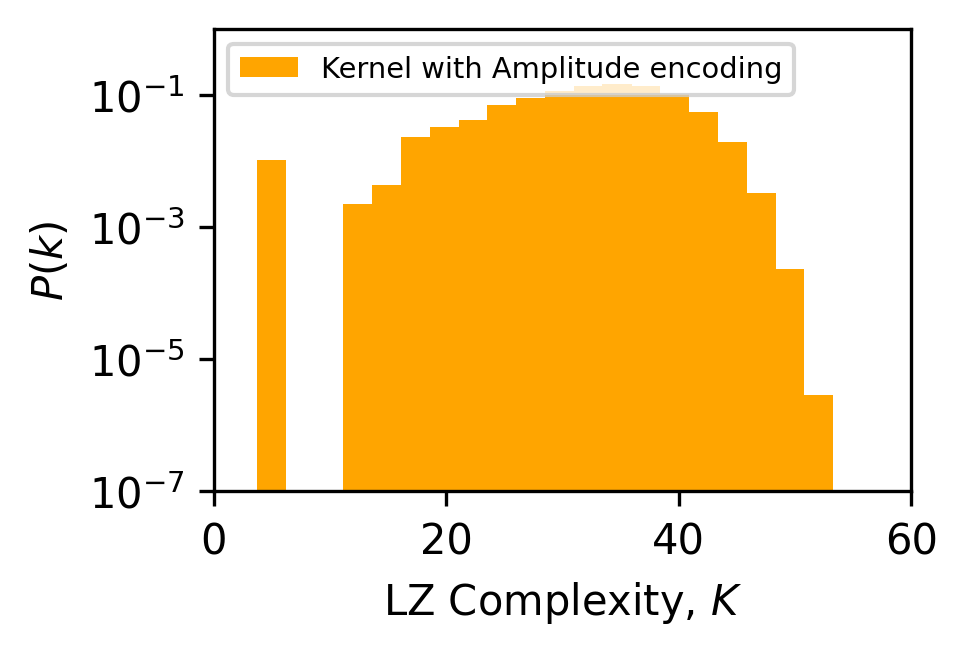}
         \caption{Prior $P(K)$ for the quantum kernel with amplitude encoding}
         \label{fig:kernel_pk_e4}
     \end{subfigure}
     \hfill
     \begin{subfigure}[t]{0.3\textwidth}
         \centering
         \includegraphics[width=\textwidth]{figures/fig_qnn_pk_vs_k/complexity_vs_kc_prob_bar_DNN.png}
         \caption{Prior $P(K)$ for the classical DNN}
         \label{fig:kernel_pk_dnn}
     \end{subfigure}
     \hfill
     \begin{subfigure}[t]{0.3\textwidth}
         \centering
         \includegraphics[width=\textwidth]{figures/fig_qnn_pk_vs_k/complexity_vs_kc_prob_random_learner.png}
         \caption{Prior $P(K)$ for a random learner}
         \label{fig:kernel_pk_random_learner}
     \end{subfigure}
        \caption{\textbf{Probability to obtain a function of Lempel-Ziv complexity $\mathbf{K}$ versus its complexity for five data qubits and $\mathbf{10^7}$ samples.} $P(f)$ versus Lempel-Ziv complexity, $K$ for a quantum kernel with basis encoding (red), ZZ feature map (blue), random relu transform (pink), amplitude encoding (orange), a DNN (green), and a random learner (purple). $P(f)$ is calculated by generating $10^7$ samples of functions from the quantum kernel by using multivariate normal in numpy and $10^5$ samples for the random learner.}
        \label{fig:pk}
\end{figure*}

We also plot the probability of obtaining boolean functions with complexity $K$ versus the complexity $K$ in Figure \ref{fig:pk} as described in Section \ref{sec:qnn_pk}. The trends in Figure \ref{fig:pk} are also similar to those of the QNN shown in Section \ref{sec:qnn_pk}.

\subsubsection{Probability of boolean function vs rank}
\begin{figure*}
     \centering
     \begin{subfigure}[t]{0.49\textwidth}
         \centering
         \includegraphics[width=0.6\textwidth]{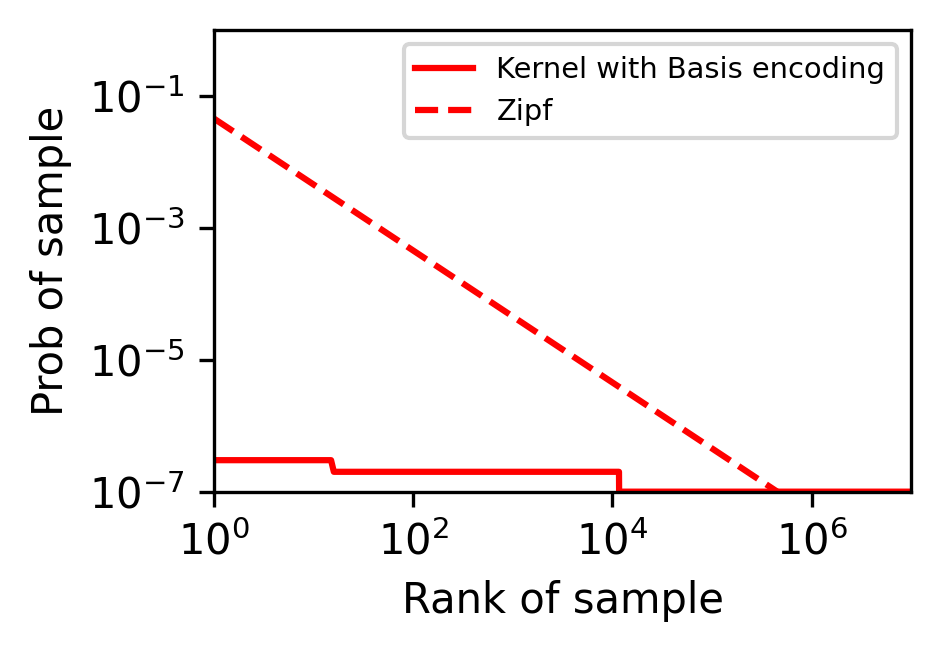}
         \caption{$P(f)$ vs Rank for the quantum kernel with basis encoding}
         \label{fig:kernel_rank_e0}
     \end{subfigure}
     \hfill
     \begin{subfigure}[t]{0.49\textwidth}
         \centering
         \includegraphics[width=0.6\textwidth]{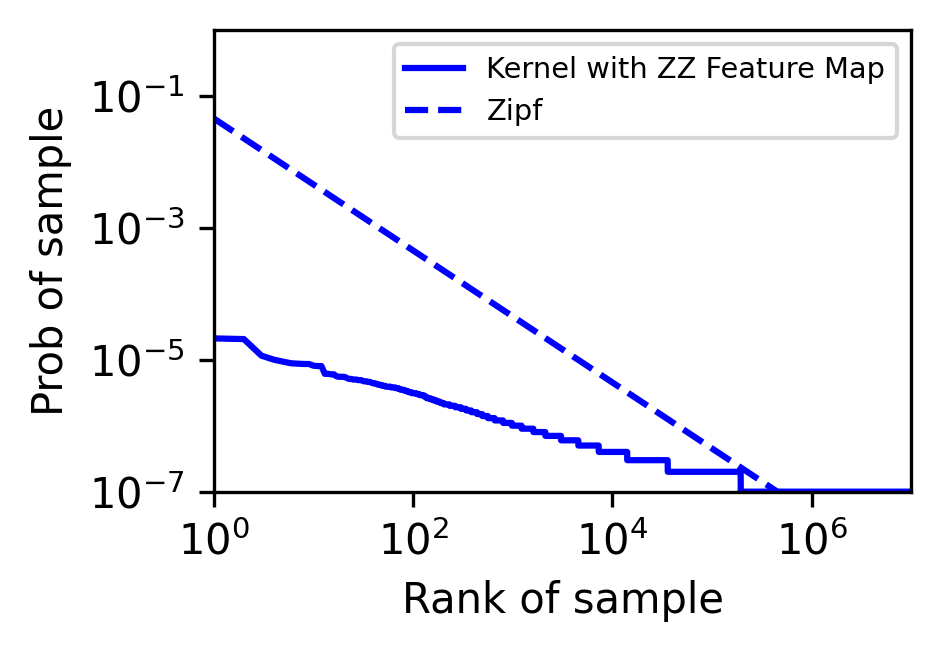}
         \caption{$P(f)$ vs Rank for the quantum kernel with ZZ feature map}
         \label{fig:kernel_rank_e1}
     \end{subfigure}
      \hfill
     \begin{subfigure}[t]{0.49\textwidth}
         \centering
         \includegraphics[width=0.6\textwidth]{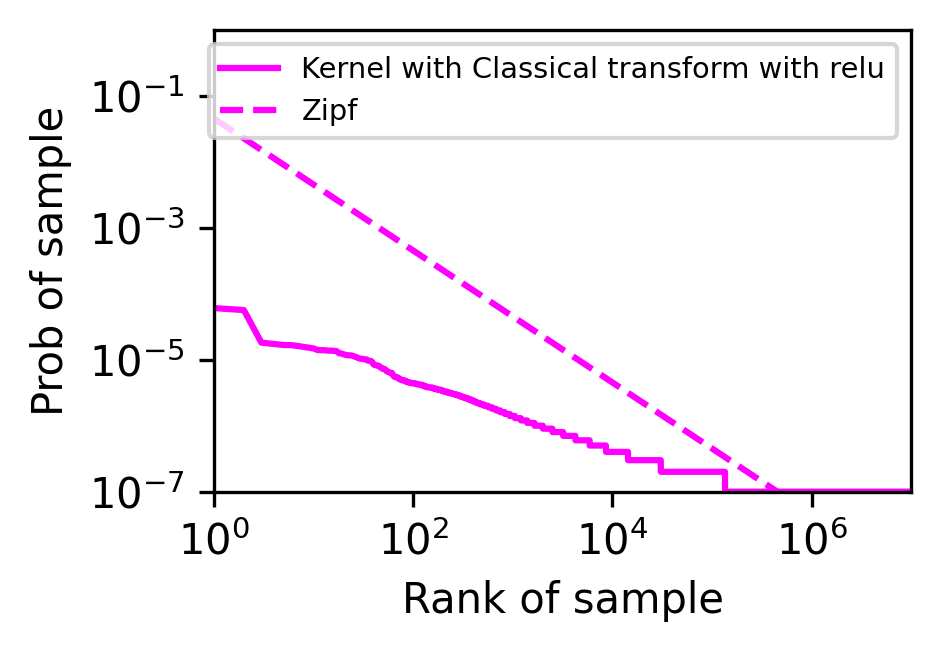}
         \caption{$P(f)$ vs Rank for the quantum kernel with random relu transform}
         \label{fig:kernel_rank_e3}
     \end{subfigure}
     \hfill
     \begin{subfigure}[t]{0.49\textwidth}
         \centering
         \includegraphics[width=0.6\textwidth]{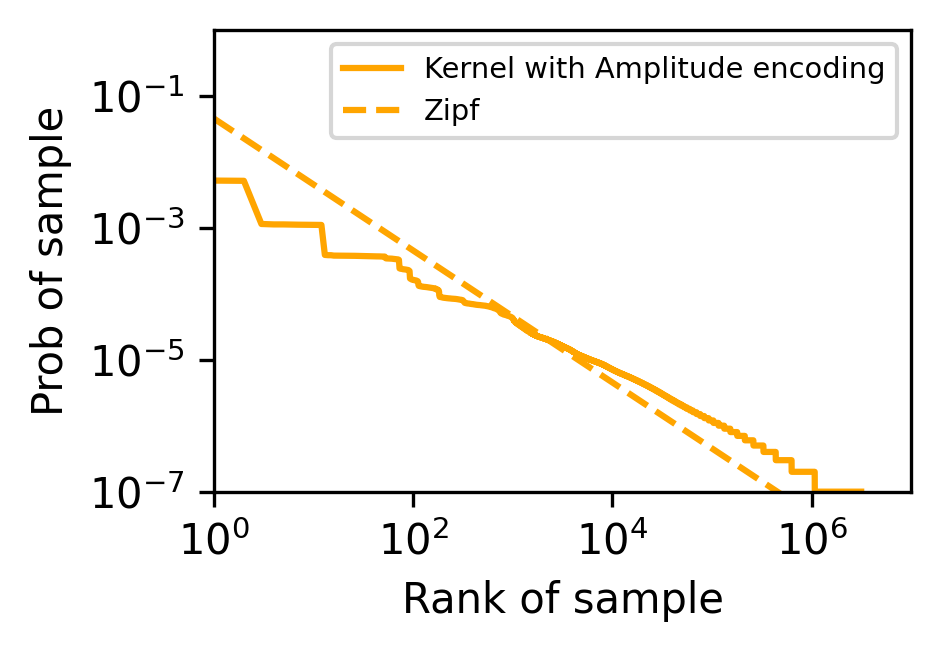}
         \caption{$P(f)$ vs Rank for the quantum kernel with amplitude encoding}
         \label{fig:kernel_rank_e4}
     \end{subfigure}
     \caption{\textbf{Probability of a boolean function versus its rank for five data qubits and $\mathbf{10^5}$ samples}. $P(f)$ ranked by probability of individual functions, generated from $10^7$ samples from the kernel. The dotted line is Zipf's law $P(f) = \frac{1}{(32 ln 2) \text{Rank}(f)}$}
        \label{fig:kernel_rank}
\end{figure*}

We plot the probability of a boolean function versus the rank of the sample in Figure \ref{fig:kernel_rank}. We compare the kernel data with the normalized Zipf law $P(r) = \frac{1}{ln(N_O)r}$ where $r$ is the rank and $N_O = 2^{2^n}$ where $n$ is the number of data qubits. Zipf's law tells us that the frequency of a sample is inversely proportional to its rank. In Figure \ref{fig:kernel_rank} we plot the data for $n=5$ so $N_O = 2^{32}$.

\subsubsection{Entropy vs complexity plots}
\begin{figure*}
     \centering
     \begin{subfigure}[t]{0.49\textwidth}
         \centering
         \includegraphics[width=0.6\textwidth]{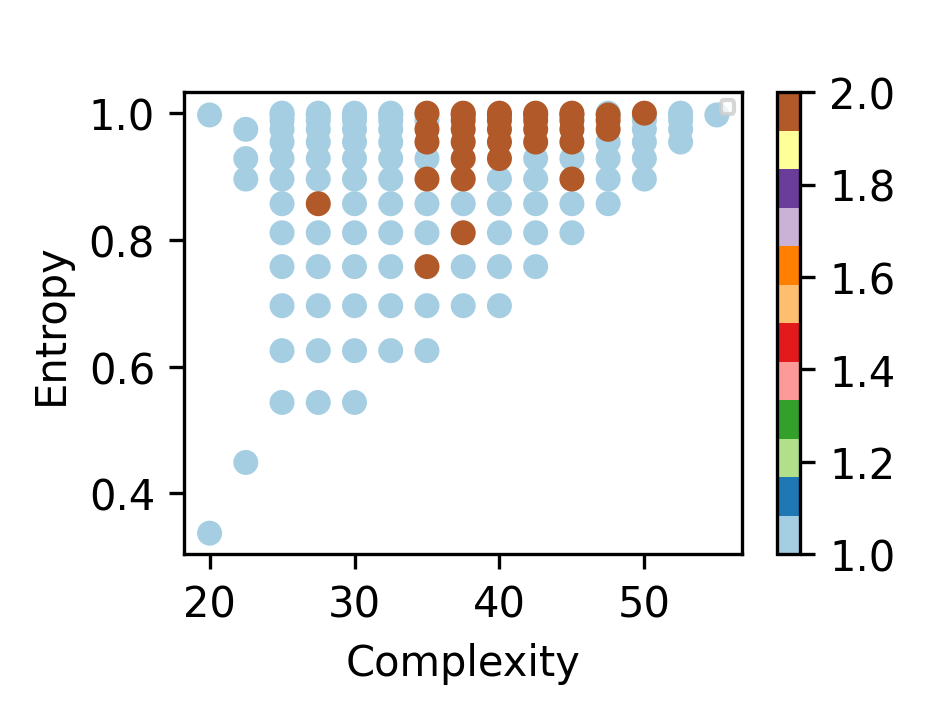}
         \caption{Entropy vs complexity for the quantum kernel with basis encoding}
         \label{fig:kernel_entropy_vs_complexity_basis_encoding}
     \end{subfigure}
     \begin{subfigure}[t]{0.49\textwidth}
         \centering
         \includegraphics[width=0.6\textwidth]{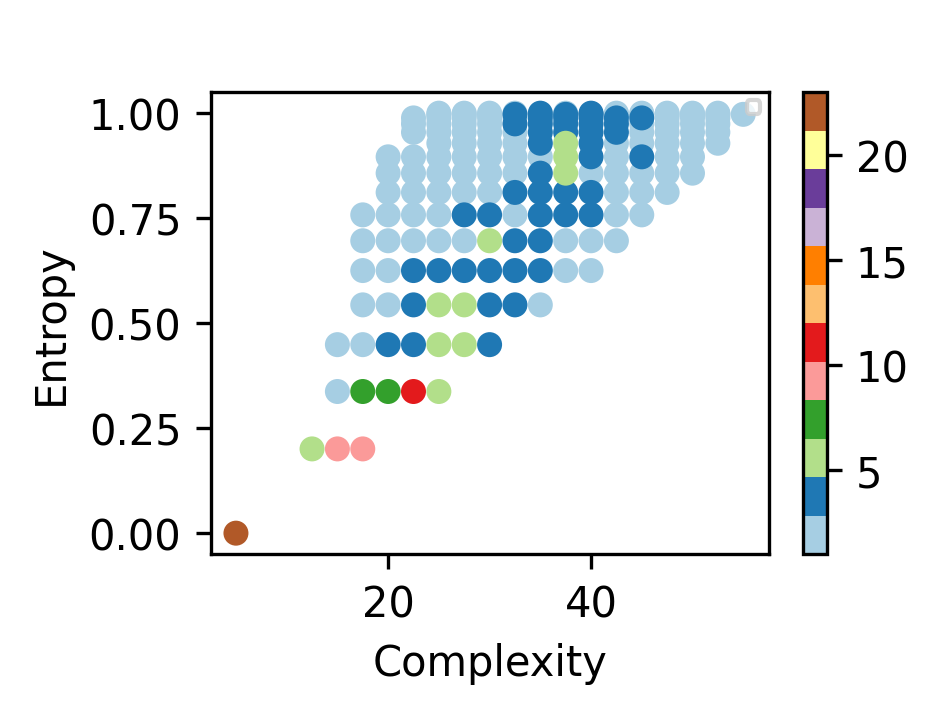}
         \caption{Entropy vs complexity for the quantum kernel with ZZ feature map}
         \label{fig:kernel_entropy_vs_complexity_zz_feature_map}
     \end{subfigure}
     \begin{subfigure}[t]{0.49\textwidth}
         \centering
         \includegraphics[width=0.7\textwidth]{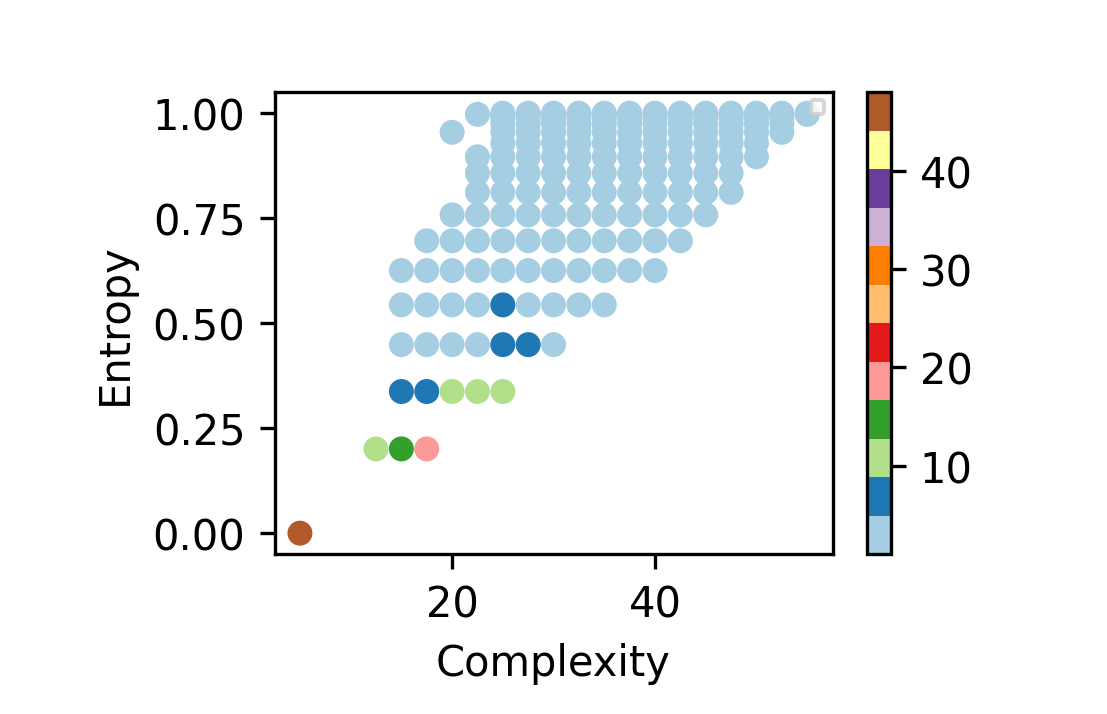}
         \caption{Entropy vs complexity for the quantum kernel with random relu transform}
         \label{fig:kernel_entropy_vs_complexity_classical_transform}
     \end{subfigure}
     \begin{subfigure}[t]{0.49\textwidth}
         \centering
         \includegraphics[width=0.7\textwidth]{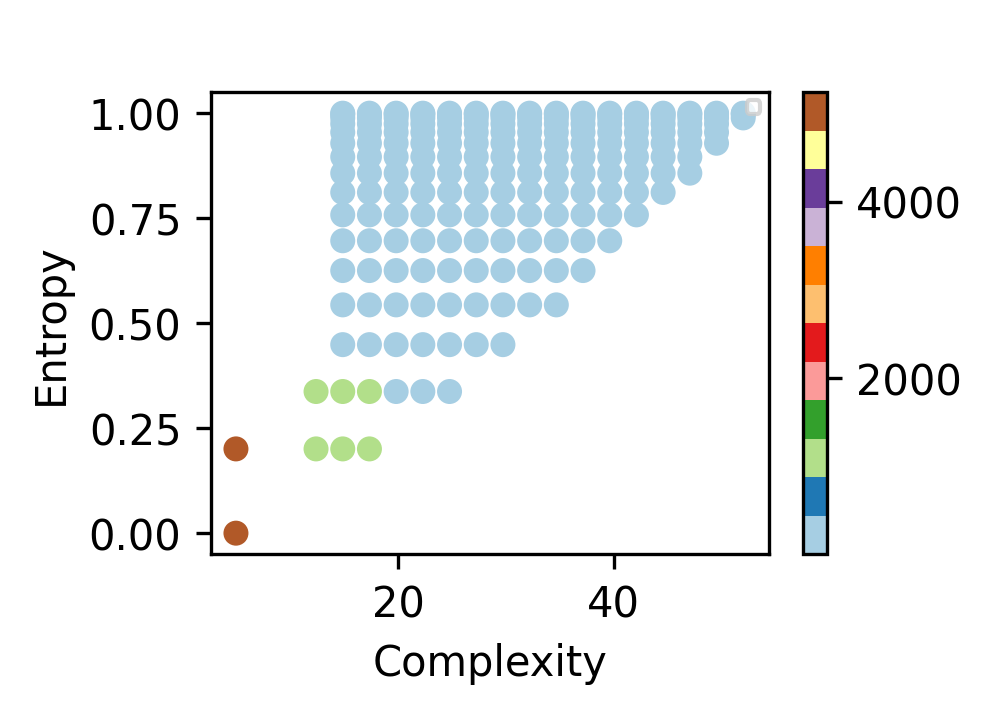}
         \caption{Entropy vs complexity for the quantum kernel with amplitude encoding}
         \label{fig:kernel_entropy_vs_complexity_amplitude_encoding}
     \end{subfigure}
        \caption{\textbf{Entropy of a boolean function $\mathbf{(f)}$ versus its complexity for five data qubits and $\mathbf{10^6}$ samples.} The colour of the datapoint corresponds to the maximum count of a function that has that complexity and entropy. The boolean functions produced by the quantum kernel were grouped and for each entropy-complexity pair, there were multiple functions, so the maximum count of a function is displayed.}
        \label{fig:kernel_entropy_vs_complexity}
\end{figure*}

Another way to visualise whether the quantum kernels have a simplicity or entropy bias is to plot the entropy vs complexity and look at the count of each datapoint. Figure \ref{fig:kernel_entropy_vs_complexity} shows the entropy vs complexity where the colour of the datapoint corresponds to the maximum count of a function that has that complexity and entropy. 

Figure \ref{fig:kernel_entropy_vs_complexity_basis_encoding} for basis encoding shows that there is neither entropy nor complexity bias as the blue and brown datapoints correspond to one and two counts of the boolean functions respectively. The difference between the counts is small and does not indicate any bias.

\FloatBarrier
\subsection{Generalisation error of QNNs}
\label{sec:generalisation_error_qnns}

\subsubsection{Target functions}
For training the QNNs to calculate the generalisation error, we use constructed and random target functions. A list of the target functions we use are shown in Table \ref{tab:target_functions}. Table \ref{tab:target_functions_detail} includes further details about these target functions. 

\begingroup \renewcommand{\arraystretch}{2}
\begin{table}
\begin{tabularx}{\textwidth}{
  | p{2cm}
  | p{2cm}
  | p{6cm}
  | p{7.3cm} |}
 \hline
    \textbf{No.} & \textbf{Symbol} & \textbf{Function} & \textbf{Description} \\
 \hline
  1 & \includegraphics[width=5mm]{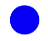} & 00000000000000000000000000000000 & all 0s \\
  \hline
  2 & \includegraphics[width=5mm]{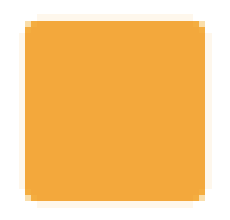} & 01001110100010110111100001100010 & random \\
  \hline
  3 & \includegraphics[width=5mm]{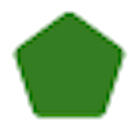} & 11111111111111111111111111111111  & all 1s \\
  \hline
  4 & \includegraphics[width=5mm]{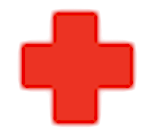} & 00000000000000001000000000000000 & one 1 \\
  \hline
  5 & \includegraphics[width=5mm]{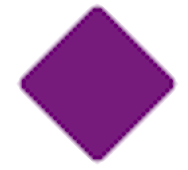} & 11110111111111111111111111011111  & two 0s \\
  \hline
  6 & \includegraphics[width=5mm]{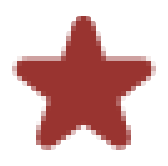} & 10000001110101011000101111010000 & ZZ sample \\
  \hline
  7 & \includegraphics[width=5mm]{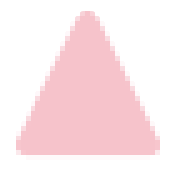} & 01101001100101101001011001101001 & parity \\
  \hline
  8 & \includegraphics[width=5mm]{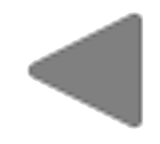} & 01010101010101010101010101010101 & 01 repeat \\
    \hline
  9 & \includegraphics[width=5mm]{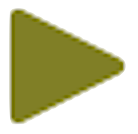} & 10101010101010101010101010101010 & 10 repeat \\
  \hline
 10 & \includegraphics[width=5mm]{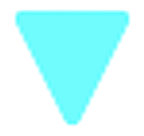} & 11111111111111110000000000000000 & halfs 1 half 0s \\
    \hline
 11 & \includegraphics[width=5mm]{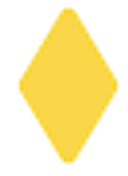} & 00000000000000001111111111111111 & half 0s half 1s \\
    \hline
 12 & \includegraphics[width=5mm]{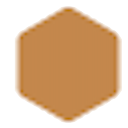} & 11101110111011101110111011101110 & random length 4 function repeated 8 times \\
    \hline
 13 & \includegraphics[width=5mm]{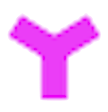} & 01011111010111110101111101011111 & random length 8 function repeated 4 times \\
    \hline
\end{tabularx} 
 \caption{\textbf{Target functions for five qubits}. The first column is the number of the target function for identifying it. The second column is the symbol used for the target function. The third column is the target function itself and the fourth column is a description of the target function.}
 \label{tab:target_functions}
\end{table}
\endgroup

\begingroup \renewcommand{\arraystretch}{2}
\begin{table}
\begin{tabularx}{\textwidth}{ 
  | >{\centering\arraybackslash}X 
  | >{\centering\arraybackslash}X
  | >{\centering\arraybackslash}X
  | >{\centering\arraybackslash}X
  | >{\centering\arraybackslash}X
  | >{\centering\arraybackslash}X
  | >{\centering\arraybackslash}X
  | >{\centering\arraybackslash}X 
  | >{\centering\arraybackslash}X
  | >{\centering\arraybackslash}X
  | >{\centering\arraybackslash}X
  | >{\centering\arraybackslash}X
  | >{\centering\arraybackslash}X
  | >{\centering\arraybackslash}X
  |}
  \hline
   &  &  \multicolumn{2}{c|}{\textbf{Value}} & \multicolumn{2}{c|}{\textbf{Category}} & \multicolumn{4}{c|}{\textbf{QNN $P(f)$}} & \multicolumn{4}{c|}{\textbf{Quantum kernel $P(f)$}} \\
 \hline
  \textbf{No.} &  \textbf{Symbol} & \textbf{K} & \textbf{E} & \textbf{K} & \textbf{E} & \textbf{Basis} & \textbf{ZZ} & \textbf{Relu} & \textbf{Amp.} & \textbf{Basis} & \textbf{ZZ} & \textbf{Relu} & \textbf{Amp.}  \\
 \hline
  1 & \includegraphics[width=5mm]{figures/markers/circle.png} & 5 & 0 & Low & Low & 0 & $3\cdot10^{-2}$ & $8\cdot10^{-4}$ & $3\cdot10^{-2}$ & 0 & 0 & 0 & 0 \\
  \hline
  2 & \includegraphics[width=5mm]{figures/markers/square.png} & 55 & 1 & High & High & 0 & 0 & 0 & 0 & 0 & 0 & 0 & 0 \\
  \hline
  3 & \includegraphics[width=5mm]{figures/markers/pentagon.png} & 5 & 0 & Low & Low & 0 & $3\cdot10^{-2}$ & $8\cdot10^{-4}$ & $3\cdot10^{-2}$ & 0 & 0 & 0 & 0 \\
  \hline
  4 & \includegraphics[width=5mm]{figures/markers/cross.png} & 15 & 0.2 & Low & Low & 0 & $5\cdot10^{-3}$ & $9\cdot10^{-5}$ & $3\cdot10^{-3}$ & 0 & $8\cdot10^{-6}$ & $1\cdot10^{-5}$ & 0  \\
  \hline
  5 & \includegraphics[width=5mm]{figures/markers/large_diamond.png} & 25 & 0.33 & Med. & Low & 0 & $7\cdot10^{-4}$ & $2\cdot10^{-5}$ & 0 & 0 & $7\cdot10^{-7}$ & $1\cdot10^{-6}$ & 0 \\
  \hline
  6 & \includegraphics[width=5mm]{figures/markers/star.png} & 45 & 0.99 & High & High & 0 & 0 & 0 & 0 & 0 & 0 & 0 & 0 \\
  \hline
  7 & \includegraphics[width=5mm]{figures/markers/up_triangle.png} & 45 & 1 & High & High & 0 & 0 & 0 & 0 & 0 & $1\cdot10^{-8}$ & 0 & 0 \\
  \hline
  8 & \includegraphics[width=5mm]{figures/markers/left_triangle.png} & 15 & 1 & Low & High & 0 & 0 & 0 & $2\cdot10^{-4}$ & 0 & 0 & 0 & $1\cdot10^{-4}$\\
    \hline
  9 & \includegraphics[width=5mm]{figures/markers/right_triangle.png} & 15 & 1 & Low & High & 0 & 0 & 0 & $2\cdot10^{-4}$ & $1\cdot10^{-8}$ & 0 & 0 & $1\cdot10^{-4}$ \\
    \hline
 10 & \includegraphics[width=5mm]{figures/markers/down_triange.png} & 15 & 1 & Low & High & 0 & 0 & 0 & $1\cdot10^{-4}$ & 0 & 0 & 0 & $1\cdot10^{-4}$ \\
    \hline
 11 & \includegraphics[width=5mm]{figures/markers/thin_diamond.png} & 15 & 1 & Low & High & 0 & 0 & 0 & $1\cdot10^{-4}$ & 0 & 0 & 0 & $1\cdot10^{-4}$ \\
    \hline
 12 & \includegraphics[width=5mm]{figures/markers/hexagon.png} & 17.5 & 0.81 & Low & High & 0 & 0 & 0 & 0 & 0 & 0 & 0 & $4\cdot10^{-6}$ \\
    \hline
 13 & \includegraphics[width=5mm]{figures/markers/three.png} & 22.5 & 0.81 & Low & High & 0 & 0 & 0 & $2\cdot10^{-4}$ & 0 & 0 & 0 & $8\cdot10^{-5}$ \\
    \hline
\end{tabularx} 
 \caption{\textbf{Target functions for five qubits}. We show the LZ complexity (K) and the entropy (E) of the target function in the third and fourth columns respectively. We categorise the complexity and entropy in the fifth and sixth columns. For the following columns, we show the probability of obtaining the target function $f$ from the QNN and Quantum kernel with the different encoding methods.}\label{tab:target_functions_detail}
\end{table} \endgroup

\subsubsection{Generalisation error vs complexity, entropy, and probability of target function}

\begin{figure*}
     \centering
     \begin{subfigure}[t]{0.3\textwidth}
         \centering
         \includegraphics[width=\textwidth]{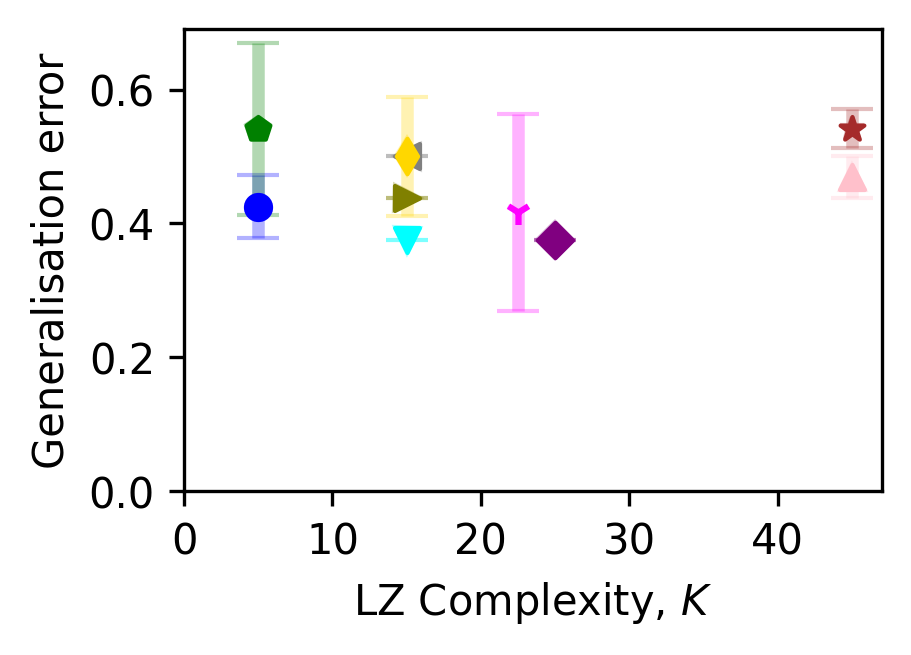}
         \caption{Generalisation error vs complexity for QNN with basis encoding}
         \label{fig:fig:qnn_error_vs_complexity_e0}
     \end{subfigure}
     \begin{subfigure}[t]{0.3\textwidth}
         \centering
         \includegraphics[width=\textwidth]{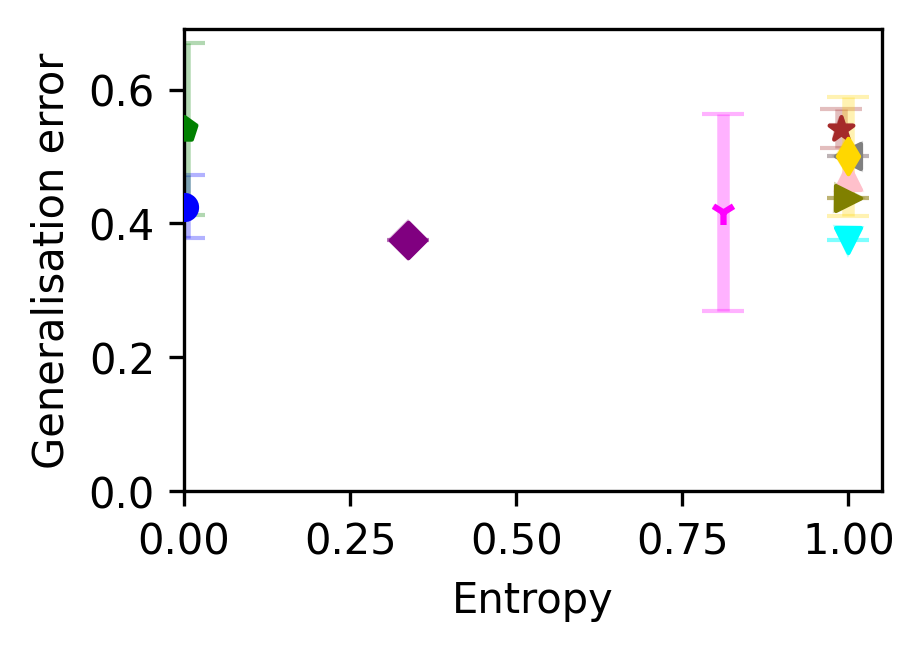}
         \caption{Generalisation error vs entropy for QNN with basis encoding}
         \label{fig:qnn_error_vs_entropy_e1}
     \end{subfigure}
     \begin{subfigure}[t]{0.3\textwidth}
         \centering
         \includegraphics[width=\textwidth]{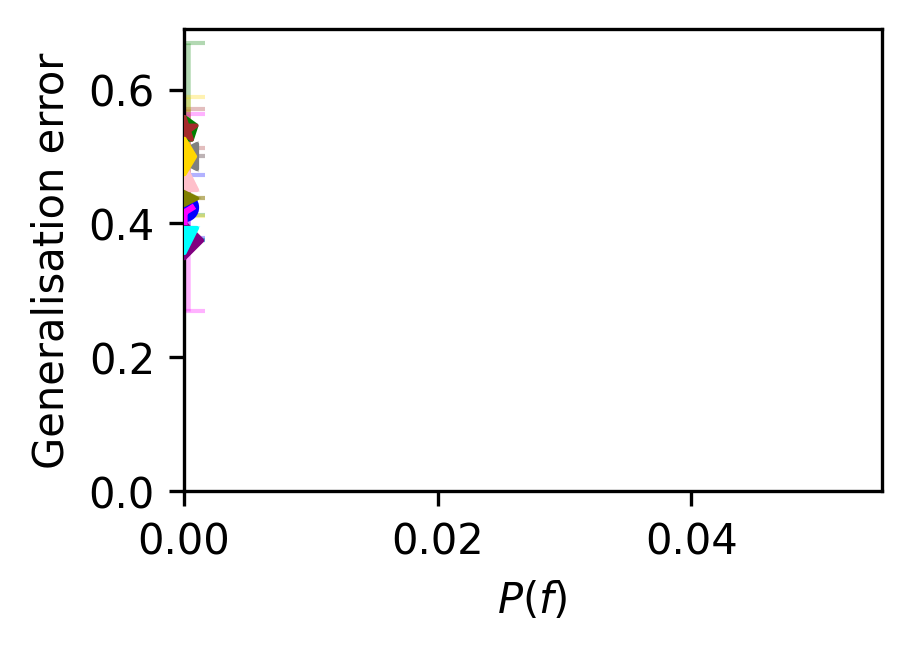}
         \caption{Generalisation error vs probability for QNN with basis encoding}
         \label{fig:fig:qnn_error_vs_prob_e0}
     \end{subfigure}

     \begin{subfigure}[t]{0.3\textwidth}
         \centering
         \includegraphics[width=\textwidth]{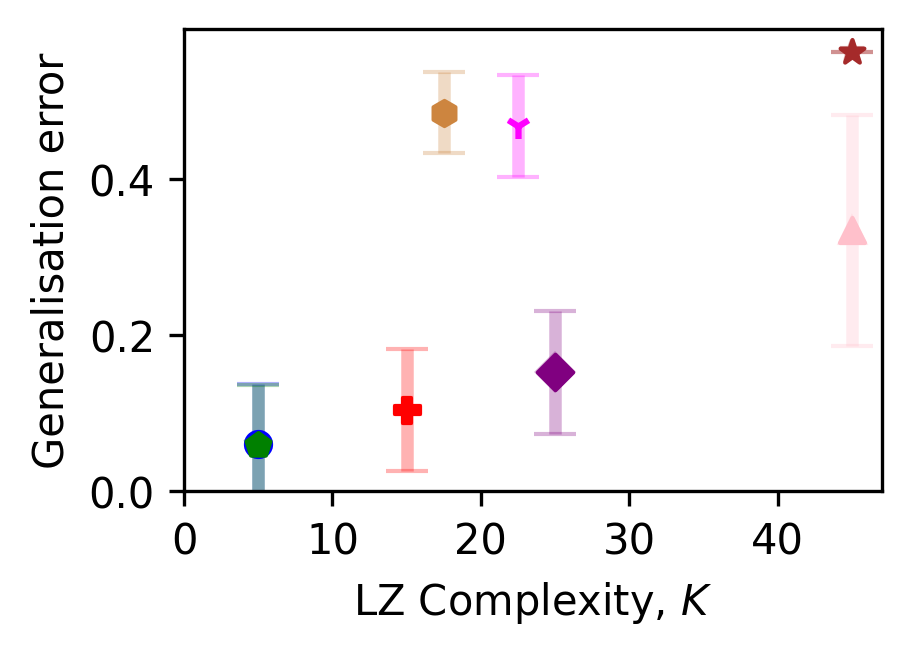}
         \caption{Generalisation error vs complexity for QNN with ZZ Feature Map}
         \label{fig:fig:qnn_error_vs_complexity_e1}
     \end{subfigure}
     \begin{subfigure}[t]{0.3\textwidth}
         \centering
         \includegraphics[width=\textwidth]{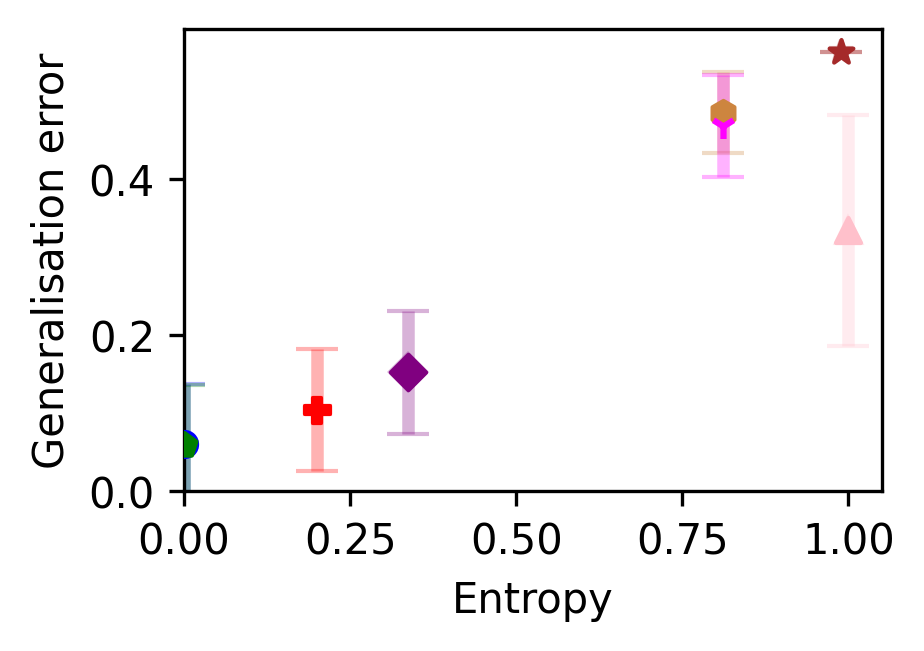}
         \caption{Generalisation error vs entropy for QNN with ZZ Feature Map}
         \label{fig:fig:qnn_error_vs_entropy_e1}
     \end{subfigure}
     \begin{subfigure}[t]{0.3\textwidth}
         \centering
         \includegraphics[width=\textwidth]{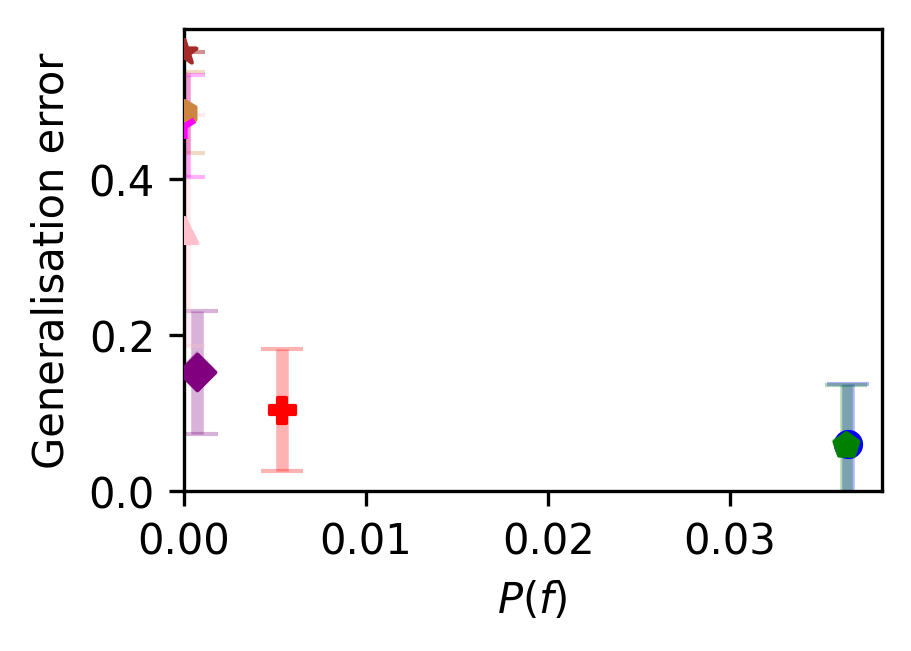}
         \caption{Generalisation error vs probability for QNN with ZZ Feature Map}
         \label{fig:fig:qnn_error_vs_prob_e1}
     \end{subfigure}

     \begin{subfigure}[t]{0.3\textwidth}
         \centering
         \includegraphics[width=\textwidth]{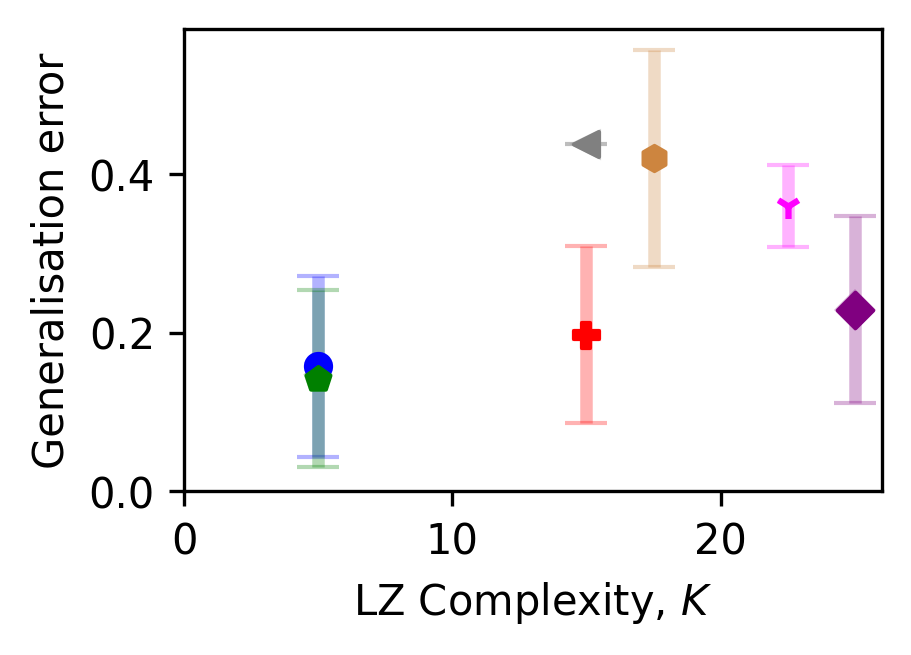}
         \caption{Generalisation error vs complexity for QNN with relu transform}
         \label{fig:qnn_error_vs_complexity_e3}
     \end{subfigure}
     \begin{subfigure}[t]{0.3\textwidth}
         \centering
         \includegraphics[width=\textwidth]{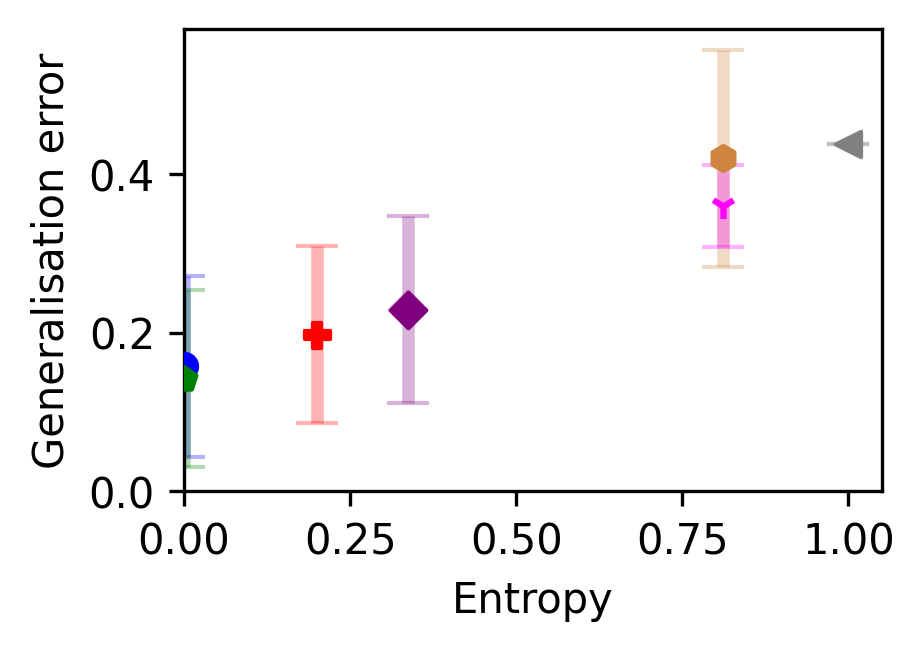}
         \caption{Generalisation error vs entropy for QNN with relu transform}
         \label{fig:qnn_error_vs_entropy_e3}
     \end{subfigure}
     \begin{subfigure}[t]{0.3\textwidth}
         \centering
         \includegraphics[width=\textwidth]{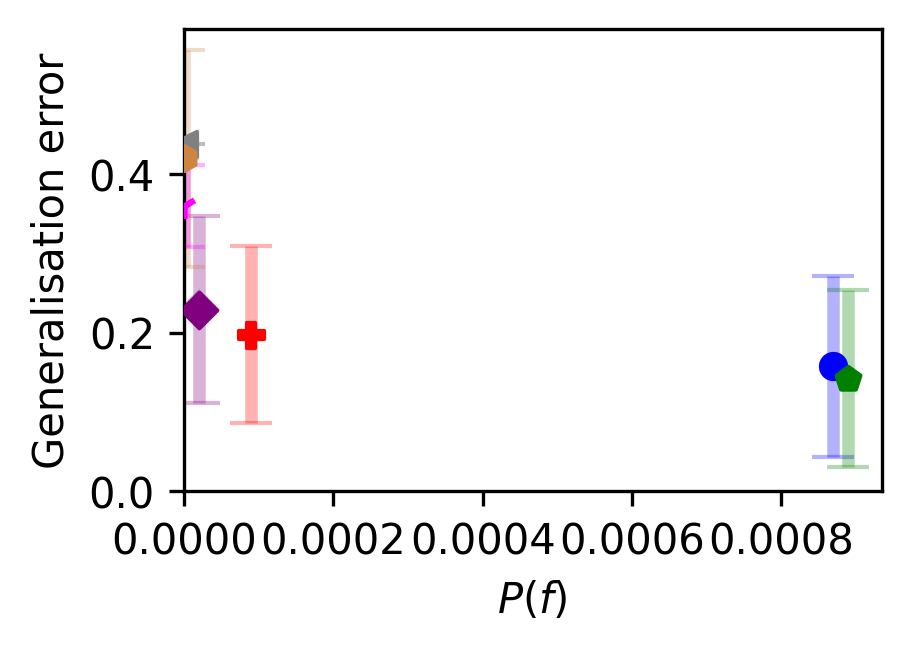}
         \caption{Generalisation error vs probability for QNN with relu transform}
         \label{fig:qnn_error_vs_prob_e3}
     \end{subfigure}

     \begin{subfigure}[t]{0.3\textwidth}
         \centering
         \includegraphics[width=\textwidth]{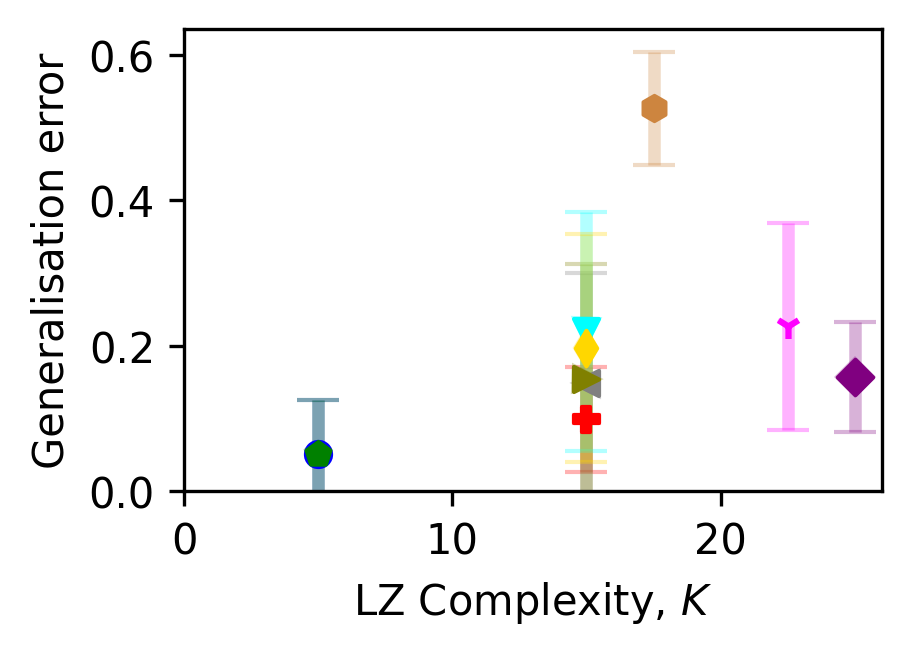}
         \caption{Generalisation error vs complexity for QNN with amplitude encoding}
         \label{fig:qnn_error_vs_complexity_e4}
     \end{subfigure}
     \begin{subfigure}[t]{0.3\textwidth}
         \centering
         \includegraphics[width=\textwidth]{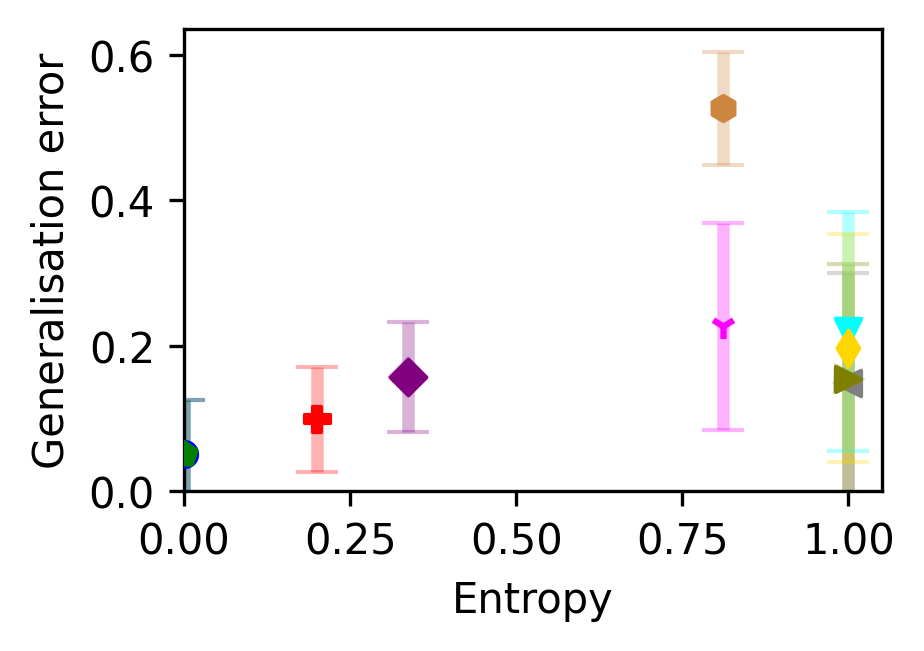}
         \caption{Generalisation error vs entropy for QNN with amplitude encoding}
         \label{fig:qnn_error_vs_entropy_e4}
     \end{subfigure}
     \begin{subfigure}[t]{0.3\textwidth}
         \centering
         \includegraphics[width=\textwidth]{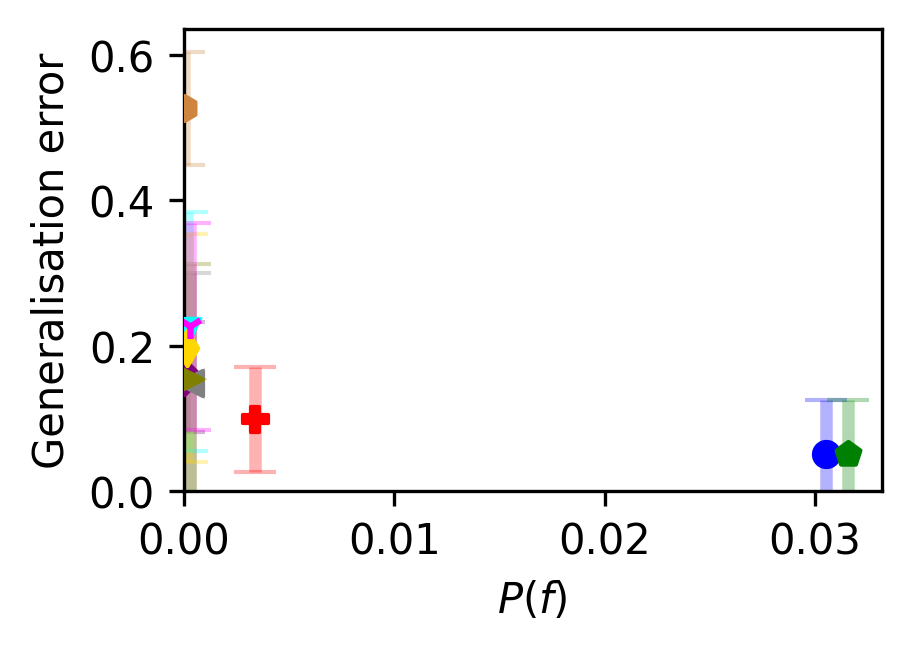}
         \caption{Generalisation error vs probability for QNN with amplitude encoding}
         \label{fig:qnn_error_vs_prob_e4}
     \end{subfigure}
        \caption{\textbf{Generalisation error of a fully expressive QNN with different encoding methods for five data qubits} \textbf{(a)} Generalisation error versus Lempel-Ziv complexity of the target function for a QNN with different encoding methods. The QNN is trained to zero training error on a training set $S$ of size $m = 16$ and the generalisation error is calculated on the remaining $16$ functions. The QNNs are trained by randomly sampling parameters $10^5$ times and finding the generalisation error on the set of parameters that obtain zero training error. Error bars are one standard deviation. \textbf{(b)} Generalisation error versus entropy of the target function with the same specifications listed in (a). \textbf{(c)} Generalisation error vs probability of a target function for a QNN with different encoding methods. $10^5$ samples of functions are generated from the QNN by using random samples of parameters. The probability of each function is calculated and its generalisation error is calculated.}
        \label{fig:qnn_error}
\end{figure*}

In Figure \ref{fig:qnn_error}, we chose constructed and random target functions (as shown in Table \ref{tab:target_functions} and Table \ref{tab:target_functions_detail}) and train the QNN with different encoding methods on these target functions and calculate the generalisation error. We then plot the generalisation error versus the complexity of the target function, the entropy of the target function, and the probability of obtaining the target function when randomly sampling the QNN. Other details about how the QNNs were trained are in Section \ref{sec:generalisation_error_qnns_simple}.

These results show that the QNN with basis encoding has the equivalent performance of a random learner as across all target functions, it has a 0.5 generalisation error, regardless of the complexity or entropy of the target function. 

The QNN with the ZZ feature map generalises well on low-complexity low-entropy functions, such as the all 0s function. However, we can see two target functions that break the trend of low-complexity low-generalisation error. Those are the target functions 12 and 13 which have low complexity but high entropy. The QNN with the ZZ feature map generalises poorly on these, with around a 0.5 error.

If we look at the entropy plot for the QNN with the ZZ feature map in Figure \ref{fig:qnn_error_vs_entropy_e1}, we see that this method has a bias towards low entropy functions. 

This pattern is similar for the random relu transform, which is a random unitary matrix applied to the input data and then put through a relu layer. Target functions 8 and 12 have low complexity but high entropy and the relu transform generalises poorly on these target functions. The entropy plot, Figure \ref{fig:qnn_error_vs_entropy_e3}, shows that this transform also has a bias towards low entropy functions.

The QNN with amplitude encoding shows a different story. For target functions with low complexity and high entropy, target functions 8 - 13, we see that the generalisation error is low, showing that amplitude encoding has a bias towards low complexity functions, similar to the DNN.

\subsubsection{Generalisation error vs complexity of output function}

\begin{figure*}
     \centering
     \begin{subfigure}[t]{0.3\textwidth}
         \centering
         \includegraphics[width=\textwidth]{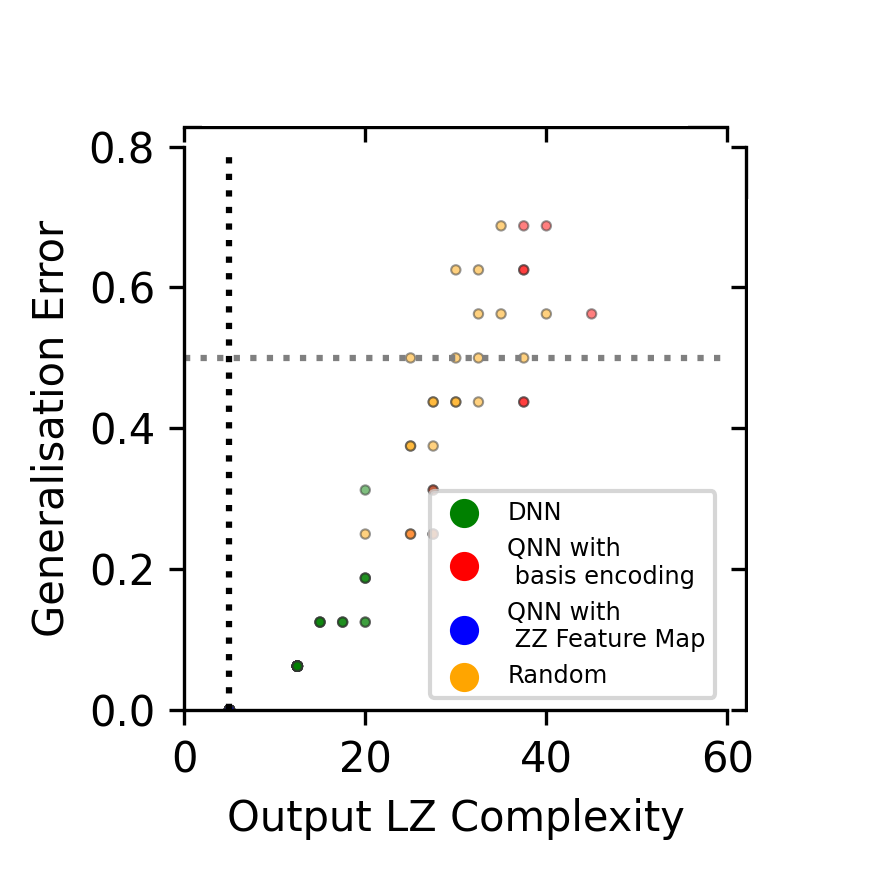}
         \caption{Target function LZ = 5}
         \label{fig:generalisation_error_vs_complexity_qnn_scatterplots_target_5}
     \end{subfigure}
     \hfill
     \begin{subfigure}[t]{0.3\textwidth}
         \centering
         \includegraphics[width=\textwidth]{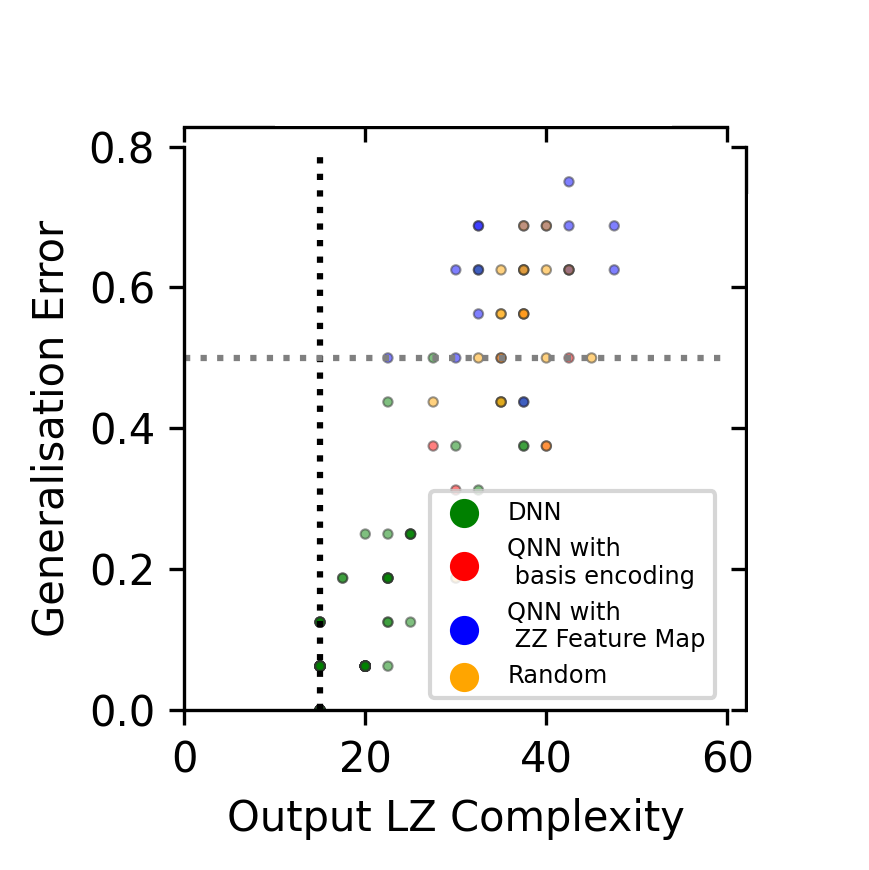}
         \caption{Target function LZ = 15}
         \label{fig:generalisation_error_vs_complexity_qnn_scatterplots_target_15}
     \end{subfigure}
     \hfill
     \begin{subfigure}[t]{0.3\textwidth}
         \centering
         \includegraphics[width=\textwidth]{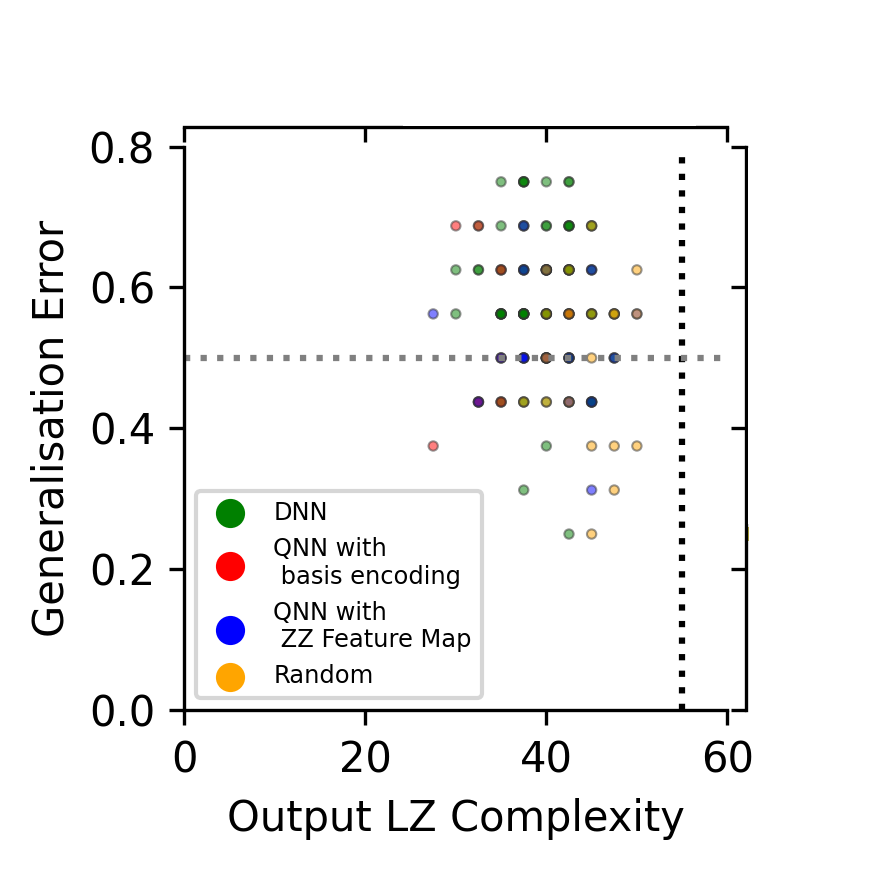}
         \caption{Target function LZ = 55}
         \label{fig:generalisation_error_vs_complexity_qnn_scatterplots_target_55}
     \end{subfigure}
        \caption{\textbf{Generalisation error versus output function LZ complexity scatterplots for $n=5$ and for 20 random initialisations for three target functions}. The output LZ complexity is the complexity of the target function that the neural network produces after it has been trained to zero training error. A neural network that produces an output target function that is equivalent to the specified target function will have zero generalisation error and the same output LZ complexity as the specified target function. Each datapoint corresponds to one iteration. The dashed vertical line denotes the target function complexity.}
        \label{fig:generalisation_error_vs_complexity_qnn_scatterplots}
\end{figure*}

Each figure in Figure \ref{fig:generalisation_error_vs_complexity_qnn_scatterplots} shows the generalisation error vs output LZ complexity for a specified target function.  

Figure \ref{fig:generalisation_error_vs_complexity_qnn_scatterplots_target_5} shows the generalisation error vs output LZ complexity for a simple target function with Lempel-Ziv complexity $LZ = 5$. The plot shows that the DNN datapoints remain in the lower left of the plot as it produces target functions with low complexity and low generalisation error. The random learner produces random target functions (i.e. random boolean functions), and given that there are exponentially more complex functions than simple functions, the output target functions it produces are more likely to be complex and therefore have high generalisation error. The QNN with basis encoding is similar to the random learner and also produces high complexity output target functions even when trained on a low complexity target function.

Figures \ref{fig:generalisation_error_vs_complexity_qnn_scatterplots_target_15} and \ref{fig:generalisation_error_vs_complexity_qnn_scatterplots_target_55} show the results for target functions with higher complexities. We see that the datapoints for the DNN shift towards the upper right of the graph as the DNN produces target functions with higher complexity and higher generalisation error. The random learner and QNN with basis encoding continue to produce high complexity output functions, which have high generalisation error.

\FloatBarrier
\FloatBarrier
\subsection{Quantum generalisation performance on parity function}

\begin{figure*}[h]
     \centering
     \begin{subfigure}[t]{0.4\textwidth}
         \centering
         \includegraphics[width=\textwidth]{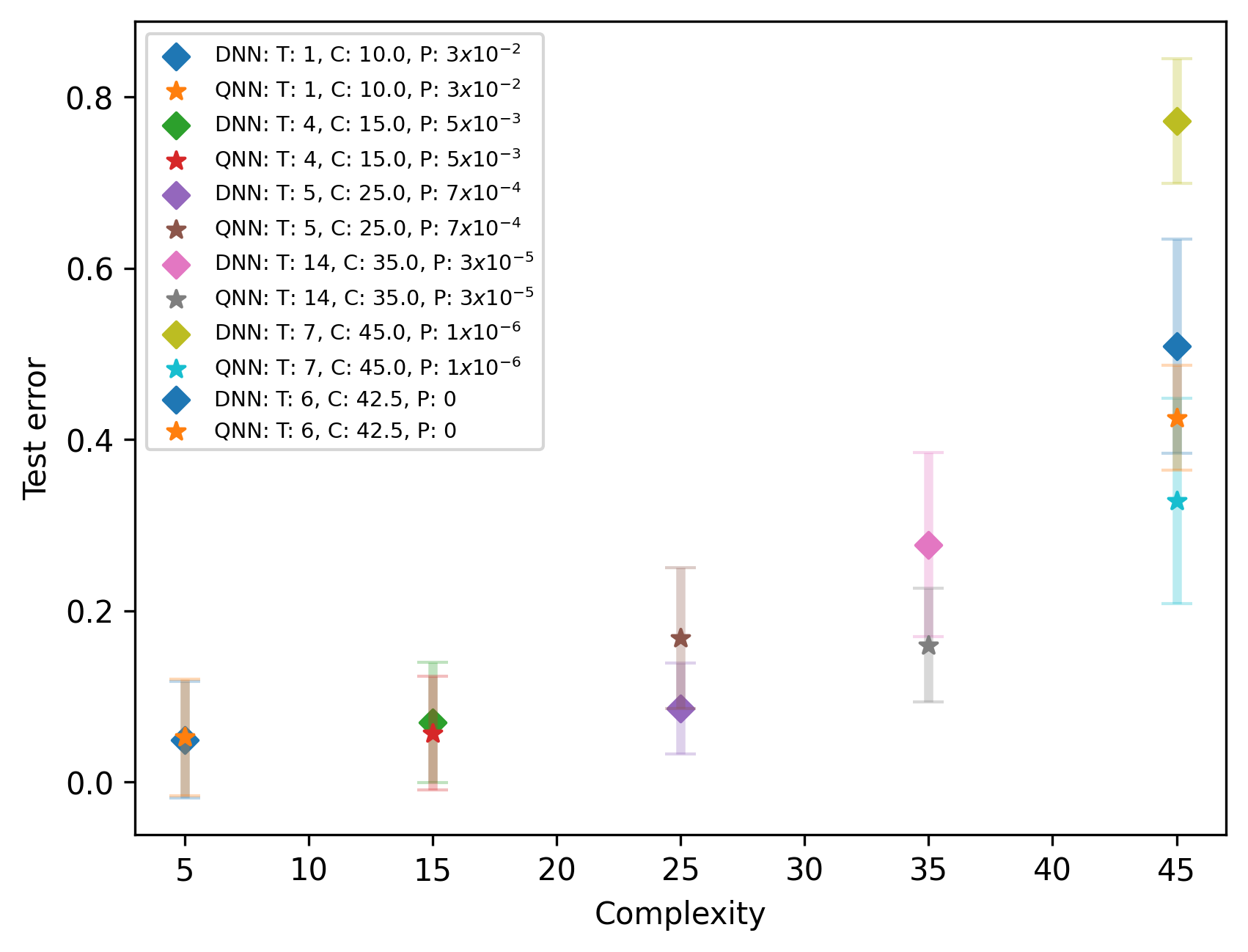}
         \caption{Generalisation error vs complexity. of the QNN with ZZ Feature Map vs DNN}
         \label{fig:generalisation_error_vs_complexity_qnn_zz_feature_map}
     \end{subfigure}
        \caption{\textbf{Generalisation error versus Lempel-Ziv complexity} for $n=5$ for different target functions $T$ training to zero error on a training set $S$ of size $m = 16$, with the error calculated on the remaining $|T| = 16$ functions. The target functions correspond to those in Table \ref{tab:target_functions} and Table \ref{tab:target_functions_detail}. An additional target function was added in Table \ref{tab:target_function_additional}. The networks were trained with COBYLA optimiser on MSE loss for 20 random initialisations. Error bars are one standard deviation. The stars are the datapoints for the QNN with the ZZ feature map and the diamonds are the datapoints for the DNN.The legend indicates the LZ complexity (C) of the target function and the probability of obtaining the target function in $10^6$ samples.}
        \label{fig:generalisation_error_vs_complexity}
\end{figure*}

Target function 7 in Figure \ref{fig:generalisation_error_vs_complexity_qnn_zz_feature_map} is the parity function (whose value is one if and only if the input vector has an odd number of ones). The classical perceptron can not express the XOR function (the parity function on two inputs). In Figure \ref{fig:generalisation_error_vs_complexity_qnn_zz_feature_map}, the QNN with the ZZ feature map has a lower generalisation error for the parity function compared to the DNN. The parity function has high complexity ($LZ$ complexity = 45) and therefore we'd expect a high generalisation error for the DNN as the DNN has a bias towards simple functions. One might also expect the QNN to have a high generalisation error on the parity function as it has a high complexity. We see the opposite however -- the QNN with the ZZ feature map has a low generalisation error on the parity function.

This is because the QNN with the ZZ feature map has a stronger bias towards the parity function (it was engineered to do so) \cite{havlicek_supervised_2019}. The QNN with the ZZ feature map has a low generalisation error (around 0.25) for the parity function. This can also be seen in comparison with the other target functions with complexity 45, which have generalisation errors around 0.5. 

It should be noted, however, that another DNN could be constructed to have a strong bias towards the parity function. Although the standard DNN has a poor generalisation performance on the parity function and the QNN has a strong generalisation performance on the parity function, which could seem like a quantum advantage, another type of DNN could be constructed to also have a strong generalisation performance on the parity function. 

\begingroup \renewcommand{\arraystretch}{2}
\begin{table}[h]
\begin{tabularx}{\textwidth}{
  | >{\centering\arraybackslash}X
  | >{\centering\arraybackslash}X
  | >{\centering\arraybackslash}X
  | >{\centering\arraybackslash}X |}
 \hline
    \textbf{No.} & \textbf{Function} & \textbf{Description} \\
 \hline
  14 & 01000001000000101000000000100000 & ZZ sample \\
  \hline
\end{tabularx}
\caption{\textbf{Additional target function for five qubits: } this target function with LZ complexity = 35 is used in Figure \ref{fig:generalisation_error_vs_complexity}. }\label{tab:target_function_additional}
\end{table}\endgroup

\newpage
\FloatBarrier
\subsection{Training QNNs}
\label{sec:training_qnns_results}
In this section, we show the plots of the training process of the QNNs using an optimizer. It was difficult to train via this method because it was hard to obtain zero training error for certain encoding methods. Therefore, we trained via random sampling to obtain the results in the main text of the paper (see Section \ref{sec:generalisation_error_qnns_simple}). In this section, we show the results of the training process using the optimizer. 

We want to train the QNNs to learn a particular boolean function, which is the target function. Further details about the training method can be read in Section \ref{sec:training_qnns_simple}. We used Qiskit's EstimatorQNN to train the QNNs using the COBYLA and SPSA optimizers. The training score is obtained by running the training inputs through the QNN with the training parameters and obtaining the predicted $y$ labels. The accuracy score $s$ is calculated by the fraction of positions where the predicted label $y_{\text{predict}}$ and the training label $y_{\text{train}}$ match. The error is $1-s$. We ensure that the QNN trains to zero training error (i.e. that the predicted label $y_{\text{predict}}$ exactly matches the training label $y_{\text{train}}$). 

With zero training error, we can then compute the test error (also called the generalisation error). The test score $g$ is obtained by running the test inputs through the QNN with the finalised set of parameters (that obtained zero training error) and obtaining the predicted $y_{\text{predict}}$ labels. The test accuracy is the fraction where the predicted label $y_{\text{predict}}$ and the test label $y_{\text{test}}$ match. The generalisation error is $1-g$. 

\begin{figure*}
     \centering
     \begin{subfigure}[t]{0.3\textwidth}
         \centering
         \includegraphics[width=\textwidth]{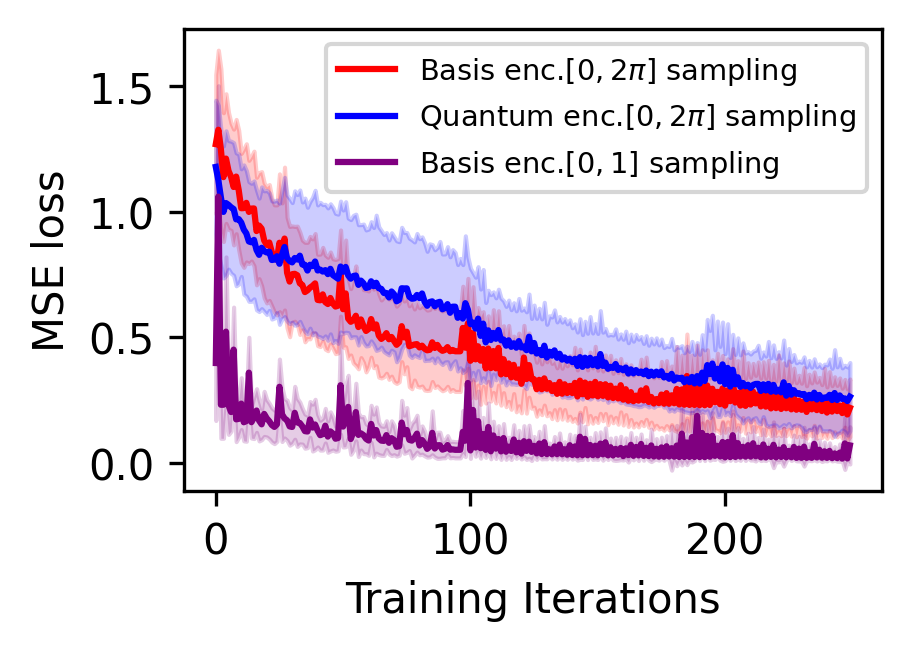}
         \caption{MSE Loss vs iterations for target function 0 (all 0s), complexity 5.0}
         \label{fig:qnn_loss_vs_iteration_t0}
     \end{subfigure}
     \hfill
     \begin{subfigure}[t]{0.3\textwidth}
         \centering
         \includegraphics[width=\textwidth]{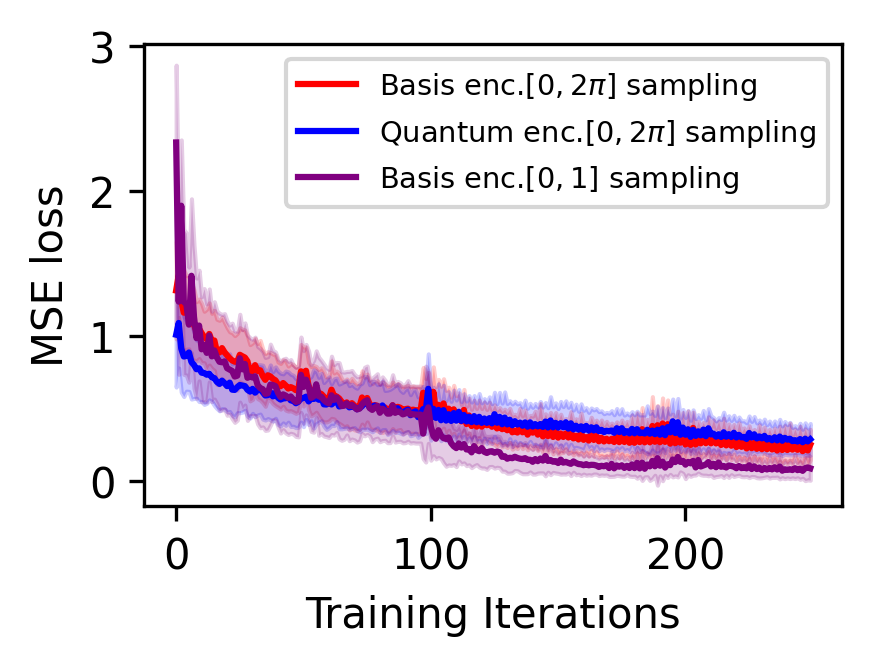}
         \caption{MSE Loss vs iterations for target function 6 (all 1s), complexity 5.0}
         \label{fig:qnn_loss_vs_iteration_t6}
     \end{subfigure}
     \hfill
     \begin{subfigure}[t]{0.3\textwidth}
         \centering
         \includegraphics[width=\textwidth]{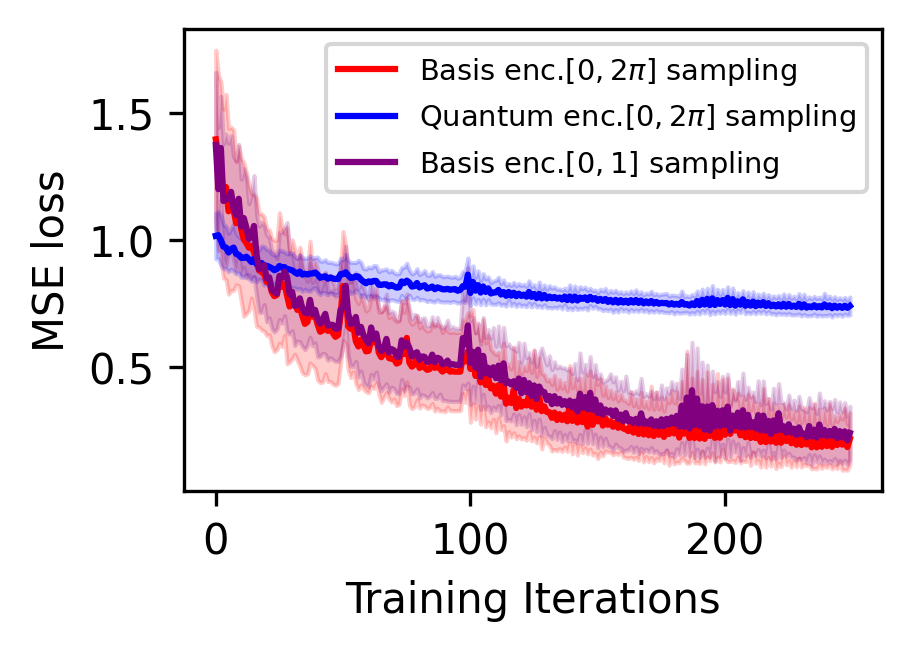}
         \caption{MSE Loss vs iterations for target function 5, complexity 55.0}
         \label{fig:qnn_loss_vs_iteration_t5}
     \end{subfigure}
     \begin{subfigure}[t]{0.3\textwidth}
         \centering
         \includegraphics[width=\textwidth]{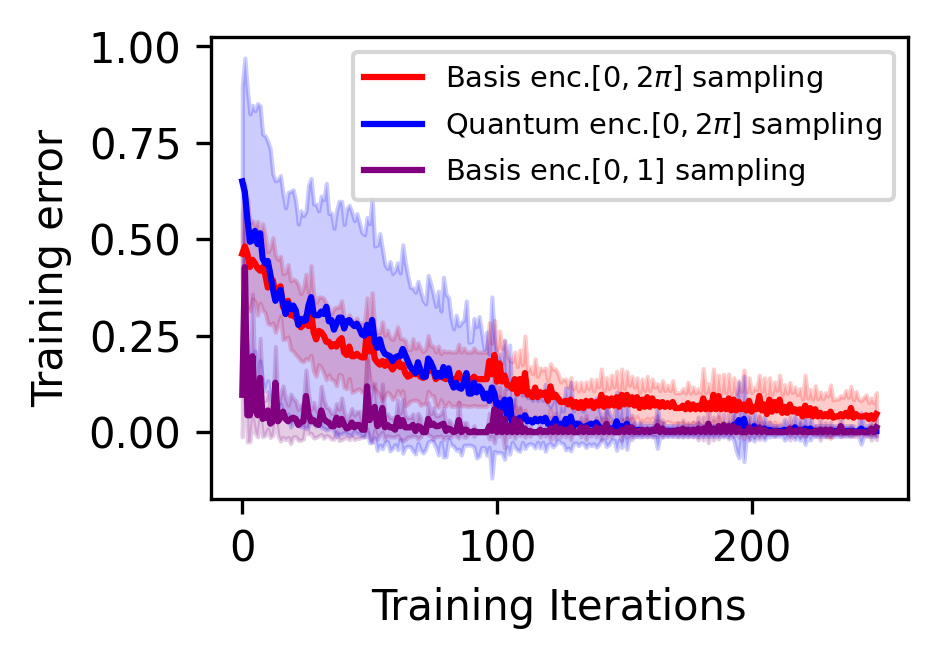}
         \caption{Training error vs iterations for target function 0 (all 0s), complexity 5.0}
         \label{fig:qnn_training_error_vs_iteration_t0}
     \end{subfigure}
     \hfill
     \begin{subfigure}[t]{0.3\textwidth}
         \centering
         \includegraphics[width=\textwidth]{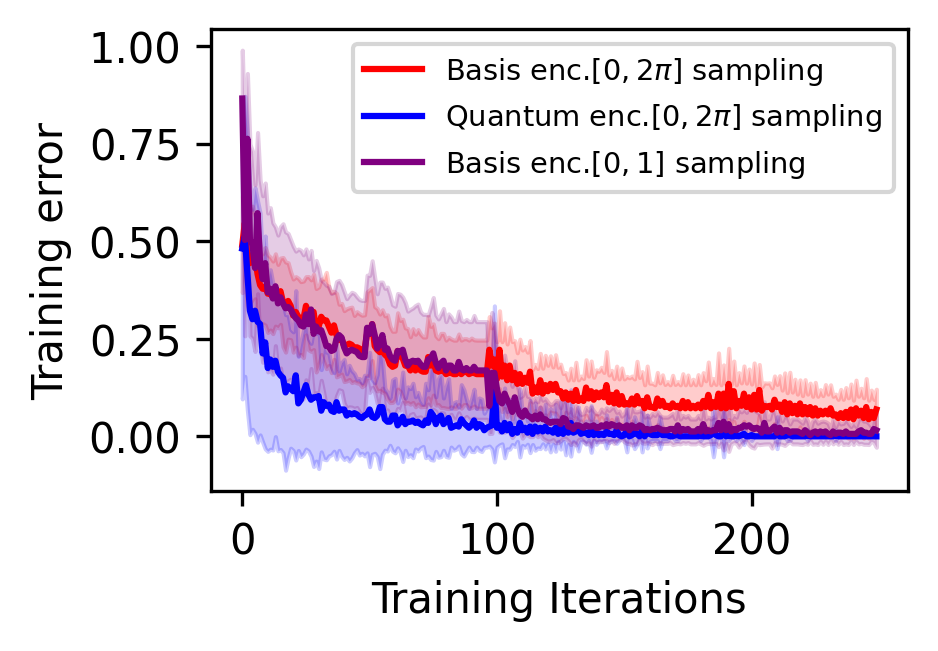}
         \caption{Training error vs iterations for target function 6 (all 1s), complexity 5.0}
         \label{fig:qnn_training_error_vs_iteration_t6}
     \end{subfigure}
     \hfill
     \begin{subfigure}[t]{0.3\textwidth}
         \centering
         \includegraphics[width=\textwidth]{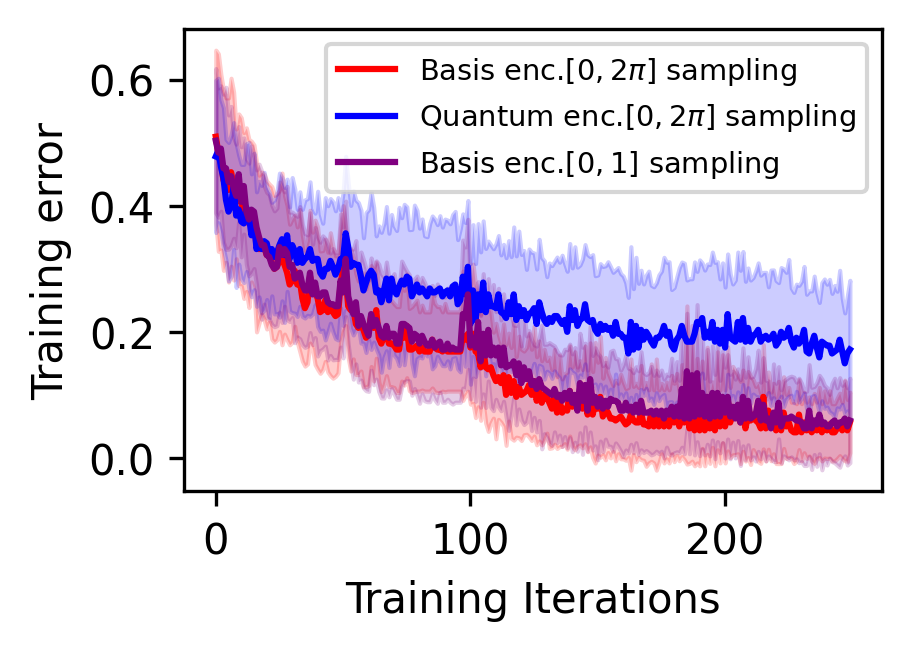}
         \caption{Training error vs iterations for target function 5, complexity 55.0}
         \label{fig:qnn_training_error_vs_iteration_t5}
     \end{subfigure}
        \caption{MSE loss and training error vs iterations for different target functions and different encoding methods for $n=5$}
        \label{fig:qnn_loss_vs_iteration}
\end{figure*}

In Figures \ref{fig:qnn_loss_vs_iteration_t0} - \ref{fig:qnn_loss_vs_iteration_t5} we show the mean squared error (MSE) loss vs training iterations during the training process for different target functions. 

We show the training process for two encoding methods: basis encoding and the ZZ feature map. In addition, we show the plot for basis encoding with $[0,1]$ sampling, which means that we sample the initial parameters for the QNN from the range $[0,1]$ instead of the range $[0,2\pi]$. As of the date of these experiments, IBM's Qiskit software includes, by default, the initial parameters for the QNN from the range $[0,1]$. We overrided this default to sample the initial parameters for the QNN from the range $[0,2\pi]$. We include both of these parameter initialisations to show how restricting the range of the parameters affects the training.

Figure \ref{fig:qnn_loss_vs_iteration_t0} shows the MSE loss versus training iterations for target function $f_t = 0$ which is the all $0$s function (e.g. for $n=3$, $f_t = 00000000$), which has complexity $LZ = 5$ . Note that the plots are for $n=5$. We see that the basis encoding and ZZ feature map with $[0,2\pi]$ sampling perform similarly well, whereas the basis encoding with $[0,1]$ sampling achieves lower MSE loss quicker. One hypothesis for this is because restricting the initial parameter range restricts the states that the QNN can express. Sampling from $[0,1]$ restricts the range of states for the $U_3$ gate to states where its classification will be zero, which is the function we are learning -- therefore, the basis encoding with $[0,1]$ sampling is initialised with the solution.

\begin{figure*}[h]
     \centering
     \begin{subfigure}[t]{0.475\textwidth}
         \centering
         \includegraphics[width=0.7\textwidth]{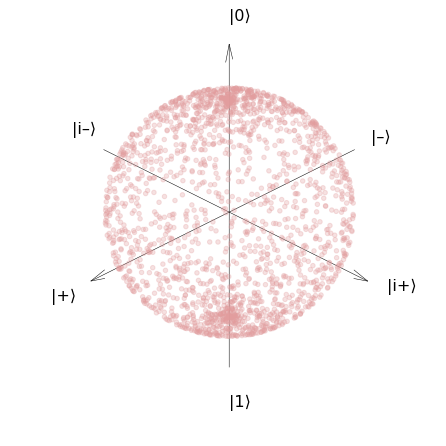}
         \caption{Bloch sphere for $U_3$ gate applied to a qubit with parameters sampled from $[0,2\pi]$}
         \label{fig:bloch_sphere_fully_expressive}
     \end{subfigure}
     \hfill
     \begin{subfigure}[t]{0.475\textwidth}
         \centering
         \includegraphics[width=0.7\textwidth]{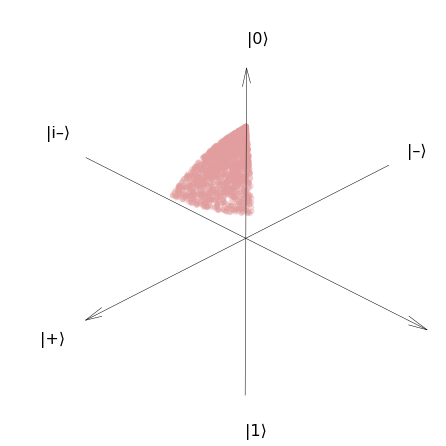}
         \caption{Bloch sphere for $U_3$ gate applied to a qubit with parameters sampled from $[0,1]$}
         \label{fig:bloch_sphere_zero_one_sampling}
     \end{subfigure}
        \caption{Bloch sphere for the $U_3$ gate applied to one qubit with parameters sampled from different distributions}
        \label{fig:bloch_sphere}
\end{figure*}

Figure \ref{fig:bloch_sphere_fully_expressive} shows a Bloch sphere with the range of states that the $U_3$ gate can express with $[0,2\pi]$ sampling and Figure \ref{fig:bloch_sphere_zero_one_sampling} shows a Bloch sphere with the range of states that the $U_3$ gate can express with $[0,1]$ sampling -- as one can see the states are restricted to the top hemisphere of the Bloch sphere and the top hemisphere is the zero classification (whereas the bottom hemisphere is the one classification). We believe it is important to be aware of this as IBM's Qiskit software starts with this default. 

\newpage
Figure \ref{fig:qnn_loss_vs_iteration_t6} shows the MSE loss versus training iterations for target function $f_t = 6$ which is the all $1$s function (e.g. for $n=3$, $f_t = 11111111$), which has complexity $LZ = 5$. We see that all methods perform similarly. This highlights the difference that the $[0,1]$ sampling made for training the QNN as when we train on the all 1s function, the QNN with $[0,1]$ sampling does not perform any better than with $[0,2\pi]$ sampling, whereas when we train on the all 0s function, the QNN with $[0,1]$ sampling does perform better than $[0,2\pi]$ sampling. 

Figure \ref{fig:qnn_loss_vs_iteration_t5} shows the MSE loss versus training iterations for target function $f_t = 5$ which has complexity $LZ = 55$. The ZZ feature map performs worse than the basis encoding. Intuitively, this is because we have skewed the data using the feature map, and therefore it has become harder to learn. Therefore, introducing certain inductive biases into a QNN via an encoding method can make it harder for the QNN to train and learn. The reason that we didn't see this for Figures \ref{fig:qnn_loss_vs_iteration_t0} and \ref{fig:qnn_loss_vs_iteration_t5} is because the QNN with the ZZ feature map has a bias towards the all 0s and all 1s function. So the QNN with the ZZ feature map will train well on functions it has a bias towards. When it encounters functions it has a weak bias towards, however, the training suffers. 

Figures \ref{fig:qnn_training_error_vs_iteration_t0} - \ref{fig:qnn_training_error_vs_iteration_t5} show the same trends but for training error instead of MSE loss.

\FloatBarrier
\subsection{Inductive bias of Quantum Convolutional Neural Networks (QCNNS)}
\label{sec:qcnns_background}
In this section we present some experiments on the inductive bias of QCNNs. 

\subsubsection{Definition of QCNNs}
\begin{figure*}
    \centering
    \adjustbox{scale=1}{%
     \begin{tikzcd}
        \lstick{$q_0$} & \gate[4,style={fill=green!20}]{\text{Convolutional Layer}} & \gate[4,style={fill=blue!20}]{\text{Pooling Layer}} & \qw & \qw & \qw\\
        \lstick{$q_1$} & \qw & \qw  & \qw & \qw & \qw \\
        \lstick{$q_2$} & \qw & \qw & \gate[2,style={fill=green!20}]{\text{Convolutional Layer}} & \gate[2,style={fill=blue!20}]{\text{Pooling Layer}} & \qw \\
        \lstick{$q_3$} & \qw & \qw & \qw & \qw & \qw \rstick{$q_r$}\\
    \end{tikzcd}}
\caption{General circuit diagram for a QCNN with four qubits}
\label{fig:qcnn}
\end{figure*}
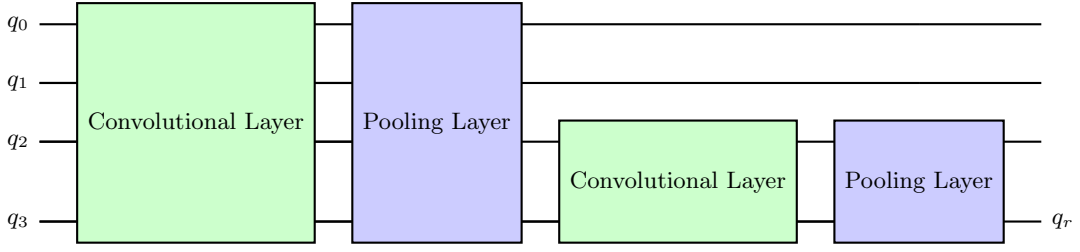

The Quantum Convolutional Neural Network (QCNN) was first proposed by Cong et. al. \cite{cong_quantum_2019}. The QCNN still has the general structure of the QNN explained in Section \ref{sec:qnn_overview} and the variational structure is more specific in the QCNN. In particular, alternating convolutional and pooling layers are applied for the variational circuit, which reduce the dimension of the circuit until one qubit is left -- the readout qubit $q_r$. In Figure \ref{fig:qcnn} we show an example of a QCNN for $n=4$. The first pooling layer reduces the dimension of the QCNN from four to two qubits by disregarding the first two qubits. The second pooling layer reduces the dimension of the QCNN from two qubits to one qubit $q_r$. 

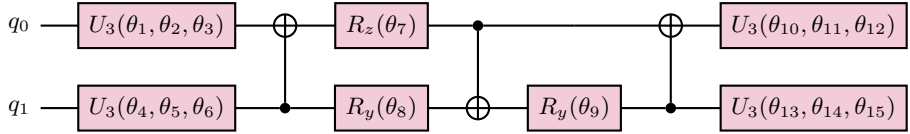
\begin{figure*}
    \centering
    \adjustbox{scale=1}{%
     \begin{tikzcd}
        \lstick{$q_0$} & \gate{U_3(\theta_1,\theta_2,\theta_3)} & \targ{} & \gate{R_z(\theta_7)} & \ctrl{1} & \qw & \targ{} & \gate{U_3(\theta_{10}, \theta_{11}, \theta_{12})} \\
        \lstick{$q_1$} & \gate{U_3(\theta_4,\theta_5,\theta_6)} & \ctrl{-1} & \gate{R_y(\theta_8)} & \targ{} & \gate{R_y(\theta_9)} & \ctrl{-1} & \gate{U_3(\theta_{13}, \theta_{14}, \theta_{15})}
    \end{tikzcd}}
\caption{Circuit for general two-qubit rotation}
\label{fig:two_qubit_rotation_circuit}
\end{figure*}

We wish to use a general two-qubit rotation gate in the layers. Vatan and Williams \cite{vatan_optimal_2004} propose an optimal quantum circuit to construct a general two-qubit rotation gate by using 15 parameters. This circuit is shown in Figure \ref{fig:two_qubit_rotation_circuit}.

For the convolutional layer, the general two-qubit unitary is applied to neighbouring qubits: first to all even pairs, then to all odd pairs (the first and final qubits are also a pair). For the pooling layer, the general two-qubit unitary is applied to all pairs of qubits. Then one qubit from each pair is ignored for the rest of the QCNN. 

\subsubsection{Experiment details}
\begin{figure*}
    \centering
    \adjustbox{scale=1}{%
     \begin{tikzcd}
        \lstick{$q_0$} & \qw & \targ{} & \gate{R_z(\theta_7)} & \ctrl{1} & \qw & \targ{} & \gate{R_z(\frac{\pi}{2})} & \qw \\
        \lstick{$q_1$} & \gate{R_z(-\frac{\pi}{2})} & \ctrl{-1} & \gate{R_y(\theta_8)} & \targ{} & \gate{R_y(\theta_9)} & \ctrl{-1} & \qw & \qw
    \end{tikzcd}}
\caption{Restricted two-qubit rotation}
\label{fig:two_qubit_rotation_circuit_qiskit}
\end{figure*}
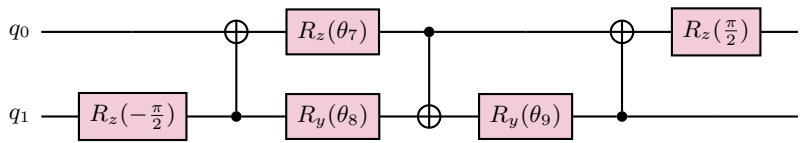

We use Qiskit's QCNN framework. In their example, they use a restricted version of the general two-qubit rotation unitary (as shown in Figure \ref{fig:two_qubit_rotation_circuit_qiskit}) which means that it is not able to express all states in the Hilbert space. Our work has demonstrated that the expressivity of the QNN can affect its bias so we make plots for both the restricted two-qubit rotation unitary (Figures \ref{fig:qcnn_fully_expressive_e0} - \ref{fig:qcnn_fully_expressive_e5}) and the fully expressive two-qubit rotation unitary (Figures \ref{fig:qcnn_non_fully_expressive_e0} - \ref{fig:qcnn_non_fully_expressive_e5}).

We sample from the QCNN in the same manner as the QNN and plot the probability of a boolean function versus its $LZ$ complexity for three encoding methods: basis encoding, ZZ feature map, Z feature map. The Z feature Map and ZZ feature Map are derived from the Pauli Expansion Circuit as defined in Equation \ref{eq:pauli_expansion_circuit} from \cite{noauthor_pauli_nodate}.

\begin{equation}
    U_{\Phi(\vec{x})}=\exp\left(i\sum_{S\subseteq [n]}
\phi_S(\vec{x})\prod_{i\in S} P_i\right)
\label{eq:pauli_expansion_circuit}
\end{equation}

where $P_i \in \{ I, X, Y, Z \}$ are the Pauli matrices. $S$ describes connectivities between different qubits $S \in \{\binom{n}{k},\ k = 1,... n \}$ and the feature map is defined in Equation \ref{eq:pauli_feature_map}.

\begin{equation}
    \phi_S(\vec{x})=\left\{\begin{array}{l}x_0 \text { if } k=1 \\ \prod_{j \in S}\left(\pi-x_j\right) \text { otherwise }\end{array}\right.
\label{eq:pauli_feature_map}
\end{equation}

\subsubsection{Probability of boolean function vs complexity}
\begin{figure*}
     \centering
     \begin{subfigure}[t]{0.3\textwidth}
         \centering
         \includegraphics[width=\textwidth]{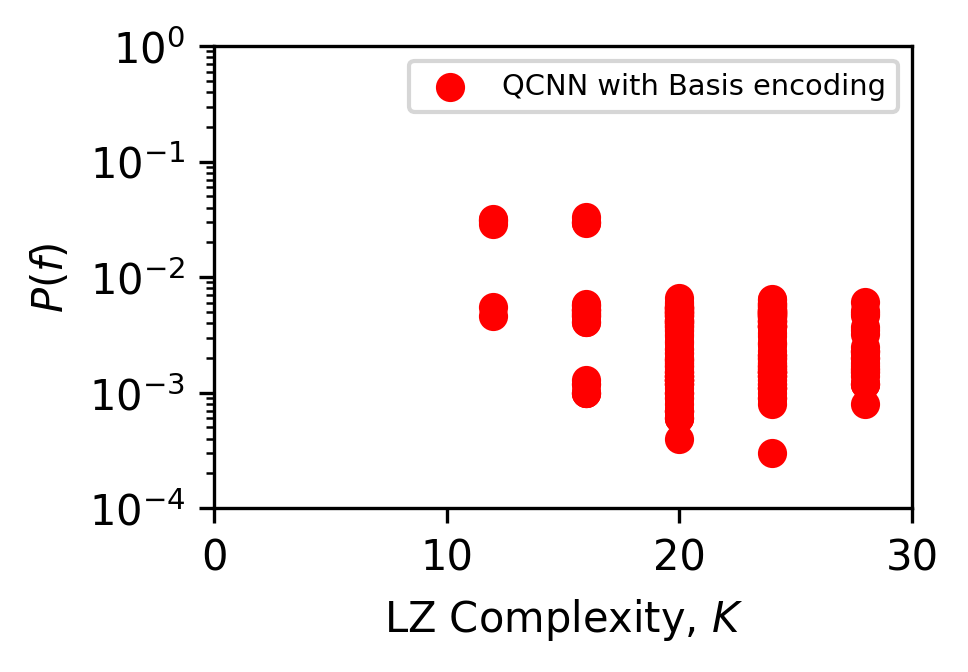}
         \caption{ $P(f)$ versus LZ complexity for non-fully expressive unitary in QCNN with basis encoding}
         \label{fig:qcnn_non_fully_expressive_e0}
     \end{subfigure}
     \hfill
     \begin{subfigure}[t]{0.3\textwidth}
         \centering
         \includegraphics[width=\textwidth]{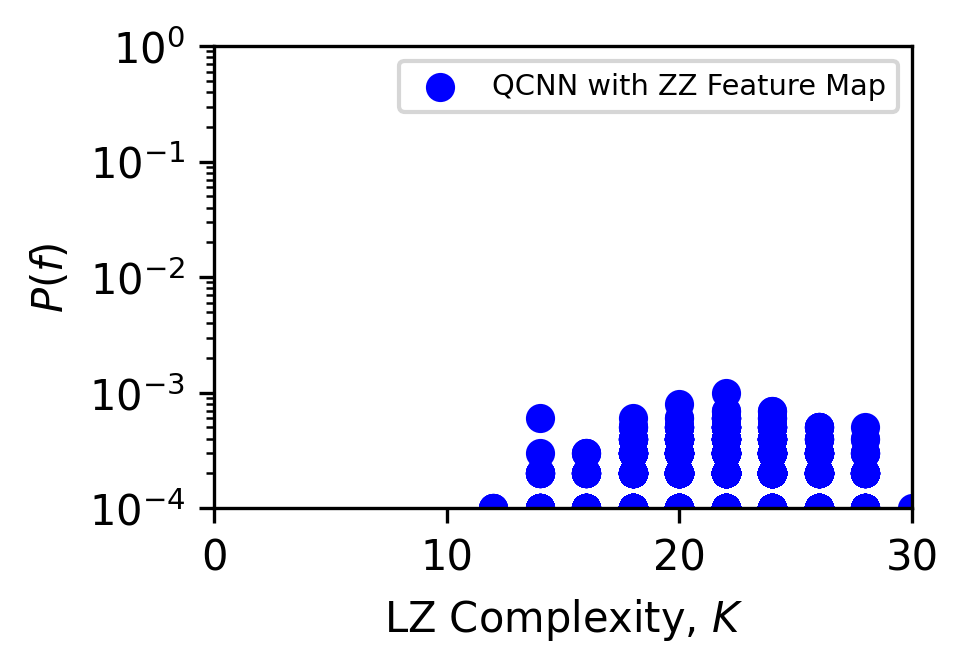}
         \caption{$P(f)$ versus LZ complexity for non-fully expressive unitary in QCNN with ZZ Feature Map}
         \label{fig:qcnn_non_fully_expressive_e1}
     \end{subfigure}
     \hfill
     \begin{subfigure}[t]{0.3\textwidth}
         \centering
         \includegraphics[width=\textwidth]{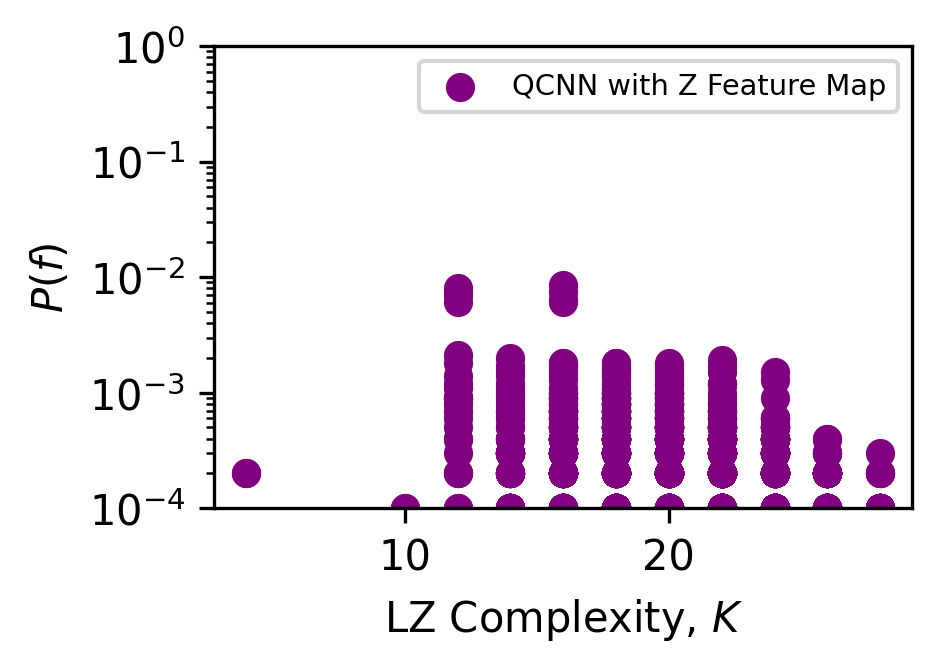}
         \caption{$P(f)$ versus LZ complexity for non-fully expressive unitary in QCNN with Z Feature Map}
         \label{fig:qcnn_non_fully_expressive_e5}
     \end{subfigure}
     \begin{subfigure}[t]{0.3\textwidth}
         \centering
         \includegraphics[width=\textwidth]{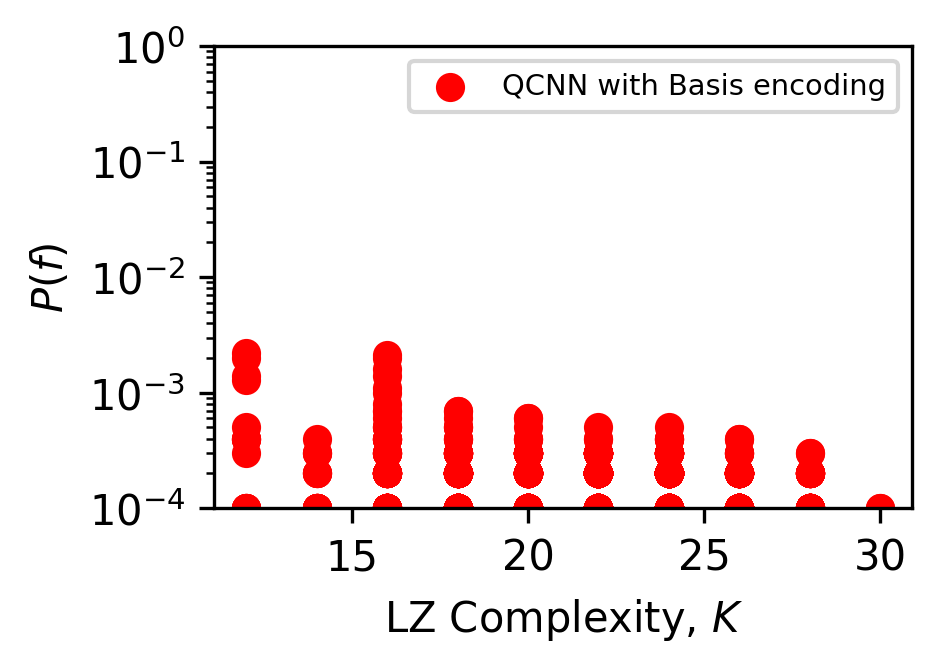}
         \caption{$P(f)$ versus LZ complexity for fully expressive unitary in QCNN with Z Feature Map}
         \label{fig:qcnn_fully_expressive_e0}
     \end{subfigure}
     \hfill
     \begin{subfigure}[t]{0.3\textwidth}
         \centering
         \includegraphics[width=\textwidth]{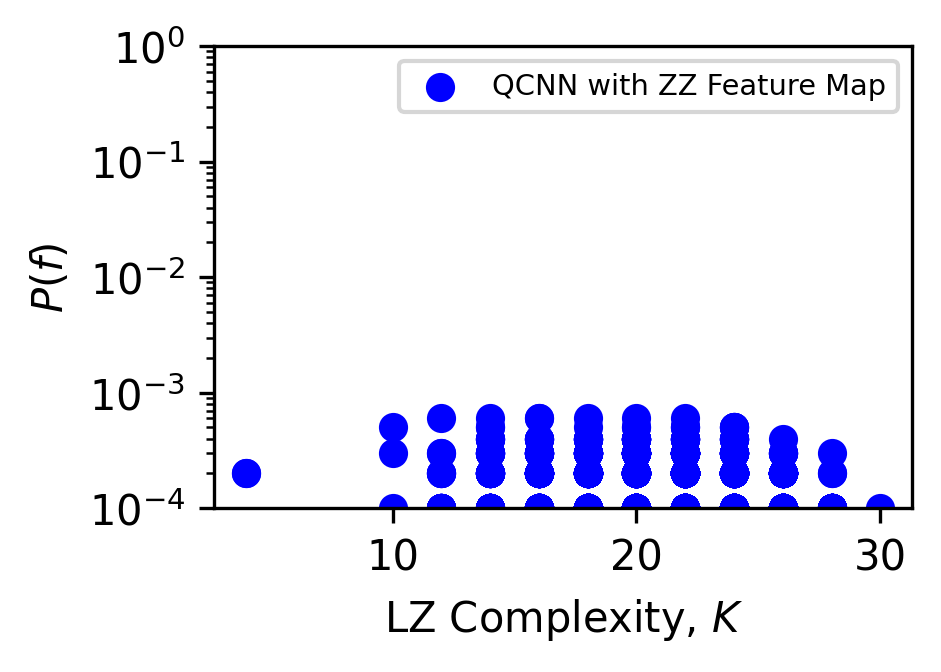}
         \caption{$P(f)$ versus LZ complexity for fully expressive unitary in QCNN with Z Feature Map}
         \label{fig:qcnn_fully_expressive_e1}
     \end{subfigure}
     \hfill
     \begin{subfigure}[t]{0.3\textwidth}
         \centering
         \includegraphics[width=\textwidth]{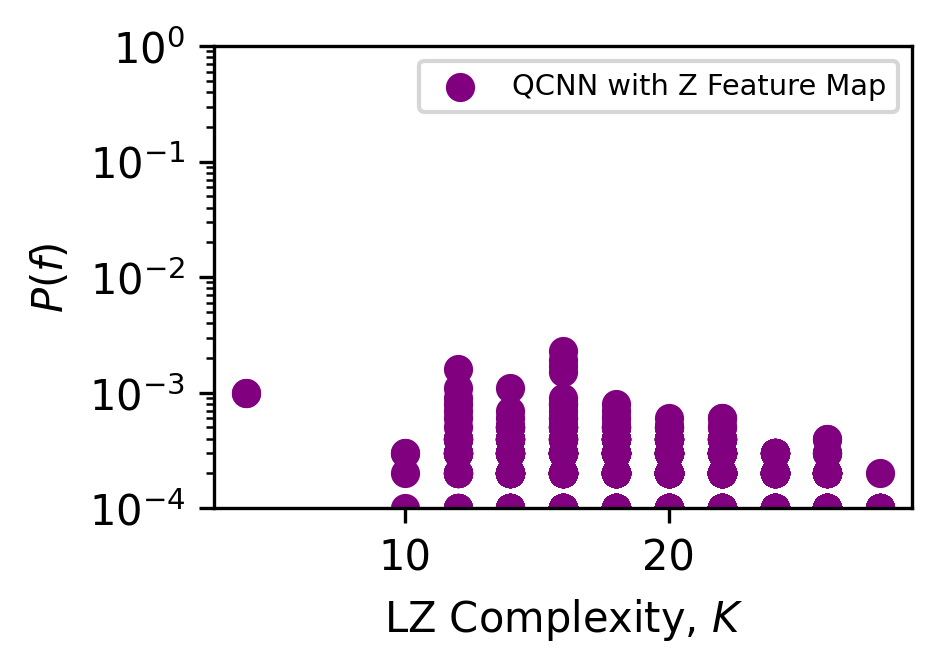}
         \caption{$P(f)$ versus LZ complexity for fully expressive unitary in QCNN with Z Feature Map}
         \label{fig:qcnn_fully_expressive_e5}
     \end{subfigure}
        \caption{The prior probability $P(f)$ versus LZ complexity $K$ for QCNNs with different encoding methods and expressivity for $n=4$ and $10^4$ samples}
        \label{fig:qcnn_bias}
\end{figure*}

Figure \ref{fig:qcnn_bias} shows the probability of the boolean function vs the function's complexity as described in Section \ref{sec:pf_vs_k_qnn}. Any bias seems to be further reduced in the fully expressive case, which aligns with our results that non-fully expressive QNNs can exhibit an artificial inductive bias. 

\FloatBarrier

\newpage
\section{Background Information}
\FloatBarrier
\subsection{The boolean system}
\label{sec:boolean_system}
The Boolean system is a key object of study in computer science. The boolean dataset consists of all possible permutations of zeros and ones up to a length $n$. In other words, the dataset covers $\{0,1\}^n$. For example, if $n = 3$, then the boolean dataset $\vec{x}$ is $x_0 = 000, x_1 = 001, x_2 = 010, x_3 = 011, x_4 = 100, x_5 = 101, x_6 = 110, x_7 = 111$. To obtain the boolean function, for each input $x_i$, we obtain an output $y_i$ where $y_i = H(x_i) \in \{0,1\}$ where $H$ is some function that maps the input $x_i$ to the output $y_i$. Therefore, the boolean function is 

\begin{equation*}
    f:\{0,1\}^n\rightarrow \{0,1\}
\end{equation*}

There are $2^{2^n}$ possible different functions $f$. The n-dimensional boolean dataset is $\mathcal{B}(f)_n=\{(x_1,f(x_1))\dots(x_{2^n},f(x_{2^n}))\}$, where $x_i$ covers $\{0,1\}^n$ and $f(x_i)\in\{0,1\}$.

An example of a Boolean function is shown in Table \ref{tab:quantum_boolean_function_system}.

\begin{table}[h]
\centering
\begin{tabularx}{0.2\textwidth}{ 
  | >{\centering\arraybackslash}X 
  | >{\centering\arraybackslash}X | }
 \hline
    $x$ & $f(x)$ \\
 \hline
  000 & 0 \\
  \hline
  001 & 0 \\
  \hline
  010 & 1 \\
  \hline
  011 & 0 \\
  \hline
  100 & 1 \\
  \hline
  101 & 0 \\
  \hline
  110 & 0 \\
  \hline
  111 & 1 \\
  \hline
\end{tabularx}
\caption{\textbf{Example of a Boolean function}: $x$ consists of all the possible combinations of a binary string with length three. Each input $x_i$ is mapped to an output $f(x_i)$, which is the boolean function $00101001$.}
\label{tab:quantum_boolean_function_system}
\end{table}

In this example $f = 00101001$. $f$ is therefore a binary string of length $2^n$ and there are $2^{2^n}$ possible different functions.

\subsubsection{Boolean system with deep neural networks (DNNs)}

In relation to DNNs, the inputs to the DNN are the boolean inputs, $\vec{x}$ and the prediction of the DNN is the boolean output $y = H(\vec{x})$ where $H$ represents the function of the DNN. When all the boolean inputs are run through the DNN during the feedforward process with the same set of weights, a continuous output is obtained. The continuous output is then thresholded to obtain a boolean function $f$. If the output $y \geq 0.5$, then the binary output is $1$. If the output $y < 0.5$, then the binary output is $0$. By choosing another random set of weights and running the feedforward process with the boolean inputs again, another boolean function $f$ is obtained. This is repeated with $m$ samples of random set of weights (i.e. parameters) that are chosen from some parameter initialisation distribution. Therefore, we obtain $m$ boolean functions and can then calculate $P(f)$.

The complexity of the boolean function $K(f)$ can be computed using an approximation to the Kolomogorov complexity, such as the Lempel-Ziv complexity $\tilde{K}(f)$. With $P(f)$ and $\tilde{K}(f)$, one can see how $P(f)$ changes as $\tilde{K}(f)$ changes, which shows the bias of the DNN. 
\FloatBarrier
\newpage
\subsection{Quantum Neural Networks}
\label{sec:qnns}

An overview of Quantum Neural Networks (QNNs) and the encoding methods are given in Section \ref{sec:qnn_overview}. Further details about the QNNs used in this work are given in Section \ref{sec:qnn_details}. In this section, we provide background information on the measurement operators and learning process of QNNs. 

\subsubsection{Measurement operators}
After the variational quantum circuit, we need to make measurements in order to gain information so that we can evaluate the loss function. In particular, on the readout qubit, we measure a Pauli operator, such as $\sigma_z = \begin{psmallmatrix} 1 & 0 \\ 0 & -1 \end{psmallmatrix}$, which we can call $Z_{n+1}$. This gives +1 or -1. We have to run the circuit multiple times to get the useful statistics from the circuit; in particular we want the expectation value of $Z_{n+1}$ The expectation value of an operator $A$ in the state $\psi$ is $\bra{\psi} A \ket{\psi}$. Our state $\psi$ in the variational circuit is equivalent to the unitary matrix of the circuit $U(\vec{\theta})$ applied to the initial state of the circuit $\ket{z,1}$ where $\ket{z} = \ket{z_1z_2 \cdots z_n}$ is the conversion of the classical data string $z = z_1z_2 \cdots z_n$ and $\ket{1}$ is the state of the readout qubit, which gives us the initial state of the circuit $\ket{z,1}$. Thus $\ket{\psi} = U(\vec{\theta}) \ket{z,1}$ and the expectation of the measurement operator is $\bra{z,1} U^\dagger(\vec{\theta}) Z_{n+1} U(\vec{\theta}) \ket{z,1}$. This is equivalent to the average of the observed outcomes of $Z_{n+1}$ Our loss function is shown in Equation \ref{eq:loss} \cite{farhi_classification_2018}.

\begin{equation}
    \text{loss}(\vec{\theta},z) = 1 - l(z) \bra{z,1} U^\dagger(\vec{\theta}) Z_{n+1} U(\vec{\theta}) \ket{z,1}
    \label{eq:loss}
\end{equation}

where $l(z)$ is the correct label of the input string. 

To achieve, with probability greater than 99\%, an estimate of the sample loss that is within $\delta$ of the true sample loss we need to make at least $2/\delta^2$ measurements. The expectation value of the measurement operator should correspond to the correct label $l(z)$ of the input string. In order to get our predicted label (the expectation value of the measurement operator) as close as possible to the true label, we have to find the parameters $\vec{\theta}$ in the variational quantum circuit that will get us there.

\subsubsection{Learning}
The learning procedure is as follows:
\begin{enumerate}[noitemsep]
    \item Start with random parameters $\vec{\theta}$ in the variational quantum circuit
    \item Pick a string $z^1$ from the training set
    \item Encode the string $z^1$ into the circuit via the encoder circuit
    \item Measure $\sigma_z$ on the readout qubit multiple times to obtain the expectation value of $Z_{n+1}$
    \item Compute $\text{loss}(\vec{\theta},z^1)$ as shown in Equation \ref{eq:loss}
    \item Update the parameters via some classical optimizer (such as Adam), giving us new parameters $\vec{\theta^1}$
    \item Pick a new string $z^2$ from the training set
    \item Repeat steps 3-4 and then compute $\text{loss}(\vec{\theta^1}{z^2})$
    \item Repeat the above steps generating a sequence $\vec{\theta^1}, \vec{\theta^1}, \cdots \vec{\theta^S}$
\end{enumerate}

The final version of the unitary matrix of the circuit $U(\vec{\theta^S})$ with parameters $\vec{\theta^S}$ acting on the state $\ket{z,1}$ will result in measurements that will give the correct label $l(z)$.

\FloatBarrier
\newpage
\subsection{Quantum Kernels}
\label{sec:quantum_kernels}

In this section, we provide further details on the calculations for quantum kernels, which we use for investigating the inductive bias. There are three main steps to using a quantum kernel: 
\begin{enumerate}
    \item Define a quantum feature map
    \item Construct the kernel matrix
    \item Train the kernel
\end{enumerate}

\subsubsection{Define a quantum feature map} 

A kernel function $k$ maps the input data into a higher dimensional space $k(\vec{x}_i, \vec{x}_j) = \langle f(\vec{x}_i), f(\vec{x}_j) \rangle$ where $\vec{x}_i, \vec{x}_j$ are $n$ dimensional inputs, $f$ is a map from $n$-dimension to $m$ dimension space, and $\langle a, b \rangle$ denotes the inner product between $a$ and $b$. A kernel matrix $K$ can be constructed from all the input data: $K_{ij} = k(\vec{x}_i,\vec{x}_j)$.

A quantum feature map $\phi(\vec{x})$ maps classical data, represented as a vector $\vec{x}$, to a quantum Hilbert space and the Kernel matrix is $K_{ij} = \left| \langle \phi^\dagger(\vec{x}_j)| \phi(\vec{x}_i) \rangle \right|^{2}$.

The quantum feature maps are described in Section \ref{sec:encoding_methods}.

\subsubsection{Construct the kernel matrix}

To obtain the Kernel matrix from a QNN, we encode the $x_i$ data into the quantum circuit that encodes the data and then obtain the final statevector $\ket{\phi}$. We can then compute $\left| \langle \phi^\dagger(\vec{x}_j)| \phi(\vec{x}_i) \rangle \right|^{2}$ for all pairs of $i,j$.

For example, we can compute the kernel matrices on the boolean dataset for $n = 2$ for different encoding methods. The boolean dataset for $n = 2$ is equivalent to $\vec{x} = (x_0, x_1, x_2, x_3)$ where $x_0 = 00, x_1 = 01, x_2 = 10, x_3 = 11$. 

\begin{figure}[h]
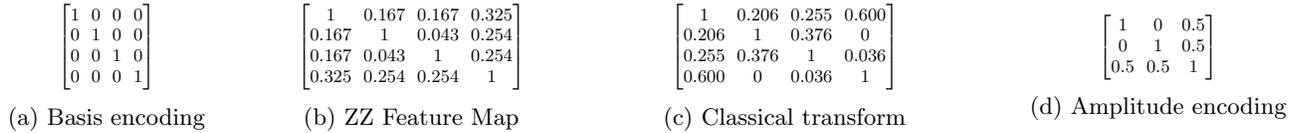

     \centering
     \captionsetup[subfigure]{justification=centering}
     \begin{subfigure}[t]{0.22\textwidth}
         \centering
        \adjustbox{scale=0.75}{%
         $\begin{bmatrix}
1 & 0 & 0 & 0 \\
0 & 1 & 0 & 0 \\
0 & 0 & 1 & 0 \\
0 & 0 & 0 & 1 \\
\end{bmatrix}$}
         \caption{Basis encoding}
         \label{fig:matrix_e0}
     \end{subfigure}
     \begin{subfigure}[t]{0.22\textwidth}
         \centering
         \adjustbox{scale=0.75}{%
         $\begin{bmatrix}
1 & 0.167 & 0.167 & 0.325 \\
0.167 & 1 & 0.043 & 0.254 \\
0.167 & 0.043 & 1 & 0.254 \\
0.325 & 0.254 & 0.254 & 1 \\
\end{bmatrix}$}
         \caption{ZZ Feature Map}
         \label{fig:matrix_e1}
     \end{subfigure}
      \hfill
     \begin{subfigure}[t]{0.22\textwidth}
         \centering
         \adjustbox{scale=0.75}{%
         $\begin{bmatrix}
1 & 0.206 & 0.255 & 0.600 \\
0.206 & 1 & 0.376 & 0 \\
0.255 & 0.376 & 1 & 0.036 \\
0.600 & 0 & 0.036 & 1 \\
\end{bmatrix}$}
         \caption{Classical transform}
         \label{fig:matrix_e3}
     \end{subfigure}
     \hfill
     \begin{subfigure}[t]{0.22\textwidth}
         \centering
         \adjustbox{scale=0.75}{%
         $\begin{bmatrix}
1 & 0 & 0.5 \\
0 & 1 & 0.5 \\
0.5 & 0.5 & 1 \\
\end{bmatrix}$}
         \caption{Amplitude encoding}
         \label{fig:matrix_e4}
     \end{subfigure}
\caption{\textbf{Kernel matrices for $n=2$:} kernel matrices calculated using $K_{ij} = \left| \langle \phi^\dagger(\vec{x}_j)| \phi(\vec{x}_i) \rangle \right|^{2}$ where $\vec{x_i}$ consist of all the combinations of binary strings of length two.}
\label{fig:kernel_matrices}
\end{figure}

Figure \ref{fig:kernel_matrices} shows the kernel matrices for each encoding method for two qubits ($n=2$). It should be noted that amplitude encoding can not encode the all 0s boolean function. Therefore, its kernel matrix is $(2^n-1 \times 2^n-1)$ instead of $(2^n \times 2^n)$ . In the following sections, we explain how to calculate each kernel matrix using its encoding method. 

\FloatBarrier

\subsubsection{Basis encoding kernel matrix}
\label{sec:basis}
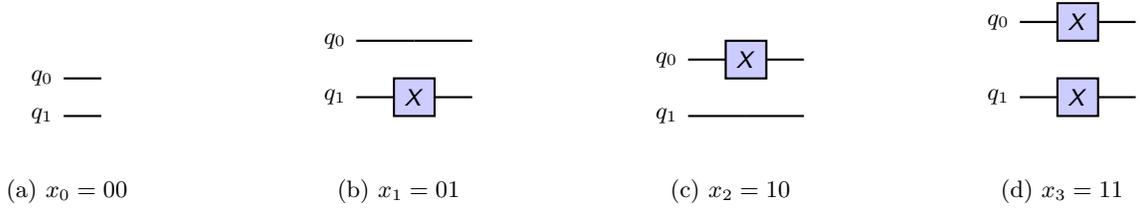
\begin{figure}[h]
    \centering
    \captionsetup[subfigure]{justification=centering}
     \begin{subfigure}[b]{0.24\textwidth}
     \centering
        \adjustbox{scale=1}{%
         \begin{tikzcd}
            \lstick{$q_0$} & \qw\\
            \lstick{$q_1$} & \qw \\
        \end{tikzcd}
        }
         \caption{$x_0 = 00$}
         \label{fig:basis_x0}
     \end{subfigure}
     \begin{subfigure}[b]{0.24\textwidth}
     \centering
        \adjustbox{scale=1}{%
         \begin{tikzcd}
            \lstick{$q_0$} & \qw & \qw \\
            \lstick{$q_1$} & \circuitX & \qw \\
        \end{tikzcd}
        }
         \caption{$x_1 = 01$}
         \label{fig:basis_x1}
     \end{subfigure}
     \begin{subfigure}[b]{0.24\textwidth}
     \centering
        \adjustbox{scale=1}{%
         \begin{tikzcd}
            \lstick{$q_0$} & \circuitX & \qw \\
            \lstick{$q_1$} & \qw & \qw \\
        \end{tikzcd}
        }
         \caption{$x_2 = 10$}
         \label{fig:basis_x2}
     \end{subfigure}
     \begin{subfigure}[b]{0.24\textwidth}
     \centering
        \adjustbox{scale=1}{%
         \begin{tikzcd}
            \lstick{$q_0$} & \circuitX & \qw \\
            \lstick{$q_1$} & \circuitX & \qw \\
        \end{tikzcd}
        }
         \caption{$x_3 = 11$}
         \label{fig:basis_x3}
     \end{subfigure}
\caption{\textbf{Quantum circuits for basis encoding:} these quantum circuits encode the data using basis encoding. (a) this circuit which consists of no quantum gates (equivalent to Identity gates) encodes the data $x_0 = 00$ as the qubits $q_0$ and $q_1$ are initially in the state $\ket{0}\ket{0}$. (b) this circuit consists of no gate on the $q_0$ qubit and applies the $X$ gate to the qubit $q_1$ as it flips the state of this qubit to $\ket{1}$ and therefore encodes the data $x_1 = 01$. (c) this circuit applies the $X$ gate to the qubit $q_0$ as it flips the qubit to the state $\ket{1}$ and does nothing to the qubit $q_1$ and therefore encodes $x_2 = 10$. (d) this circuit applies the $X$ gate to both of the qubits $q_0$ and $q_1$ so that it encodes the data $x_3 = 11$.}
\label{fig:kernel_basis_encoding}
\end{figure}

To generate the basis encoding kernel matrix, we construct the quantum circuits that encode this boolean dataset. Figure \ref{fig:kernel_basis_encoding} shows the four quantum circuits to encode $\vec{x} = (x_0, x_1, x_2, x_3)$.

We can now obtain the statevector from the encoding quantum circuits. We calculate the statevector as $\ket{\phi} = U\ket{0^{n}}$ where $U$ is the unitary matrix of the circuit and $\ket{0^n}$ is applied to $n$ qubits. For Figure \ref{fig:basis_x0}, the unitary matrix of the circuit is equivalent to $I \otimes I$, therefore the statevector is $\ket{\phi} = \ket{00}$. For Figure \ref{fig:basis_x1}, the unitary matrix of the circuit is equivalent to $I \otimes X$, therefore the statevector is $\ket{\phi} = \ket{01}$. For Figure \ref{fig:basis_x2}, the unitary matrix of the circuit is equivalent to $X \otimes I$, therefore the statevector is $\ket{\phi} = \ket{10}$. For Figure \ref{fig:basis_x3}, the unitary matrix of the circuit is equivalent to $X \otimes X$, therefore the statevector is $\ket{\psi} = \ket{11}$.   

We can now calculate the entries of the kernel matrix as $K_{ij} = \left| \langle \phi^\dagger(\vec{x}_j)| \phi(\vec{x}_i) \rangle \right|^{2}$ for all possible $i, j$. For example, when $i = 0$ and $j = 0$, we calculate $K_{00} = \left| \langle \phi^\dagger(\vec{x}_0)| \phi(\vec{x}_0) \rangle \right|^{2} = \left| \bra{00} \ket{00} \right|^{2} = 1$. When i = 0 and j = 1, we calculate $K_{01} = \left| \langle \phi^\dagger(\vec{x}_1)| \phi(\vec{x}_0) \rangle \right|^{2} = \left| \bra{01} \ket{00} \right|^{2} = 0$. We repeat this for all possible $i, j$ and then construct the kernel matrix from its entries. For basis encoding, the matrix becomes that shown in Figure \ref{fig:matrix_e0}.

\subsubsection{ZZ feature map kernel matrix}
\label{sec:zz_feature_map}
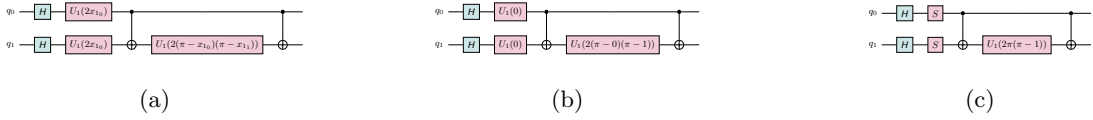
\begin{figure}[h]
    \centering
    \begin{subfigure}[b]{0.3\textwidth}
    \adjustbox{scale=0.4}{%
     \begin{tikzcd}
        \lstick{$q_0$} & \circuitH & \gate{U_1(2x_{1_0})} & \ctrl{1} & \qw & \ctrl{1} & \qw \\
        \lstick{$q_1$} & \circuitH & \gate{U_1(2x_{1_0})} & \targ{} & \gate{U_1(2(\pi-x_{1_0})(\pi-x_{1_1}))} & \targ{} & \qw \\
    \end{tikzcd}
    }
    \caption{}
    \label{fig:step1}
    \end{subfigure}
    \begin{subfigure}[b]{0.3\textwidth}
    \adjustbox{scale=0.4}{%
     \begin{tikzcd}
        \lstick{$q_0$} & \circuitH & \gate{U_1(0)} & \ctrl{1} & \qw & \ctrl{1} & \qw \\
        \lstick{$q_1$} & \circuitH & \gate{U_1(0)} & \targ{} & \gate{U_1(2(\pi-0)(\pi-1))} & \targ{} & \qw \\
    \end{tikzcd}
    }
    \caption{}
    \label{fig:step2}
    \end{subfigure}
    \begin{subfigure}[b]{0.3\textwidth}
    \adjustbox{scale=0.4}{%
     \begin{tikzcd}
        \lstick{$q_0$} & \circuitH & \gate{S} & \ctrl{1} & \qw & \ctrl{1} & \qw \\
        \lstick{$q_1$} & \circuitH & \gate{S} & \targ{} & \gate{U_1(2\pi(\pi-1))} & \targ{} & \qw \\
    \end{tikzcd}
    }
    \caption{}
    \label{fig:step3}
    \end{subfigure}
\caption{\textbf{ZZFeature Map for two qubits that encodes 01:} this figure shows the quantum circuit that encodes the data $x_1 = 01$: the $0^{th}$ element of $x_1$ is $x_{1_0} = 0$ and the $1^{st}$ element of $x_1$ is $x_{1_1} = 1$. In Figure \ref{fig:step1}, the quantum circuit consists of two $H$ (Hadamard) gates where $H = \frac{1}{\sqrt{2}} \begin{psmallmatrix}
    1 & 1 \\
    1 & - 1
\end{psmallmatrix}$ followed by two $U_1$ gates applied to both qubits, where $U_1 (\theta) = \begin{psmallmatrix}
    1 & 0 \\
    0 & e^{i \theta}
\end{psmallmatrix}$ The angle $\theta$ of the $U_1$ gates are $2x_{1_0}$ which is equal to $0$ in this example. Therefore these matrices are equivalent to $U_1(0) = \begin{psmallmatrix} 1 & 0 \\ 0 & e^{i0}\end{psmallmatrix} = \begin{psmallmatrix} 1 & 0 \\ 0 & i\end{psmallmatrix} = S$ where $S$ is the phase matrix. 
These $U_1$ gates are followed by a $CNOT$ gate and then another $U_1$ gate applied to $q_1$ where the angle of this gate is equivalent to $2(\pi - x_{1_0})(\pi - x_{1_1})$. This gate is equivalent to the matrix $\begin{psmallmatrix}
    1 & 0 \\
    0 & e^{i(2(\pi - x_{1_0})(\pi - x_{1_1})}
\end{psmallmatrix}$ = $\begin{psmallmatrix}
    1 & 0 \\
    0 & e^{i(2\pi(\pi - 1)}
\end{psmallmatrix}$. Figures \ref{fig:step2} and \ref{fig:step3} show the steps of replacing the angles with the datapoints and simplifying the form of the quantum circuit.}
\label{fig:zz_feature_map_steps}
\end{figure}

To generate the ZZ feature map kernel matrix, we construct the quantum circuits that encode this boolean dataset. Figure \ref{fig:zz_feature_map_steps} shows how we encode the data into the quantum circuit and obtain a simplified form of the quantum circuit that encodes the data. 

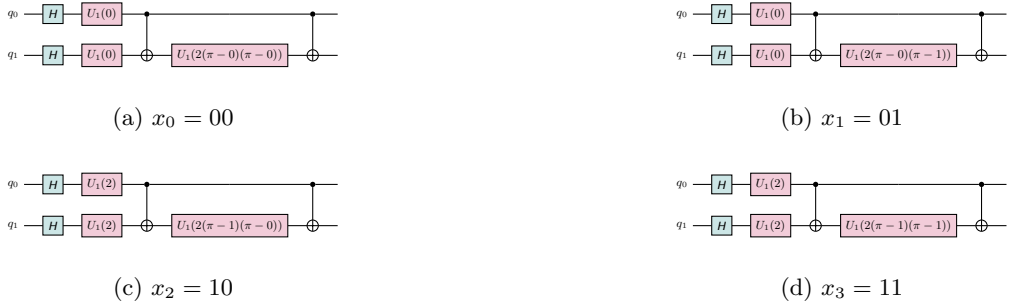
\begin{figure}[h]
    \centering
    \captionsetup[subfigure]{justification=centering}
     \begin{subfigure}[b]{0.49\textwidth}
     \centering
        \adjustbox{scale=0.5}{%
     \begin{tikzcd}
        \lstick{$q_0$} & \circuitH & \gate{U_1(0)} & \ctrl{1} & \qw & \ctrl{1} & \qw \\
        \lstick{$q_1$} & \circuitH & \gate{U_1(0)} & \targ{} & \gate{U_1(2(\pi-0)(\pi-0))} & \targ{} & \qw \\
    \end{tikzcd}
        }
         \caption{$x_0 = 00$}
         \label{fig:zz_x0}
     \end{subfigure}
     \begin{subfigure}[b]{0.49\textwidth}
     \centering
        \adjustbox{scale=0.5}{%
     \begin{tikzcd}
        \lstick{$q_0$} & \circuitH & \gate{U_1(0)} & \ctrl{1} & \qw & \ctrl{1} & \qw \\
        \lstick{$q_1$} & \circuitH & \gate{U_1(0)} & \targ{} & \gate{U_1(2(\pi-0)(\pi-1))} & \targ{} & \qw \\
    \end{tikzcd}
        }
         \caption{$x_1 = 01$}
         \label{fig:zz_x1}
     \end{subfigure}
     \par\bigskip
     \begin{subfigure}[b]{0.49\textwidth}
     \centering
        \adjustbox{scale=0.5}{%
         \begin{tikzcd}
            \lstick{$q_0$} & \circuitH & \gate{U_1(2)} & \ctrl{1} & \qw & \ctrl{1} & \qw \\
            \lstick{$q_1$} & \circuitH & \gate{U_1(2)} & \targ{} & \gate{U_1(2(\pi-1)(\pi-0))} & \targ{} & \qw \\
        \end{tikzcd}
        }
         \caption{$x_2 = 10$}
         \label{fig:zz_x2}
     \end{subfigure}
     \begin{subfigure}[b]{0.49\textwidth}
     \centering
        \adjustbox{scale=0.5}{%
         \begin{tikzcd}
            \lstick{$q_0$} & \circuitH & \gate{U_1(2)} & \ctrl{1} & \qw & \ctrl{1} & \qw \\
            \lstick{$q_1$} & \circuitH & \gate{U_1(2)} & \targ{} & \gate{U_1(2(\pi-1)(\pi-1))} & \targ{} & \qw \\
        \end{tikzcd}
        }
         \caption{$x_3 = 11$}
         \label{fig:zz_x3}
     \end{subfigure}
\caption{\textbf{Quantum circuits for the ZZ feature map:} these quantum circuits encode the data using the ZZ feature map.}
\label{fig:kernel_zz_encoding}
\end{figure}

Figure \ref{fig:kernel_zz_encoding} shows the four quantum circuits to encode $\vec{x} = (x_0, x_1, x_2, x_3)$.

Using the same method as we described for basis encoding, we obtain the statevector from the encoding quantum circuits and then calculate the entries of the kernel matrix. For the ZZ feature map, the matrix becomes that shown in Figure \ref{fig:matrix_e1}. 

\subsubsection{Random relu transform kernel matrix}
\label{sec:random_relu_transform}
\begin{figure}[h]
\centering
\includegraphics[width=0.4\textwidth]{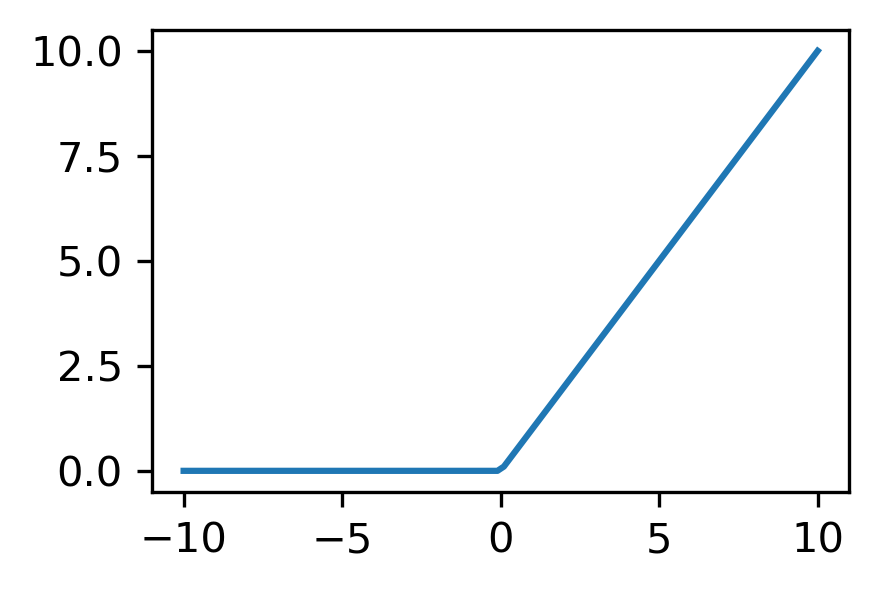}
\caption{\textbf{Relu function:} the relu function returns the input if it is positive, otherwise it returns zero.}
\label{fig:relu}
\end{figure}

In classical deep learning, a rectified linear unit (ReLU) is an activation function that introduces nonlinearity to a deep learning model. It is defined as the positive part of its argument - in other words, its output is the input if it is positive, otherwise, its output is zero as shown in Figure \ref{fig:relu} and Equation \ref{eq:relu}.

\begin{equation}
    f(x)=x^{+}=\max (0, x)=\frac{x+|x|}{2}= \begin{cases}x & \text { if } x>0 \\ 0 & \text { otherwise }\end{cases}
    \label{eq:relu}
\end{equation}

For this encoding method, we apply the relu function to the statevector resulting from the encoding circuit. The encoding circuit consists of encoding the boolean data using basis encoding and then applying a random unitary matrix to the qubits. Given that our statevector is a complex vector, we apply the relu function to the imaginary and real parts of the complex number. After applying the relu function to the statevector, we then re-normalize the statevector. Table \ref{tab:relu} shows the input and relu output with a complex number. 

\begin{table}[h]
    \centering
    \begin{tabular}{ |c|c|c| }
    \hline
    \multicolumn{2}{|c|}{Input}     & Output     \\
    \hline
    Re($x$)    & Im($x$) & Relu($x$)                    \\
    \hline
    $< 0$         & $< 0$      & $0 + 0j$                       \\
    \hline
    $< 0$         & $\geq 0$   & $0 + \text{Im}(x)j$            \\
    \hline
    $\geq 0$      & $< 0$      & $\text{Re}(x) + 0j$            \\
    \hline
    $\geq 0$      & $ \geq 0$  & $\text{Re}(x) + \text{Im}(x)j$ \\ 
    \hline 
    \end{tabular}
    \caption{\textbf{Relu function}: The relu function returns the input if it is positive, otherwise it returns zero. When applied to a complex number, we apply relu to the imaginary and real parts of the complex number.}
    \label{tab:relu}
\end{table}

\begin{figure}
    \centering
    \captionsetup[subfigure]{justification=centering}
     \begin{subfigure}[b]{0.24\textwidth}
     \centering
        \adjustbox{scale=1}{%
         \begin{tikzcd}
            \lstick{$q_0$} & \gate[2]{U} & \qw \\
            \lstick{$q_1$} & \qw & \qw \\
        \end{tikzcd}
        }
         \caption{$x_0 = 00$}
         \label{fig:relu_x0}
     \end{subfigure}
     \begin{subfigure}[b]{0.24\textwidth}
     \centering
        \adjustbox{scale=1}{%
         \begin{tikzcd}
            \lstick{$q_0$} & \qw & \gate[2]{U} & \qw \\
            \lstick{$q_1$} & \circuitX & \qw & \qw \\
        \end{tikzcd}
        }
         \caption{$x_1 = 01$}
         \label{fig:relu_x1}
     \end{subfigure}
     \begin{subfigure}[b]{0.24\textwidth}
     \centering
        \adjustbox{scale=1}{%
         \begin{tikzcd}
            \lstick{$q_0$} & \circuitX & \gate[2]{U} & \qw \\
            \lstick{$q_1$} & \qw & \qw & \qw  \\
        \end{tikzcd}
        }
         \caption{$x_2 = 10$}
         \label{fig:relu_x2}
     \end{subfigure}
     \begin{subfigure}[b]{0.24\textwidth}
     \centering
        \adjustbox{scale=1}{%
         \begin{tikzcd}
            \lstick{$q_0$} & \circuitX & \qw & \gate[2]{U} & \qw \\
            \lstick{$q_1$} & \circuitX & \qw & \qw & \qw \\
        \end{tikzcd}
        }
         \caption{$x_3 = 11$}
         \label{fig:relu_x3}
     \end{subfigure}
\caption{\textbf{Quantum circuits for random relu transform:} these quantum circuits encode the data using the random relu transform. The first step of the random relu transform is encoding the boolean data using basis encoding and then applying a random unitary matrix to the qubits. After the statevector of the quantum circuits are obtained, the relu function is applied to the state.}
\label{fig:kernel_relu}
\end{figure}
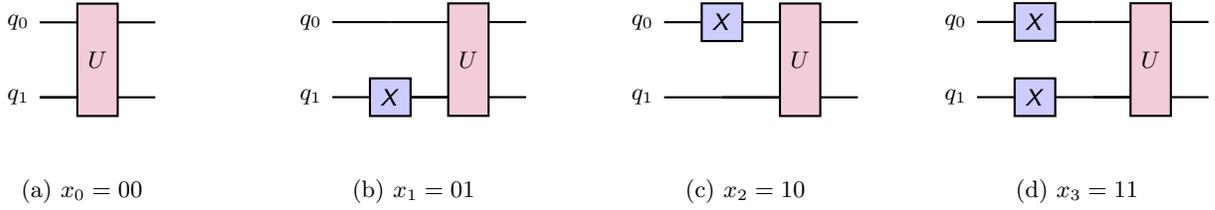

To generate the random relu transform, we construct the quantum circuits that encode this boolean dataset. Figure \ref{fig:kernel_relu} shows the four quantum circuits to encode $\vec{x} = (x_0, x_1, x_2, x_3)$.

To summarise, the random relu transform includes the following steps: 
\begin{enumerate}
    \item Construct a quantum circuit that uses basis encoding to encode the data
    \item Apply a random unitary matrix
    \item Obtain the statevector from the encoding quantum circuit
    \item Apply the relu function to the statevector
    \item Normalise the resulting statevector
\end{enumerate}

From this final statevector, we then calculate the entries of the kernel matrix. For the random relu transform, the matrix becomes that shown in Figure \ref{fig:matrix_e3}.

\subsubsection{Amplitude encoding kernel matrix}
\label{sec:amplitude_encoding}
Amplitude encoding encodes the data into the amplitudes of the quantum state, so that our quantum state becomes $\ket{\psi} = \frac{1}{\norm{x_p}} \sum_{i=0}^{n-1} x_{p_i} \ket{i}$ where our input data is $x_p = x_{p_{n-1}} x_{p_{n-2}} \cdots x_{p_0}$ and $x_{p_i}$ is the $i^{th}$ element of $x_p$ and $x_p$ is the $p^{th}$ element of $\vec{x} = (x_0, x_1, x_2, x_3)$ and $\ket{i}$ is the $i^{th}$ computational basis state. 

Given that the statevector of $n$ qubits contains $2^n$ amplitudes, we only need $\log_{2}(n)$ qubits to encode data with length $n$. When $n = 2$, this means we only need one qubit as $\log_{2}(2) = 1$. 

When $n = 2$, we can encode the data into one qubit as $x_{p_0} \cdot \ket{0} + x_{p_1} \cdot \ket{1}$. When $p = 1$, our input data is $x_1 = 01$. Therefore our statevector would be $ 0 
\cdot \ket{0} + 1 \cdot \ket{1} = \ket{1}$ as $x_{1_0} = 0$ and $x_{1_1} = 1$. When $p = 2$, our input data is $x_2 = 10$ and our statevector would be $ 1 \cdot \ket{0} + 0 \cdot \ket{1} = \ket{0}$. When $p = 3$, our input data is $x_3 = 11$ and our statevector would be $ 1 \cdot \ket{0} + 1 \cdot \ket{1}$. This is an invalid statevector so we need to normalise the input data. Therefore, when $p = 3$, our input data is $x_3 = \frac{1}{\sqrt{2}} \frac{1}{\sqrt{2}}$ and our statevector would be $ \frac{1}{\sqrt{2}} \cdot \ket{0} + \frac{1}{\sqrt{2}} \cdot \ket{1}$. When $p = 0$, our input data is $x_0 = 00$, but this would be equivalent to the statevector $ 0 \cdot \ket{0} + 0 \cdot \ket{1}$, which is invalid. Therefore, we can not encode the all 0s boolean input data with amplitude encoding. As a result, with amplitude encoding, we can only have $2^{n} - 1$ input datapoints instead of $2^{n}$. This means the kernel matrix for it would be $(2^n-1 \times 2^n-1)$ instead of $(2^n \times 2^n)$. 

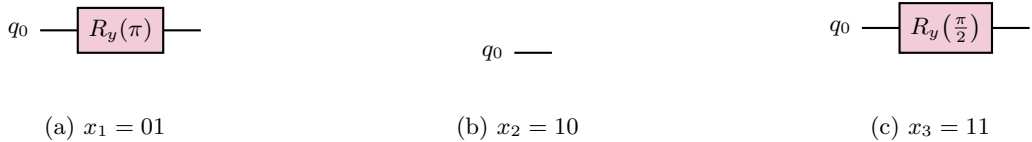
\begin{figure}[h]
    \centering
    \captionsetup[subfigure]{justification=centering}
     \begin{subfigure}[b]{0.3\textwidth}
     \centering
        \adjustbox{scale=1}{%
         \begin{tikzcd}
            \lstick{$q_0$} & \gate{R_y(\pi)} & \qw \\
        \end{tikzcd}
        }
         \caption{$x_1 = 01$}
         \label{fig:amp_x1}
     \end{subfigure}
     \begin{subfigure}[b]{0.3\textwidth}
     \centering
        \adjustbox{scale=1}{%
         \begin{tikzcd}
            \lstick{$q_0$} & \qw \\
        \end{tikzcd}
        }
         \caption{$x_2 = 10$}
         \label{fig:amp_x2}
     \end{subfigure}
     \begin{subfigure}[b]{0.3\textwidth}
     \centering
        \adjustbox{scale=1}{%
         \begin{tikzcd}
            \lstick{$q_0$} & \gate{R_y\big(\frac{\pi}{2}\big)} & \qw \\
        \end{tikzcd}
        }
         \caption{$x_3 = 11$}
         \label{fig:amp_x3}
     \end{subfigure}
\caption{\textbf{Quantum circuits for amplitude encoding:} these quantum circuits prepare the statevectors in which the amplitudes are equivalent to the input data. (a) we encode the data $x_1 = 01$, so our statevector needs to be $0 \cdot \ket{0} + 1 \cdot \ket{1}  = \ket{1}$. To get this statevector, we apply the $R_y(\pi)$ gate to the qubit. $R_y(\theta) = \begin{psmallmatrix}
    \cos(\theta/2) & -\sin(\theta/2) \\ 
    \sin(\theta/2) & \cos(\theta/2)
\end{psmallmatrix}$. Therefore, $R_y(\pi) = \begin{psmallmatrix}
    \cos(\pi/2) & -\sin(\pi/2) \\ 
    \sin(\pi/2) & \cos(\pi/2)
\end{psmallmatrix} = \begin{psmallmatrix}
    0 & -1 \\
    1 & 0
\end{psmallmatrix}$. When we apply $R_y(\pi)$ to the qubit, we obtain $\begin{psmallmatrix}
    0 & -1 \\
    1 & 0
\end{psmallmatrix} 
\begin{psmallmatrix}
    1  \\
    0 
\end{psmallmatrix} = 
\begin{psmallmatrix}
    0  \\
    1
\end{psmallmatrix} = \ket{1}$. Therefore our final statevector is $\ket{1}$ which is equivalent to encoding the data $x_1 = 01$ into the state. (b) we encode the data $x_2 = 10$, so our statevector needs to be $1 \cdot \ket{0} + 0 \cdot \ket{1} = \ket{0}$. Given that the qubits are initialised to $\ket{0}$, we do not need to apply any quantum gates to obtain this statevector. (c) the data is $x_3 = 11$, which we normalise to $x_3 = \frac{1}{\sqrt{2}} \frac{1}{\sqrt{2}}$ so our statevector needs to be $\frac{1}{\sqrt{2}} \cdot \ket{0} + \frac{1}{\sqrt{2}} \cdot \ket{1}$. To get this statevector we apply the $R_y(\frac{\pi}{2})$ gate to the qubit. $R_y(\frac{\pi}{2}) = \begin{psmallmatrix}
    \cos(\pi/4) & -\sin(\pi/4) \\ 
    \sin(\pi/4) & \cos(\pi/4)
\end{psmallmatrix} = \begin{psmallmatrix}
    \frac{1}{\sqrt{2}} & -\frac{1}{\sqrt{2}} \\
    \frac{1}{\sqrt{2}} & \frac{1}{\sqrt{2}}
\end{psmallmatrix}$. When we apply $R_y(\frac{\pi}{2})$ to the qubit, we obtain $\begin{psmallmatrix}
    \frac{1}{\sqrt{2}} & -\frac{1}{\sqrt{2}} \\
    \frac{1}{\sqrt{2}} & \frac{1}{\sqrt{2}}
\end{psmallmatrix}
\begin{psmallmatrix}
    1  \\
    0 
\end{psmallmatrix} = 
\begin{psmallmatrix}
    \frac{1}{\sqrt{2}}  \\
    \frac{1}{\sqrt{2}}
\end{psmallmatrix} = \frac{1}{\sqrt{2}} \ket{0} + \frac{1}{\sqrt{2}} \ket{1}$. Therefore this encodes $x_3 = 11$ into the state.
}
\label{fig:kernel_amp}
\end{figure}

To generate the amplitude encoding, we construct the quantum circuits that encode this boolean dataset. Figure \ref{fig:kernel_amp} shows the four quantum circuits to encode $\vec{x} = (x_0, x_1, x_2, x_3)$.

After obtaining the statevector from the encoding quantum circuits, we calculate the entries of the kernel matrix using the method described in the section for basis encoding (Appendix \ref{sec:basis}). For amplitude encoding, the matrix becomes that shown in Figure \ref{fig:matrix_e4}.
\FloatBarrier
\end{appendices}

\end{document}